\documentclass[10pt, a4paper]{article}
\usepackage{amsmath,amssymb}

\DeclareMathOperator\ric{Ric}
\usepackage{empheq}
\usepackage[bottom]{footmisc}
\usepackage{graphicx}
\usepackage{xcolor}
\usepackage{pgf,tikz}
\usepackage{float}
\usepackage{amsthm}
\usepackage{amsfonts}
\usepackage{amsthm}
\usepackage{amsmath}
\usepackage{amscd}
\usepackage{soul}
\usepackage{enumerate}
\usepackage{csquotes}
\usepackage{fixmath}
\newtheorem{theorem}{Theorem}[section]
\newtheorem*{theorem*}{Theorem}

\newtheorem*{definition*}{Definition}
\newtheorem{lemma}[theorem]{Lemma}
\newtheorem*{lemma*}{Lemma}
\newtheorem{corollary}[theorem]{Corollary}
\newtheorem{prop}[theorem]{Proposition}
\newtheorem*{prop*}{Proposition}
\newtheorem{remark}[theorem]{Remark}
\newtheorem{claim}[theorem]{Claim}
\newtheorem*{claim*}{Claim}

\graphicspath{ {Writings/} }
\usepackage[margin=1in]{geometry}

\usepackage[affil-it]{authblk}
\usepackage[linktoc=all]{hyperref}
\hypersetup{
    colorlinks = true,
    linkcolor = {black},
    citecolor = {black}
}

\usepackage{verbatim}
\usepackage{cleveref}

\begin{document}

\title{The Stable Trapping Phenomenon for Black Strings and Black Rings and its Obstructions on the Decay of Linear Waves}
\author{Gabriele Benomio\thanks{g.benomio16@imperial.ac.uk}}
\affil{\small Imperial College London, Department of Mathematics, \\ South Kensington Campus, London SW7 2AZ, United Kingdom }

\maketitle

\begin{abstract}
The geometry of solutions to the higher dimensional Einstein vacuum equations presents aspects that are absent in four dimensions, one of the most remarkable being the existence of stably trapped null geodesics in the exterior of asymptotically flat black holes. This paper investigates the stable trapping phenomenon for two families of higher dimensional black holes, namely black strings and black rings, and how this trapping structure is responsible for the slow decay of linear waves on their exterior. More precisely, we study decay properties for the energy of solutions to the scalar, linear wave equation $\Box_{g_{\textup{ring}}} \Psi=0$, where $g_{\textup{ring}}$ is the metric of a fixed black ring solution to the five-dimensional Einstein vacuum equations. For a class $\mathfrak{g}$ of black ring metrics, we prove a logarithmic lower bound for the uniform energy decay rate on the black ring exterior $(\mathcal{D},g_{\textup{ring}})$, with $g_{\textup{ring}}\in\mathfrak{g}$. The proof generalizes the perturbation argument and quasimode construction of Holzegel--Smulevici \cite{SharpLogHolz} to the case of a non-separable wave equation and crucially relies on the presence of stably trapped null geodesics on $\mathcal{D}$. As a by-product, the same logarithmic lower bound can be established for any five-dimensional black string.  

Our result is the first mathematically rigorous statement supporting the expectation that black rings are dynamically unstable to generic perturbations. In particular, we conjecture a new \textit{nonlinear} instability for five-dimensional black strings and thin black rings which is already present at the level of scalar perturbations and clearly differs from the mechanism driven by the well-known Gregory--Laflamme instability.
\end{abstract}

\tableofcontents

\section{Introduction}

The Einstein vacuum equations of general relativity with cosmological constant $\Lambda$,
\begin{equation} \label{EVE}
\ric(g)= \Lambda   g \, ,
\end{equation}
form, after gauge fixing, a system of \textit{nonlinear} partial differential equations, with a Lorentzian metric $g$ as the unknown. It is a remarkable fact that the problem of solving the Einstein equations can be reformulated as a locally well-posed initial value problem for quasilinear wave equations in the metric $g$, with initial data defined by suitable geometric quantities and satisfying specific constraints \cite{CB52, CB69}. The initial value formulation turns out to be the correct framework for the mathematically rigorous study of the Einstein equations. 

The mathematical question at the core of the present paper concerns the \textit{stability} of solutions to the Einstein vacuum equations, meaning how small perturbations of the initial data for a typically well-known solution evolve under equations \eqref{EVE}. 

The first major contribution to answer this question has been provided by Christodoulou and Klainerman in \cite{StabMink}, where the \textit{nonlinear} stability of Minkowski spacetime is proven (see also alternative proof in \cite{Lind_Rod_Stab_Mink}). 

In the context of black hole spacetimes, a natural stability notion to study is that of the \textit{exterior region}. In the case of zero cosmological constant, it has been conjectured that the exterior region of the two-parameter family of Kerr black holes $g_{a,M}$ is \textit{nonlinearly} stable for $a<M$.\footnote{On the other hand, extremal Kerr black holes ($a=M$) are affected by the so-called \textit{Aretakis instability} \cite{Aretakis_instab_Kerr, Aretakis_nonlinear_scalar_instab,  Lucietti_Reall_extreme_Kerr}, which suggests that these spacetimes should not be included in a natural formulation of the Kerr stability conjecture.} The conjecture asserts that any slightly perturbed initial data relative to some member of the Kerr family eventually settle down, in the exterior region, to a nearby (but possibly different) member of the Kerr family, where the evolution of such perturbed data follows the fully nonlinear system \eqref{EVE}. See the introduction of \cite{Scattering_Daf_Holz_Rod} for a more precise formulation of the conjecture. 

Much progress has been made towards a mathematical proof of the Kerr stability conjecture. Recent contributions include a proof of \textit{linear} stability of Schwarzschild spacetime ($a=0$) to gravitational perturbations \cite{LinStabSchw, Wang_Schwarz, Lin_Stab_Schw_wave_gauge} and its \textit{nonlinear} stability to polarized perturbations \cite{Klainerman_Schw}. A complete understanding of the problem would, more in general, shed some light on the dynamics of vacuum black holes settling down to Kerr (see, for instance, the role of the Kerr exterior stability in \cite{Daf_Luk_Stab_Kerr_Cauchy_Hor}).  

For the Einstein vacuum equations with positive cosmological constant, the \textit{nonlinear} stability of the corresponding de Sitter and slowly rotating Kerr--de Sitter spacetimes has been fully understood \cite{Friedrich_stab_dS, StabKerrdS}. In contrast, for negative cosmological constant, the maximally symmetric solution, Anti-de Sitter, has been conjectured to be dynamically \textit{unstable} \cite{Daf_Holz_Conj_AdS_instab, Daf_talk_AdS, Anderson_AdS} (see also numerical work \cite{Bizon_Rot_Instab_AdS} and recent progress in \cite{Moschidis_AdS}), the same being true for the Kerr-Anti-de Sitter family of black hole spacetimes (see Section 1.4 of \cite{Decay_KG_Kerr-AdS} and related work \cite{SharpLogHolz}, but also \cite{Santos_Stab_AdS} for a stability perspective).

The question of stability of black hole exteriors can be extended to the higher dimensional Einstein equations, i.e. for a number of dimensions greater than four. Higher dimensional general relativity is a major area of research in gravitational physics. The flourishing of this subject has been mainly motivated by the study of unifying theories, such as sting theory, whose formulation requires a high number of dimensions. However, what is more intriguing for us is that higher dimensional general relativity is not just a formal extension of the four dimensional theory, but it presents novel mathematical features, the most striking being the failure of the rigidity and exterior stability properties of stationary black hole solutions that one expects in four dimensions. See reviews \cite{ReallRev, Horowitz_book} and references therein.

The crucially new aspects of the higher dimensional theory are already manifest in \textit{five} dimensions, for which a number of explicit families of black holes have been discovered \cite{Tang, MPBH, Emp_Reall_Gen_Weyl, EmpReallBR, Mishima_Ig_Ring, Figueras_2sph_ring, DoubleBR, LuciettiBlackLenses, BlackSaturn}. Indeed, the geometry and classification of solutions to five-dimensional Einstein vacuum equations is the better understood among all possible higher dimensions, and this is one of the reasons why this paper focuses on dimension five.

For what will be later discussed, we are particularly interested in three of these families of five-dimensional black holes. The first originates from the somehow straightforward observation that one can produce new solutions by adding a flat direction to a four dimensional black hole \cite{Greg_Laflamme_Hypercyl_BH}. In this way, \textit{black strings} of the form $\textup{Schw}_4\times\mathbb{R}$ and $\textup{Kerr}_4\times\mathbb{R}$ can be constructed, where points along the new extended direction are usually periodically identified. Black strings suffer from a linear instability to gravitational perturbations, the so-called \textit{Gregory--Laflamme instability} \cite{GLinstability}. The nonlinear dynamics of such instability has been largely investigated numerically and seems to suggest a possible violation of the weak cosmic censorship \cite{Lehner_Pret_Instab_String}.

While black string spacetimes are not asymptotically flat, there exists a number of independent families of stationary, asymptotically flat black hole solutions in five dimensions, including \textit{Myers--Perry black holes} \cite {MPBH} and \textit{black rings} \cite{EmpReallBR, DoubleBR}. The former are a five-dimensional generalization of the Kerr family to black holes with two planes of rotation and event horizon topology $S^3$, while black rings form a completely new family, with no analogue in four dimensions. 

The most remarkable property of black rings is the non-spherical horizon topology $S^1\times S^2$. Black rings can be \textit{singly-spinning}, with rotation along the $S^1$, or \textit{doubly-spinning}, with rotation along both the $S^1$ and the $S^2$. The mere existence of these additional solutions seriously questions the possibility to recover any rigidity result in five dimensions (see review \cite{RingsRev}). From the stability point of view, both heuristic and numerical works show that every member of this family is affected by linear instabilities to gravitational perturbations \cite{Figuerasrings, Santosrings}. In particular, rings whose radius is much greater than the $S^2$-radius at the event horizon, often called \textit{thin black rings}, resemble black strings and suffer from Gregory--Laflamme instabilities \cite{Elv_Emp_String, Hovd_Myers_String_Ring_GL}. The nonlinear, numerical evolution of such instabilities strongly suggests that \textit{black rings are nonlinearly unstable to generic perturbations} and possibly lead to a violation of the weak cosmic censorship conjecture \cite{FiguerasEndpointRings}.

\textbf{The aim of this paper is to provide a first mathematically rigorous result in the context of the stability problem for black rings.}

\subsection{Scalar, linear wave equation}

Some of the main difficulties in the mathematical study of the Einstein equations originate from the nonlinear, and together tensorial, character of the equations. Given the hyperbolicity of \eqref{EVE}, the scalar, linear wave equation
\begin{equation} \label{WE_Intro}
\Box _g \Psi=0
\end{equation}
on a \textit{fixed} Lorentzian background manifold $(\mathcal{D},g)$ is a preliminary (and, in principle, simpler) toy model to consider. Typically, one chooses the spacetime $(\mathcal{D},g)$ whose nonlinear stability is under investigation and studies properties of solutions to \eqref{WE_Intro}, such as \textit{uniform boundedness} and \textit{decay in time} of the \textit{energy} associated to $\Psi$.\footnote{One is ultimately interested in proving pointwise decay for $\Psi$. The methods of proving pointwise decay by means of estimates involving positive definite, $L^2$-based quantities (energies) are frequently called \textit{energy methods}. These provide a robust framework to deal with both linear and nonlinear problems.} Uniform boundedness and sufficiently strong decay of the energy at the linear level are typically essential to hope for nonlinear stability of $(\mathcal{D},g)$.

In this regard, we are ultimately interested in \textit{uniform} energy statements of the form 
\begin{displayquote}
\textit{There exists a universal constant $C>0$ (independent of time) such that, for any given smooth, compactly supported initial data\footnote{This is the class of initial data that we consider throughout the introduction, even if not specified.} at time $t=0$, solutions to \eqref{WE_Intro} satisfy
\begin{align}
&E[\Psi](t) \leq C  E[\Psi](0)  & \text{(uniform boundedness)}  \nonumber \\
&E[\Psi](t) \leq C  \delta(t) E[\Psi](0) & \text{(decay)}    \label{Unif_En_decay}
\end{align}
for any $t>0$, where $\delta(t)\rightarrow 0$ as $t\rightarrow\infty$.}
\end{displayquote}
The \textit{energy} $E[\Psi](t)$ is a positive definite quantity of the form  
\begin{equation*}
E[\Psi](t)\sim \sum_{|\alpha|=1}\int _{\Sigma_t}|\partial^{\alpha}\Psi|^2(t,x)\, dx
\end{equation*}
for some suitable spacelike hypersurface $\Sigma_t $.\footnote{The reader should refer to Section \ref{Notation_section} for multi-index notation and the meaning of $\sim$. In Section \ref{sect_metrics} we define hypersurfaces $\Sigma_t $, while in Section \ref{sect_energy_currents} we give a rigorous definition of the energy.} A \textit{k-th (higher) order energy} involves higher derivatives of the solution
\begin{equation*}
E_{k}[\Psi](t)\sim \sum_{1\leq |\alpha| \leq k} \int _{\Sigma_t}|\partial^{\alpha}\Psi|^2(t,x)\, dx \, .
\end{equation*}
We sometimes refer to \textit{local} energy statements, for which the \textit{local energy} of the solution
\begin{equation*} 
E_{loc}[\Psi](t) \sim \sum_{|\alpha|=1}\int _{\Sigma_t\cap\Omega}|\partial^{\alpha}\Psi|^2(t,x)\, dx
\end{equation*}
is considered, with $\Sigma_t\cap\Omega$ some bounded, non-empty set. We will also denote it as $E_{\Omega}[\Psi](t)$. The function $\delta(t)$ appearing in \eqref{Unif_En_decay} is the \textit{uniform energy decay rate}.

It is crucial that our energy statements are \textit{uniform}, in the sense that they hold for \textit{any solution} to the wave equation and \textit{for all times} $t>0$. If no uniform energy decay with decay rate faster than $\delta(t)$ can possibly hold, we say that the uniform decay rate is \textit{sharp}.

Of particular interest for us is the case when $(\mathcal{D},g)$ \textit{is the exterior region of a black ring spacetime}. Before moving to that, we first briefly outline the state of the art for linear waves on some other black hole exteriors of relevance for our discussion. 

A series of works \cite{Kay_Wald_Schw, Blue_Sterbenz_Schw, DecaySchw, LukImprSchw, BoundSlowKerr, DecaySlowKerr, LukImprKerr, AndBlue, DecayTataru, TataruTo}, culminating in \cite{FullKerr}, has shown that the energy of any solution to \eqref{WE_Intro} (and of all its higher order derivatives) is uniformly bounded and, in fact, decays (fast) \textit{polynomially} in time on the exterior of sub-extremal Kerr black holes, with a higher order energy on the right hand side of \eqref{Unif_En_decay}.\footnote{For reasons that we shall discuss later, uniform energy decay on black hole exteriors cannot have the same right hand side of \eqref{Unif_En_decay}. When we refer to decay on such spacetimes, we always implicitly consider an inequality like \eqref{Unif_En_decay}, but with a higher order energy on the right hand side.} This remains true up to the event horizon, in contrast with extremal Kerr black holes, for which some derivatives do not decay along the horizon \cite{Aretakis_instab_Kerr}. 

For Kerr-AdS black holes, the energy of solutions is uniformly bounded and decays \textit{logarithmically} in time when the black hole parameters satisfy certain bounds \cite{Holzegel_KG_Slow_KerrAdS, Decay_KG_Kerr-AdS}. In fact, such uniform energy decay rate is \textit{sharp} \cite{SharpLogHolz}. In some other parameter regime, superradiance allows linear waves to grow exponentially in time \cite{Dold_KerrAdS}. 

For higher dimensional spacetimes, works \cite{Laul_Metcalfe_Higher_Schw, SchlueSchw} show polynomial decay on Schwarzschild black holes in any dimension, while \cite{MorawetzMP} proves integrated local energy decay on  five-dimensional Myers--Perry black holes with small angular momenta. 

The wave equation on the exterior of black strings has been mainly investigated from the numerical point of view. For Schwarzschild black strings, the numerical analysis in \cite{Cross-section} suggests the absence of growing mode solutions, even when the black string gets boosted. The expectation is different for Kerr black strings, for which superradiant instabilities have been numerically shown \cite{Cardoso_New_Instab_String, Cardoso_Instab_String, Rosa_Black_String_Bomb}.

Scalar perturbations of black rings will be the main topic of the present paper. To the best of the author's knowledge, there is no rigorous study of the wave equation on these spacetimes in literature (apart from an application of a general result by Moschidis \cite{LogDecay}, we will come back to this later). In view of the fact that the near-horizon geometry of large radius, thin black rings approximates that of a boosted black string, heuristic arguments suggest a correlation between properties of linear waves on black strings and on this class of black rings \cite{Cross-section, Inst_DS_BR}. In this respect, one could expect uniform boundedness and decay on \textit{singly-spinning} thin black rings, which approximate \textit{Schwarzschild} boosted black strings, and possible instabilities on \textit{doubly-spinning} thin black rings, because the geometry is now close to the one of a \textit{Kerr} boosted black string. Our paper exploits, to some extent, this heuristic. 

It is important to note that the problem of studying the linear wave equation on a spacetime is different (but, of course, related) from investigating the linear stability of such spacetime to gravitational perturbations. In fact, the presence of an instability at the level of \textit{gravitational} perturbations, such as the Gregory--Laflamme growing modes for thin black rings, might be an effect that is not manifest for \textit{scalar} perturbations. In other words, a gravitational instability does not give, in principle, much information about boundedness or decay of scalar waves. This in part motivates our analysis.

\subsection{Stable trapping and slow energy decay}

One of the various geometric aspects of $(\mathcal{D},g)$ interacting with wave propagation is \textit{null geodesic trapping}. High frequency solutions (or high frequency components of solutions) to the wave equation approximately propagate along null geodesics for very long time. In the presence of trapped null geodesics, this causes an obstruction to decay. A manifestation of this obstruction is that uniform energy decay estimates of the form \eqref{Unif_En_decay} cannot hold \cite{Ralston_trap, GaussBeam}. 

Furthermore, the \textit{structure} of trapping is crucial when one proves energy decay. Trapping that occurs at the Schwarzschild photon sphere $r=3M$ and in the  Kerr exterior region is \textit{unstable}. Even though the high frequency components of a solution are trapped, the good structure of trapping allows an \textit{integrated} local energy decay estimate of the form
\begin{equation*}
\int_0^{\infty} \chi E_{\text{loc}}[\Psi](s)\, ds \lesssim  E[\Psi](0) \, ,
\end{equation*}
which degenerates at the trapping set, on which $\chi=0$.\footnote{See Section \ref{Notation_section} for the meaning of $\lesssim$.}\textsuperscript{,}\footnote{Here we are omitting another important property of Kerr black holes, namely the fact that superradiant components of solutions are not trapped. Together with the instability of trapping, this is a fundamental ingredient for the proof in \cite{FullKerr}.} This degeneracy can be removed by losing a derivative at initial time
\begin{equation}  \label{ILED_Intro}
\int_0^{\infty}  E_{\text{loc}}[\Psi](s)\, ds \lesssim  E_2[\Psi](0) \, .
\end{equation}

In contrast, the presence of \textit{stable} trapping in Kerr-AdS spacetimes prevents from proving integrated energy decay for high frequencies, as first shown in \cite{Decay_KG_Kerr-AdS}. This difference is what originates the different uniform energy decay rate: While integrated local energy decay \eqref{ILED_Intro} permits to establish \textit{polynomial} decay on Kerr black holes \cite{FullKerr, new_method} 
\begin{equation}  \label{En_decay_Kerr_intro}
E[\Psi](t) \lesssim \frac{1}{t^2} \,  E_{3,w}[\Psi](0)
\end{equation}
for a suitable choice of spacelike hypersurfaces $\Sigma_t$,\footnote{The energy on the right hand side contains some weights independent of $t$. If one restricts to compactly supported initial data, then those weights can be neglected. To prove decay, hypersurfaces $\Sigma_t$ need to approach future null infinity.} one can only prove \textit{logarithmic} decay for Kerr-AdS spacetimes \cite{Decay_KG_Kerr-AdS} 
\begin{equation*}
E[\Psi](t) \lesssim \frac{1}{[\log (2+t)]^2} \, E_{2}[\Psi](0) \, .
\end{equation*}
Logarithmic decay is usually regarded as \textit{slow} decay, the reason being the comparison with polynomial rates as the one appearing in \eqref{En_decay_Kerr_intro} and their application to nonlinear problems (see Section \ref{sect_nonlinear_instab}). 

From a more general point of view, these results suggest that a bad trapping structure on black hole exteriors generically leads to slow uniform energy decay rates. Furthermore, one might wonder whether arbitrarily bad trapping can produce arbitrarily slow decay or, alternatively, whether there exists a universal minimal decay rate that linear waves always satisfy. This question has been answered in the context of obstacle problems for the wave equation on Minkowski space \cite{Burq_obst}, for which the local energy decays logarithmically in time without any assumption on the geometry of the trapping obstacle. Partly motivated by this result is the conjecture that, provided uniform boundedness holds, the same universal minimal decay rate should hold for waves on the exterior of any stationary black hole, again without requiring any good structure for trapping. This has been proven for a general class of stationary, asymptotically flat (black hole) spacetimes in \cite{KeirLogLog} and \cite{LogDecay}. In particular, \textit{the class of spacetimes considered in \cite{LogDecay} includes all black rings}. Works \cite{Eperon_Reall_Santos_Microstate, MicroKeir} show that for some non-black hole geometries this expectation fails to be true.

A separate, but equally relevant question is whether the presence of stable trapping further imposes that uniform decay rates are \textit{sharp}. This is indeed the case for logarithmic decay on Kerr-AdS black holes \cite{SharpLogHolz}, on some static, spherically symmetric spacetimes considered in \cite{KeirLogLog} and for the example constructed in \cite{LogDecay} (which is a modification of a counterexample of Rodnianski and Tao in \cite{Rod_Tao}). Sharpness of the decay rate has also been shown for the Dirac equation on Schwarzschild-AdS \cite{Idelon-Riton_Lower_Bound_Dirac} and on certain microstate geometries \cite{MicroKeir}. 

The technique employed in all these works consists in constructing approximate solutions to the wave equation, called \textit{quasimodes}, which are localized along stably trapped null geodesics. While a proof of decay usually involves the \textit{global} structure of a spacetime, quasimode constructions only rely on the \textit{local} geometry in proximity of the trapping set.    
 
Before presenting the main ideas behind this technique, let us remark that uniform energy decay rates are crucial for nonlinear applications. (Fast) polynomial decay gives hope to be able to prove stability results for nonlinear problems, whereas logarithmic decay is not enough in this regard and strongly suggests nonlinear instabilities. See Section \ref{sect_nonlinear_instab}.

\subsection{Quasimodes and lower bound on the uniform energy decay rate} \label{Quasimodes_to_lower_bound_intro}

We give an overview of how quasimodes can be used to contradict uniform energy decay statements for solutions to the wave equation.\footnote{The presentation here is far from being rigorous, we leave technical details for Section \ref{sect_proof_main_th}. The exposition is very much inspired by \cite{NI_Smulevici}.}

Consider some stationary, $(d+1)$-dimensional Lorentzian manifold $(\mathcal{D},g)$ and a coordinate system $(t,x)$, with $x=(x^1, \ldots, x^d)$.  An informal definition of quasimodes can be the following:
\begin{displayquote}
\textit{We define quasimodes as complex-valued, time-periodic functions of the form
\begin{equation*}
\Psi_{m}(t,x)=e^{i\omega_{m} t} u_{m} (x),
\end{equation*}
with frequency parameters $\omega_{m}\in\mathbb{R}$ and $m\in\mathbb{Z}$, such that
\begin{enumerate}[(i)]
\item $\Psi_{m}$ is in some energy space,  
\item $\Psi_{m}$ is localized in space, i.e. $u_{m}$ is compactly supported, 
\item $\Psi_{m}$ is localized in frequency, i.e. $\left\lVert \partial^2\Psi_{m} \right\rVert \sim \omega_{m}^2 \left\lVert \Psi_{m} \right\rVert$ in some energy norm,
\item $\Psi_{m}$ is an approximate solution to the wave equation.
\end{enumerate}}
\end{displayquote}
In general, quasimodes do not satisfy the wave equation \eqref{WE_Intro}. In fact,
\begin{equation*}
\Box_g\Psi_{m}=\mathfrak{Err}_m(\Psi_{m}) \, ,
\end{equation*}
where $\mathfrak{Err}_m(\Psi_{m})$ is the \textit{error}. Functions $\Psi_{m}$ are \textit{approximate solutions to the wave equation} when $\mathfrak{Err}_m$ is small, in a sense that of course needs to be specified. In particular, we suppose that 
\begin{equation*}
\left\lVert  \mathfrak{Err}_m \right\rVert \rightarrow 0 \quad \text{as} \quad m\rightarrow\infty 
\end{equation*}
(again in some energy norm), i.e. $\Psi_{m}$ is an approximate solution to the wave equation for high frequency $m$. Typically, one is also interested in the \textit{rate} in $m$ at which the error tends to zero. So let us further assume that
\begin{equation} \label{exp_approx_WE_intro}
\left\lVert  \mathfrak{Err}_m \right\rVert = \mathcal{O}\left(e^{-Cm}\right)
\end{equation}
for $m$ large, where $C>0$ is a constant independent of $m$. 

To see what quasimodes can tell about energy decay of solutions to the wave equation, consider the initial value problem
\begin{equation} \label{WE_quasimdoes_intro}
\begin{cases}
\Box_g\Psi^H_{m}=0  \\
\Psi^H_{m}|_{t=0}=\Psi_{m}|_{t=0}  \\
\partial_t\Psi^H_{m}|_{t=0}=\partial_t\Psi_{m}|_{t=0}
\end{cases} \, ,
\end{equation}
where $\Psi^H_{m}$ denotes a solution to the \textit{homogeneous} wave equation (in contrast with $\Psi_{m}$, that solves the wave equation with inhomogeneous term $\mathfrak{Err}_m$). Initial data for $\Psi^H_{m}$ are the quasimodes. Duhamel's formula\footnote{Here we use the Duhamel's formula that holds for Minkowski space. For a general Lorentzian metric $g$, one needs to correct the formula with some factors that we omit.} gives
\begin{equation*}
\Psi_{m}(t)=\Psi^H_{m}(t)+\int^t_0 \xi (s;t,x)\, ds \, ,
\end{equation*}
with $\xi(t,x)$ solution of
\begin{equation}  \label{Duhamel_intro}
\begin{cases}
\Box_g\xi=0  \\
\xi|_{t=s}=0  \\
\partial_t\xi|_{t=s}=\mathfrak{Err}_m(\Psi_m)|_{t=s}
\end{cases} \, .
\end{equation}
From Duhamel's formula, one has
\begin{equation*}
\left\lVert \Psi^H_{m} -\Psi_{m}  \right\rVert \leq t \sup_{s\in [0,t]} \left\lVert \xi(s)  \right\rVert \, .
\end{equation*}
Suppose now that \textit{the energy of any solution to the homogeneous wave equation is uniformly bounded by the initial data for all times}, then  
\begin{align*}
\left\lVert \Psi^H_{m} -\Psi_{m}  \right\rVert &\leq C t \left\lVert \mathfrak{Err}_m  \right\rVert  \quad \quad \quad  \text{(from \eqref{Duhamel_intro})}  \\
&\leq C t e^{-C m}
\end{align*}
with $m$ sufficiently large and $C>0$ universal constant. For $t\leq e^{C m}/2C$, the reverse triangle inequality gives 
\begin{align*}
\left\lVert \Psi^H_{m}  \right\rVert(t) \geq \frac{1}{2} \left\lVert \Psi_{m}  \right\rVert(t) &= \frac{1}{2} \left\lVert \Psi_{m}  \right\rVert(0)  \\
&= \frac{1}{2} \left\lVert \Psi^H_{m}  \right\rVert(0) \, ,
\end{align*}
where the first equality holds by time-periodicity of quasimodes and the second from \eqref{WE_quasimdoes_intro}. We conclude
\begin{equation}  \label{ineq1_intro}
\left\lVert \Psi^H_{m}  \right\rVert(t) \geq \frac{1}{2} \left\lVert \Psi^H_{m}  \right\rVert(0) \quad   \text{for $t\leq \frac{e^{C m}}{2C}$ and $m$ large} \, .
\end{equation}

If 
\begin{equation*}
\left\lVert \,\cdot\, \right\rVert=E_{\text{loc}}[\,\cdot\,]
\end{equation*}
with local energy over some bounded set containing the spatial support of any $\Psi_m|_{t=0}$, then \eqref{ineq1_intro} gives
\begin{equation}  \label{ineq2_intro}
E_{\text{loc}}\left[\Psi^H_{m}\right](t) \geq \frac{1}{2} E\left[\Psi^H_{m}\right](0)  \quad   \text{for $t\leq \frac{e^{C m}}{2C}$ and $m$ large} \, ,
\end{equation}
where $E_{\text{loc}}\left[\Psi^H_{m}\right](0)=E\left[\Psi^H_{m}\right](0)$ by the localization in space of the quasimodes. 

A sequence 
\begin{equation}  \label{sequence_quasimodes}
\left\lbrace\left(\Psi^H_{m},t_{m}\right)\right\rbrace_{m\in\mathbb{Z}} 
\end{equation}
with $t_{m}=e^{C m}/2C$ contradicts a uniform energy decay statement of the form
\begin{displayquote}  
\textit{There exists a universal constant $C>0$ (independent of time) such that all smooth, compactly supported solutions $\Psi$ to the wave equation $\Box_g\Psi=0$ satisfy
\begin{equation*}
E_{\text{loc}}\left[\Psi\right](t) \leq C \delta(t) E\left[\Psi\right](0)
\end{equation*}
for all $t>0$, with $\delta(t)\rightarrow 0$ as $t\rightarrow\infty$.} 
\end{displayquote}

\begin{proof}
To see this, note that the proposition above implies that for any $\varepsilon>0$, there exists $T>0$ such that $E_{\text{loc}}\left[\Psi\right](t)\leq\varepsilon E\left[\Psi\right](0)$ for all $t\geq T$, this being true for any solution $\Psi$. We choose $\varepsilon=1/4$. We then choose $m_*$ sufficiently large, say $e^{C m_*}/2C>2T$. In view of \eqref{ineq2_intro}, there exists a solution $\Psi^H_{m_*}$ to the wave equation such that $E_{\text{loc}}\left[\Psi^H_{m_*}\right](t) \geq \frac{1}{2} E\left[\Psi^H_{m_*}\right](0)$ for $T\leq t\leq 2T$. This leads to a contradiction.
\end{proof}

Using the frequency localization of $\Psi^H_{m}(0)$, i.e. the frequency localization of quasimodes, and system \eqref{WE_quasimdoes_intro}, inequality \eqref{ineq2_intro} gives the following
\begin{equation*}
E_{\text{loc}}\left[\Psi^H_{m}\right](t) \gtrsim \frac{1}{m^2} \, E_2\left[\Psi^H_{m}\right](0)  \quad   \text{for $t\leq \frac{e^{C m}}{2C}$ and $m$ large} \, .
\end{equation*}
Therefore, sequence \eqref{sequence_quasimodes} proves that a uniform energy decay statement of the form
\begin{displayquote}  
\textit{There exists a universal constant $C>0$ (independent of time) such that all smooth, compactly supported solutions $\Psi$ to the wave equation $\Box_g\Psi=0$ satisfy
\begin{equation}  \label{eq_prop_2_intro}
E_{\text{loc}}\left[\Psi\right](t) \leq \frac{C}{[\log(2+ t)]^2}\, E_2\left[\Psi\right](0)
\end{equation}
for all $t>0$.}
\end{displayquote}
\noindent has to be \textit{sharp}, in the sense that sequence \eqref{sequence_quasimodes} disproves any uniform energy decay statement of the form \eqref{eq_prop_2_intro} (with the same loss of derivatives) with faster uniform energy decay rate.

After this brief discussion, the reader should be convinced that lower bounds for uniform energy decay rates can be proven rather easily once quasimodes are constructed. The particular bound that one is able to produce depends on the rate at which quasimodes approximate solutions to the wave equation. In fact, the hard part of the argument is proving that quasimodes with suitably small error do actually exist.

In \cite{Ralston_trap}, quasimodes are constructed to produce lower bounds in the context of the obstacle problem on Minkowski space. The major insight of \cite{SharpLogHolz} is that such construction is still possible on Kerr-AdS spacetimes and, moreover, that stable trapping lets quasimodes satisfy \eqref{exp_approx_WE_intro}. Subsequent works proving lower bounds essentially follow the same ideas of \cite{SharpLogHolz}. Even though some additional difficulties come into play, this is also the spirit of the proof of our main result.

\subsection{The main results of the paper}

Our main theorem produces a \textit{logarithmic} lower bound for the uniform energy decay rate of solutions to the wave equation on black rings. In particular, the theorem concerns decay on the black ring exterior $(\mathcal{D},g_{\textup{ring}})$, with $g_{\textup{ring}}$ belonging to some class $\mathfrak{g}$ of black ring metrics. 

We give here a first informal version of the result. The complete statement can be found in Section \ref{section_main_th_formal}. The existence of quasimodes with suitably small error is a crucial step of the proof, so we state it as an independent theorem.

\begin{theorem}[\textbf{Quasimodes for Black Rings}] \label{quasimodes_ring_intro}
Let $\mathfrak{g}$ be a class of black ring metrics $g_{\textup{ring}}$, as defined in \eqref{def_class_g}, and consider the black ring exterior region $(\mathcal{D},g_{\textup{ring}})$, with $g_{\textup{ring}}\in \mathfrak{g}$. Let $(t,r_*,\theta_*,\phi,\psi)$ be the coordinate system on $(\mathcal{D},g_{\textup{ring}})$ introduced in Section \ref{sect_black_ring}. Then, for $\delta$ sufficiently small, there exist non-zero functions $\Psi_m: \mathcal{D}\rightarrow\mathbb{C}$ such that
\begin{enumerate}[(i)]
\item $\Psi_m\in H^k(\Sigma_{t^*})$ for any $k\geq 0$,
\item $\Psi_m(t,r_*,\theta_*,\phi,\psi)=e^{-i\,\omega_m t}e^{i(m\,\phi+\hat{J}\,\psi)}\varphi_m (r_*,\theta_*)$, with $\omega_m\in\mathbb{R}$ and $m,\hat{J}\in\mathbb{Z}$,
\item inequality $c\leq \omega_m^2/m^2 \leq C$ holds, with constants $c,C>0$ independent of $m$,
\item for any $k\geq 0$, there exists a constant $C_k>0$, independent of $m$, such that 
\begin{equation*}
\left\lVert \Box_{g_{\textup{ring}}} \Psi_m  \right\rVert_{H^k(\Sigma_{t^*})}  \leq  C_k e^{-C_k \, m}  \left\lVert  \Psi_m  \right\rVert_{L^2(\Sigma_0)}
\end{equation*}
for all $t^*>0$,
\item the support of $\varphi_m (r_*,\theta_*)$ is contained in $\Omega\subset\mathcal{D}$, where $\Sigma_0\cap\Omega$ is a bounded, non-empty set and $\Phi_{t^*}(\Sigma_0\cap\Omega)=\Phi_{t^*}(\Sigma_0)\cap\Omega$ remains bounded and non-empty for all $t^*>0$, with $\Phi_{t^*}$ the one-parameter group of diffeomorphisms generated by the Killing vector field $\partial/\partial t^*$,
\item the support of $\mathfrak{Err}_m(\Psi_m):=\Box_{g_{\textup{ring}}} \Psi_m$ is contained in $\Omega_{\delta}\subset\Omega$, where 
\begin{equation*}
\Omega_{\delta}:=\left\lbrace x\in\Omega : \exists \,y\in\partial\Omega \textup{ such that } \textup{dist}((r_*(x),\theta_*(x)),(r_*(y),\theta_*(y)))\leq \delta \right\rbrace
\end{equation*}
with $\textup{dist}(\cdot,\cdot)$ the Euclidean distance,
\end{enumerate}  
with time coordinate $t^*$ and spacelike hypersurfaces $\Sigma_{t^*}$ defined in Section \ref{sect_foliation}.
\end{theorem}

\begin{theorem}[\textbf{Lower Bound for Uniform Energy Decay Rate, First Version}]  \label{Main_th_intro}
Let $\mathfrak{g}$ be a class of black ring metrics $g_{\textup{ring}}$, as defined in \eqref{def_class_g}. Consider smooth solutions $\Psi: \mathcal{D}\rightarrow\mathbb{C}$ to the scalar, linear wave equation
\begin{equation}   \label{WE_main_th_intro}
\Box_{g_{\textup{ring}}}\Psi =0  
\end{equation}
on the black ring exterior $(\mathcal{D},g_{\textup{ring}})$, with $g_{\textup{ring}}\in \mathfrak{g}$, and assume that a uniform boundedness statement (without loss of derivatives) for the energy of solutions to \eqref{WE_main_th_intro} holds. Then, there exists a universal constant $C>0$ (independent of time) such that 
\begin{equation*}
\limsup_{t^*\rightarrow +\infty}\,\,  \sup_{\Psi\in SC_0^{\infty}(\Sigma_0),\Psi\neq 0} \,\, [\log (2+t^*)]^2 \left(\frac{E_{\Omega}[\Psi](t^*) }{ E_2[\Psi](0) }\right) > C \, ,
\end{equation*}
where the supremum is taken over all smooth, non-zero solutions to the wave equation with compactly supported initial data on $\Sigma_0$. Furthermore, for any $k\in\mathbb{N}$, there exists a universal constant $C_k>0$ such that 
\begin{equation*}
\limsup_{t^*\rightarrow +\infty}\,\,  \sup_{\Psi\in SC_0^{\infty}(\Sigma_0),\Psi\neq 0} \,\, [\log (2+t^*)]^{2k} \left(\frac{E_{\Omega}[\Psi](t^*) }{ E_{k+1}[\Psi](0) }\right) > C_k \, .
\end{equation*}
The set $\Omega$ appearing in the local energy $E_{\Omega}[\Psi](t^*)$ is the one fixed in Theorem \ref{quasimodes_ring_intro}, with time coordinate $t^*$ and spacelike hypersurfaces $\Sigma_{t^*}$ defined in Section \ref{sect_foliation}.
\end{theorem}

We present a corollary of Theorem \ref{Main_th_intro} and some important remarks.

\begin{corollary}[\textbf{Sharp-logarithmic uniform energy decay}] \label{Cor_main_th_intro}
Let $\mathfrak{g}$ be a class of black ring metrics $g_{\textup{ring}}$, as defined in \eqref{def_class_g}. Consider smooth solutions $\Psi: \mathcal{D}\rightarrow\mathbb{C}$ to the scalar, linear wave equation
\begin{equation}   \label{WE_main_th_intro2}
\Box_{g_{\textup{ring}}}\Psi =0  
\end{equation}
on the black ring exterior $(\mathcal{D},g_{\textup{ring}})$, with $g_{\textup{ring}}\in \mathfrak{g}$ and compactly supported initial data on $\Sigma_0$, and assume that a uniform boundedness statement (without loss of derivatives) for the energy of solutions to \eqref{WE_main_th_intro2} holds. Then, for any $k\in\mathbb{N}$ and $\Omega\subset\mathcal{D}$ such that $\Sigma_0\cap\Omega$ is a bounded, non-empty set and $\Phi_{t^*}(\Sigma_0\cap\Omega)=\Phi_{t^*}(\Sigma_0)\cap\Omega$ remains bounded and non-empty for all $t^*>0$, there exists a universal constant $C_{k}(\Omega)>0$ (independent of time) such that 
\begin{equation}  \label{Mosch_intro}
E_{\Omega}[\Psi](t^*) \leq \frac{C_{k}(\Omega)}{[\log (2+t^*)]^{2k}} \, E_{k+1}[\Psi](0)
\end{equation}
for all $t^*>0$, with time coordinate $t^*$ and spacelike hypersurfaces $\Sigma_{t^*}$ defined in Section \ref{sect_foliation}. Furthermore, the uniform energy decay rate of \eqref{Mosch_intro} is \ul{sharp}.
\end{corollary}

\begin{remark} [\textbf{Theorem \ref{Main_th_intro} complements Moschidis \cite{LogDecay} for $\mathfrak{g}$}]
The first part of Corollary \ref{Cor_main_th_intro} follows by a result of Moschidis (see Theorem 2.1 in \cite{LogDecay}) specialized to the class $\mathfrak{g}$ of black rings. The sharpness of the statement is a direct consequence of our Theorem \ref{Main_th_intro}. 
\end{remark}

\begin{remark}[\textbf{Uniform boundedness assumption}] \label{rmk_unif_bddness}
Our assumption on uniform boundedness is crucial to produce the logarithmic lower bound once quasimodes are constructed. However, superradiance occurs for all black rings and obstructions to uniform boundedness might be present. As for Kerr black holes, one can successfully prove uniform boundedness for axisymmetric solutions to the wave equation and for non-superradiant, fixed-frequency modes, but no uniform boundedness statement holding for all solutions is known.\footnote{Note that, in view of the aforementioned result by Moschidis \cite{LogDecay}, in order to contradict uniform boundedness for black rings, it would be enough to disprove logarithmic uniform energy decay. For instance, it would suffice to show the existence of \textit{real} mode solutions or to construct quasimodes whose errors decay faster than exponentially in frequency.} From this point of view, sharp-logarithmic energy decay appears to be the best scenario that one can hope for on the class of black rings $\mathfrak{g}$, the alternative being the failure of uniform boundedness.   
\end{remark}

\begin{remark} [\textbf{Lower bound for black strings}] \label{rmk_string_intro}
The result of Theorem \ref{Main_th_intro} also holds for five-dimensional static and boosted black strings.\footnote{For the latter, one needs a further assumption on the boost parameter, as we shall see later. Note also that, since string metrics are fully-separable, one does not need the whole machinery developed in this paper to prove an analogue of Theorem \ref{Main_th_intro} for black strings.} This can be easily inferred by treating theorems in Section \ref{sect_static_string} and Section \ref{sect_boosted_string} as independent results and carrying out the quasimode construction of Section \ref{sect_proof_main_th}. Note that, in the case of black strings, uniform boundedness does hold, so one does not need to assume it in Theorem \ref{Main_th_intro}. For static black strings, uniform boundedness follows from the same arguments presented in \cite{DafRodNotes} to prove uniform boundedness on Schwarzschild spacetime. On the other hand, boosted black strings possess an ergoregion, but they are still not affected by superradiance. From the geometric point of view, the absence of superradiance corresponds to the existence of a Killing vector field
\begin{equation}  \label{Killing_boost_string}
X=(\cosh\beta) \, \partial_t-(\sinh\beta) \, \partial_z \, ,
\end{equation}
which is causal on the whole domain of outer communication (see Section \ref{Metric_boosted_string_sect} for the definition of the quantities in \eqref{Killing_boost_string}). The vector field $X$ is the linear combination of two Killing vector fields, and correctly reduces to $X=\partial_t$ in the unboosted case $\beta=0$. As for the static black string, one can prove uniform boundedness for solutions to the wave equation in the spirit of \cite{DafRodNotes}. This would give a rigorous proof to the numerical analysis in \cite{Cross-section} and to a comment in Section 3 of \cite{Inst_DS_BR}.
\end{remark}

The proof of Theorem \ref{Main_th_intro} relies on the construction of quasimodes. \textbf{The main technical difficulty arising from such construction, which does not appear in previous works, lies in that the wave equation fails to fully-separate on black rings}. This motivates our new PDE approach to the problem. See Section \ref{sect_overview_proof} for an overview.

\subsection{Stable trapping in higher dimensional black holes} \label{sect_trapping}

Black strings possess the most basic geodesic structure that one can find in higher dimensional black holes. In this regard, it is interesting to note that \textit{null} geodesic motion on black string exteriors can be understood in terms of \textit{timelike} geodesics in the exterior of the correspondent unextended black holes. In fact, at the level of the geodesic equation, one can treat the conserved momentum of a light ray along the extended direction as the mass of a particle orbiting the unextended object. Thus, from the existence of stable \textit{timelike} orbits around four-dimensional Schwarzschild and Kerr black holes, one can immediately deduce the presence of stably trapped \textit{null} geodesics in the correspondent five-dimensional (boosted) black string exteriors (see also \cite{Geodesics_strings}). Similarly, the absence of stable orbits for massive particles around higher dimensional Schwarzschild black holes \cite{Geodesics_n_Schw} suggests that one should not expect stably trapped null geodesics for static black strings in dimensions higher than five.\footnote{In this sense, one should note that Remark \ref{rmk_string_intro} concerns \textit{five-dimensional} black strings \textit{only}.}

The class $\mathfrak{g}$ of black rings that we consider in Theorem \ref{Main_th_intro} is morally a class of thin, singly-spinning black rings, whose near-horizon geometry resembles that of five-dimensional (boosted) black strings. It turns out that, in a sense that our paper shall clarify, the trapping structure of black strings locally persists for this class of black rings. In fact, we will be able to prove the existence of null geodesics whose toroidal orbits remain in a bounded region outside the event horizon (see Theorem \ref{ring_stab_trpp_redone} and Remark \ref{rmk_stab_trapp_ring}). \textbf{The appearance of \textit{stable trapping} on \textit{asymptotically flat} solutions to the Einstein vacuum equations is a remarkable geometric property of higher dimensional black holes, which is absent in four dimensions. Stable trapping has to be regarded as the fundamental mechanism underlying the slow decay of linear waves that Theorem \ref{Main_th_intro} determines}.

Our proof of stable trapping for black rings is a by-product of the proof of Theorem \ref{Main_th_intro} and puts in mathematically rigorous relation the trapping structure of black rings with that of black strings. This stable trapping phenomenon has already been investigated in \cite{BoundRing}, where a more computational approach was adopted. Other aspects of geodesic motion in black ring exteriors are presented in \cite{Hoskisson_geod_singly_ring, Durkee_geod_doubly_ring}.

\subsection{A new instability for black strings and black rings}  \label{sect_nonlinear_instab}

Theorem \ref{Main_th_intro} suggests that one should expect nonlinear instabilities for the class $\mathfrak{g}$ of black rings and, in view of Remark \ref{rmk_string_intro}, for five-dimensional black strings, the reason being that sufficiently strong decay at the linear level is necessary to prove small data global (in time) existence for nonlinear problems. It turns out that fast polynomial decay, namely faster than $1/t$, is enough to close a classical bootstrap argument, whereas logarithmic decay is not.\footnote{Here we are thinking about nonlinearities involving both the solution and first derivatives of the solution and with some nice structure, i.e. satisfying the \textit{null condition} of Christodoulou \cite{Christ_null_cond} and Klainerman \cite{Klainerman_null_cond}.}

However, \textbf{the nature of the nonlinear instability that one can conjecture from Theorem \ref{Main_th_intro} differs from the one caused by the nonlinear dynamics of the Gregory--Laflamme modes.} To make this statement more precise, we first briefly discuss some key features of the Gregory--Laflamme instability.

Given a metric perturbation 
\begin{equation*}
g_{\mu\nu} \rightarrow g_{\mu\nu}+h_{\mu\nu} \, ,
\end{equation*}
the original work of Gregory and Laflamme \cite{GLinstability} shows that, if $g_{\mu\nu}$ is the metric of a static black \textit{string} of radius $r_0$, then there exist exponentially growing (in time) modes of the form 
\begin{equation*}
h_{\mu\nu} \sim e^{-i(\omega t - J z)}H_{\mu\nu}(r) \quad \quad \omega\in\mathbb{C},J\in\mathbb{R} \, ,
\end{equation*}
with $\text{Im}(\omega)>0$ and $z$ coordinate along the extended direction, such that $g_{\mu\nu}+h_{\mu\nu}$ solves the linearised Einstein equations. For such modes to exist, the condition
\begin{equation}   \label{GL_Inst_condition}
J< J_{\text{GL}}
\end{equation}
needs to be satisfied, where $J_{\text{GL}}\sim 1/r_0$ is a positive constant characteristic of the string, and $\omega$ has to be chosen as a function of $J$. 

If one compactifies the string by identifying $z\sim z+L$, with $L$ positive constant, then the frequency parameter $J$ becomes discrete and has minimum (non-zero) value $J_{\text{min}}=1/L$. If $L\gg r_0$, then there exists $J$ satisfying \eqref{GL_Inst_condition} and the compactified black string is \textit{unstable}. Otherwise, for $L$ sufficiently small, the instability disappears. Therefore, the Gregory--Laflamme instability characterises black strings whose compactified extra-dimension is, in some sense, sufficiently large compared to the horizon radius. This aspect has a meaningful connection with Kaluza--Klein theories, see \cite{Horowitz_book}.

Work \cite{Hovd_Myers_String_Ring_GL} argues that the same picture essentially holds for boosted black strings and very thin, singly-spinning black \textit{rings}. For the latter, one should understand $r_0$ as the radius of the $S^2$ at the horizon and $L$ as the radius of the ring.   

As one can see, the Gregory--Laflamme instability is a purely \textit{gravitational} effect, which is already manifest at the \textit{linear} level and persists in the nonlinear evolution of suitably perturbed initial data \cite{Lehner_Pret_Instab_String, FiguerasEndpointRings}. On the other hand, the nonlinear instability conjectured from Theorem \ref{Main_th_intro} would emerge in the form of time integrals of error terms (nonlinearities) becoming large in the time evolution. In this sense, it would be a genuinely \textit{nonlinear} effect originating from slow decay at the \textit{scalar}, linear level. Note also that our instability is ultimately connected to a \textit{high-frequency} phenomenon, while the Gregory--Laflamme modes are \textit{low-frequency} perturbations. 

For \textit{black strings}, our instability does not require any restrictive condition on the geometry of the spacetime, thus, in contrast with the Gregory--Laflamme instability, it would affect \textit{any} five-dimensional black string, including those with small compactified extra-dimension $L\ll r_0$. In any dimension higher than five, black strings with $L\gg r_0$ still suffer from the Gregory--Laflamme instability, but none of them would suffer from our instability (see Section \ref{sect_trapping} for motivation).   

In the case of \textit{black rings}, it is interesting to note that both the instabilities affect (very) thin black rings. In particular, we expect that at least some of the members of the class $\mathfrak{g}$ would suffer from both the instabilities.

\subsection{Overview of the proof of Theorem \ref{Main_th_intro}}  \label{sect_overview_proof}

We present an overview of the proof of Theorem \ref{Main_th_intro}. Note that each step of the proof is, to some extent, self-contained and one could jump from Step 1 or 2 to the construction of quasimodes in Step 4 (see Remark \ref{rmk_string_intro}). At the beginning of each step, we report the section where the reader can find the detailed argument, so that the overview also serves as an outline of the paper. A formal version of Theorem \ref{Main_th_intro} is Theorem \ref{Main_th_formal} of Section \ref{section_main_th_formal}.

\textbf{Step 1. (Sections \ref{sect_sep_WE} and \ref{sect_static_string})} Our proof starts by considering five-dimensional static black strings $\text{Schw}_4\times\mathbb{R}$ in the usual Schwarzschild coordinate system. The metric reads
\begin{equation*} 
ds^2=-f(r)\,dt^2+f(r)^{-1}\,dr^2+r^2\,d\Omega^2_2+dz^2 \, ,
\end{equation*}
where $f(r)=1-r_0/r$, $r_0>0$. The event horizon lies at $r=r_0$ and the $z$ coordinate is periodically identified by $z=z+2\pi R$, with $R$ some positive constant. 

With an ansatz of the form
\begin{equation*}
\Psi(t,r,\theta,\phi,z)=e^{i(-\omega t+m\,\phi+J\,z)}u(r,\theta) \, ,
\end{equation*}
with $\omega\in\mathbb{R}$ and $m,J\in\mathbb{Z}$, and after a suitable factorization of $u(r,\theta)$, the wave equation on static black strings can be fully-separated. However, for later convenience, \textit{we do not fully-separate the wave equation}, but we instead rewrite it as a \textit{two-variable} Schr\"{o}dinger-type partial differential equation of the form
\begin{equation}  \label{static_string_EVP_intro}
-\frac{g(r)}{m^2}\, \Delta_{(r_*,\theta)}u+V(r,\theta)\, u=\frac{\omega^2}{m^2}\, u \, ,
\end{equation}
where $g(r)$ and $V(r,\theta)$ are smooth, positive functions \textit{independent} of the frequency parameters $\omega$ and $m$. At this stage, we have set $J=b\,m$, choosing the constant $b$ such that potential $V$ has a local minimum on the exterior region $\left\lbrace r\geq r_0 \right\rbrace$. We will keep this scaling of $J$ throughout the proof.  

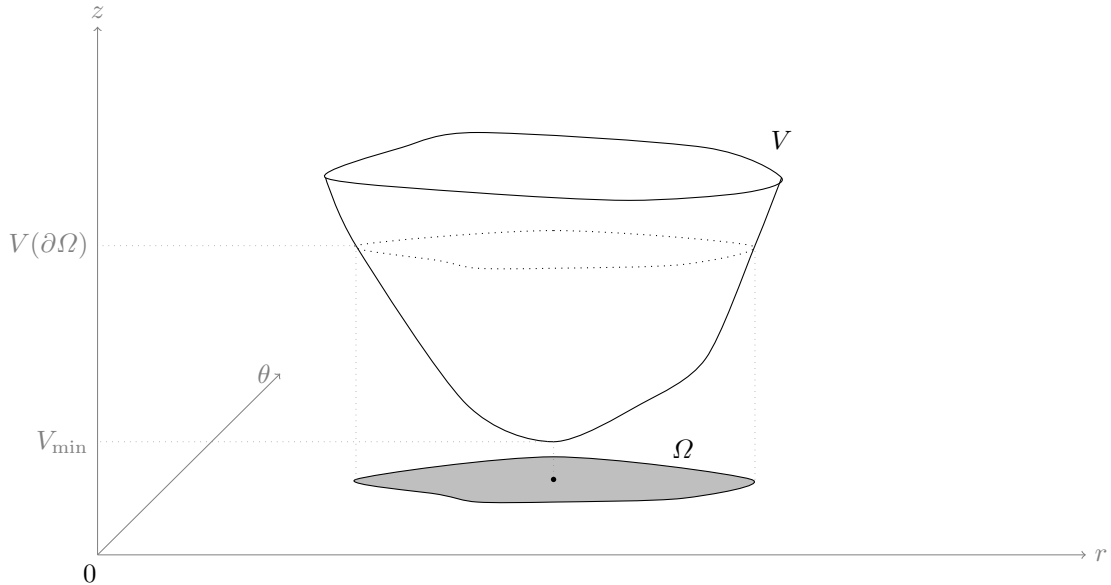
\begin{figure}[t] 

\centering
\begin{tikzpicture}

\draw plot [smooth] coordinates {(3,5) (2.65,4.1) (2,2.6) (1.15,2) (0,1.5) (-1.15,2) (-2.6,4.1) (-3,5) };

\draw  plot [smooth cycle] coordinates {(-3,5) (-2,5.4) (-1,5.6) (2,5.4) (3,5) (2.5,4.8) (1,4.7) (-1,4.8)};

\draw[fill=lightgray]  plot [smooth cycle] coordinates {(-2.6,1) (0,1.3)  (2.6,1)  (1.7,0.75) (0,0.7) (-1,0.7) (-1.5,0.8) };

\draw[dotted]  plot [smooth cycle] coordinates {(-2.6,4.1) (0,4.3)  (2.6,4.1)  (1.7,3.85) (0,3.8) (-1,3.8) (-1.5,3.9) };

\draw [dotted, help lines] (2.65,4.1)--(2.65,1);

\begin{scope}[yscale=1,xscale=-1] 
 \draw [dotted, help lines] (2.6,4.1)--(2.6,1); 
\end{scope}

\node at (1.7,1.4) {$\Omega$};
\node at (3,5.5) {$V$};
\node at (-6.1,-0.25) {$0$};

\draw[dotted, help lines] (0,1.5) -- (-6,1.5) node[left]{$V_{\textup{min}}$};
\draw[dotted, help lines] (-2.6,4.1) -- (-6,4.1) node[left]{$V(\partial\Omega)$};

\draw[dotted, help lines] (0,1.5) -- (0,1);

\fill (0,1)  circle[radius=1pt];

\draw[->,help lines] (-6,0)--(7,0) node[right]{$r$};
\draw[->,help lines] (-6,0)--(-6,7) node[above]{$z$};
\draw[->,help lines] (-6,0)--(-3.6,2.4) node[left]{$\theta$};
\end{tikzpicture}

\caption{Potential $V$ and open set $\Omega$ for the static black string eigenvalue problem \eqref{static_string_EVP_intro}.}
\label{fig:overview_proof_intro}
\end{figure}

We formulate a Dirichlet eigenvalue problem for equation \eqref{static_string_EVP_intro} on a bounded set $\Omega$ containing the local minimum of $V$ (see Figure \ref{fig:overview_proof_intro}). The set of eigenvalues for this problem is discrete, so eigenvalues $\omega^2/m^2$ cannot be freely specified. However, we will be able to prove that, for any energy level $E^2$ satisfying $V_{\text{min}}<E^2<V(\partial\Omega)$ and constant $\delta>0$ arbitrarily small, the eigenvalue problem \eqref{static_string_EVP_intro} admits an arbitrarily large number of eigenvalues in $[E^2-\delta,E^2+\delta]$ for $m^2$ sufficiently large. This is a version of Weyl's law for the Laplacian, where $1/m^2$ has to be interpreted as a semi-classical parameter. 

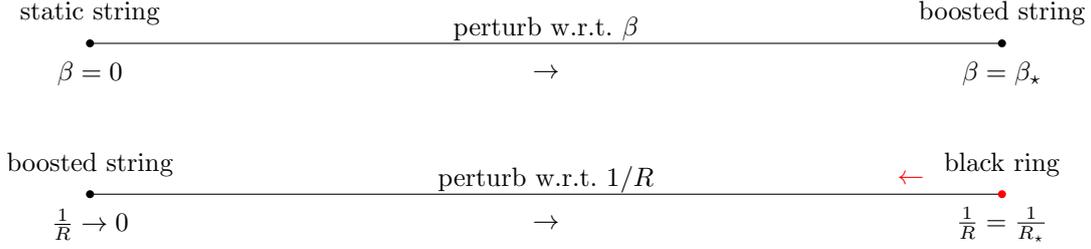
\begin{figure}[h] 

\centering

\begin{tikzpicture}

\draw (-6,1)--(6,1);
\draw (-6,-1)--(6,-1);

\fill[red] (6,-1)  circle[radius=1.5pt];
\fill (-6,-1)  circle[radius=1.5pt];
\fill (6,1)  circle[radius=1.5pt];
\fill (-6,1)  circle[radius=1.5pt];

\node at (-6,1.4) {$\textup{static string}$};
\node at (6,1.4) {$\textup{boosted string}$};
\node at (-6,0.6) {$\beta=0$};
\node at (6,0.6) {$\beta=\beta_{\star}$};
\node at (0,0.6) {$\rightarrow$};

\node at (-6,-0.6) {$\textup{boosted string}$};
\node at (6,-0.6) {$\textup{black ring}$};
\node at (-6,-1.4) {$\frac{1}{R}\rightarrow 0$};
\node at (6,-1.4) {$\frac{1}{R}=\frac{1}{R_{\star}}$};
\node at (0,-1.4) {$\rightarrow$};
\node[red] at (4.8,-0.8) {$\leftarrow$};

\node at (0,-0.8) {$\textup{perturb w.r.t.} \,\, 1/R $};
\node at (0,1.2) {$\textup{perturb w.r.t.} \,\, \beta$};

\end{tikzpicture}

\caption{The perturbation scheme adopted in our proof can be schematically divided in two steps (two segments in figure), where for each step we apply the full perturbation argument of Holzegel--Smulevici \cite{SharpLogHolz}. The first time we perturb the eigenvalue problem for the static black string (boost parameter $\beta$ equal to zero) towards the problem for the boosted black string with boost parameter $\beta_{\star}$, following the direction of the black arrow in figure. The second time we connect the problem for the boosted black string to the one for the black ring of radius $R_{\star}$. This second perturbation will generate extra errors that require an additional smallness parameter to be controlled, namely the inverse of the radius of the black ring $1/R_{\star}$. Thus, the endpoint of the second perturbation (represented by a red dot in figure) will have to be chosen sufficiently close to the starting point.}

\label{fig:perturbation scheme}

\end{figure}

For this first step we essentially adapt Section 4.1 of \cite{KeirLogLog} to our new two-variable PDE setting. The moral of the following two steps of the proof will be to replicate this construction for the abstract eigenvalue problems emerging from the separation of the wave equation on boosted black strings and black rings. By applying the perturbation scheme of Holzegel--Smulevici \cite{SharpLogHolz} \textit{twice}, we will be able to prove an analogous asymptotic property for the eigenvalues of both problems (see Figure \ref{fig:perturbation scheme}).

\textbf{Step 2. (Section \ref{sect_boosted_string})} We consider a five-dimensional \textit{boosted} black string
\begin{align*} 
ds^2=& -[1-(1-f(r))\cosh^2\beta]\,dt^2+2(1-f(r))\sinh\beta\cosh\beta \,dt \,dz \\
&+[1+(1-f(r))\sinh^2\beta]\,dz^2 +f(r)^{-1}\,dr^2+r^2\,d\Omega^2_2  
\end{align*}
with boost parameter $\beta\geq 0$. The corresponding Dirichlet eigenvalue problem 
\begin{equation}  \label{boost_static_string_EVP_intro}
-\frac{g(r)}{m^2}\Delta_{(r_*,\theta)}u+V^{\beta}_{(\omega,m)}(r,\theta)u=\frac{\omega^2}{m^2}u
\end{equation}
is now \textit{nonlinear}, since the potential depends on both $\omega$ and $m$. Note that problem \eqref{boost_static_string_EVP_intro} reduces to problem \eqref{static_string_EVP_intro} when $\beta=0$, i.e. when the boost parameter is zero.

The analysis of potential $V^{\beta}_{(\omega,m)}$ represents the main technical difficulty of this part of the proof. The potential depends on $\omega$ both linearly and quadratically, meaning that one cannot understand the structure of $V^{\beta}_{(\omega,m)}$ and energy levels $E$ independently. The idea will be to choose $E$ such that $V^{\beta}_{(Em,m)}(r,\theta)-E^2$ is negative on a bounded subset of $\Omega$ and admits a local minimum in such subset. This will implicitly reproduce the setting of Step 1. 

\textit{Note that no additional symmetry assumptions on solutions $\Psi$ to the wave equation can be made to simplify the analysis of $V^{\beta}_{(\omega,m)}$, i.e. none of the frequency parameters can be set to zero} (cf. \cite{SharpLogHolz}, where this is possible). See Lemma \ref{lemma_m_J_nonzero} for motivation.  

A perturbation argument with respect to $\beta$, based on an iterative application of the Implicit Function Theorem, will allow to use our knowledge of problem \eqref{static_string_EVP_intro} to conclude that, for $m^2$ sufficiently large, eigenvalues for the eigenvalue problem \eqref{boost_static_string_EVP_intro} exist and accumulate in the strip $[C,E^2+\delta]$, with constants $C>0$ independent of $m$ and $\delta>0$ arbitrarily small.

\textbf{Step 3. (Section \ref{sect_black_ring})} Finally, we consider a five-dimensional, singly-spinning black ring $g_{(r_0,R)}$ belonging to the class $\mathfrak{g}$.\footnote{The choice of this class is where we allow the radius $R$ of the black ring to be arbitrarily large.} Written in coordinates $(t,r,\theta,\phi,\psi)$, \footnote{These coordinates will be defined in Section \ref{sect_metrics}, but the reader should already note that they differ from the coordinates adopted for the black string metrics.} the eigenvalue problem for the black ring reads
\begin{equation} \label{ring_EVP_intro}
-\frac{g_{\text{ring}}(r,\theta)}{m^2}\Delta_{(r_*,\theta_*)}u+V^{\text{ring}}_{(\omega,m)}(r,\theta)u=0  \, ,
\end{equation} 
which is still \textit{nonlinear} and now \textit{non-separable} into two decoupled ODEs. On any fixed bounded set $\Omega$, problem \eqref{ring_EVP_intro} reduces to problem \eqref{boost_static_string_EVP_intro} in the limit $R\rightarrow \infty$. This is a manifestation of the fact that the near-horizon geometry of a large radius, thin black ring resembles that of a boosted black string.

\textit{The fact that $g_{(r_0,R)}\in \mathfrak{g}$ implies that it is possible to choose $\Omega$ and $E$ such that the potential $V^{\text{ring}}_{(Em,m)}$ has the same sign properties required for $V^{\beta}_{(Em,m)}(r,\theta)-E^2$ in Step 2}. We will then iteratively apply the Implicit Function Theorem for the second time, but now perturbing problem \eqref{boost_static_string_EVP_intro} with respect to $1/R$ and using what we already know about the eigenvalue problem for \textit{boosted} black strings. For $m^2$ sufficiently large, the eigenvalue problem \eqref{ring_EVP_intro} will be proven to admit a zero eigenvalue for values of $\omega^2/m^2$ lying in a suitable strip around $E^2$. Our choice of the energy level $E$ and set $\Omega$ will allow to conclude that such values of $\omega^2/m^2$ give the potential the correct structure to construct quasimodes with exponentially small errors. 

Note that this is the part of the proof where one can extract the existence of stably trapped null geodesics in the exterior of black rings $g_{(r_0,R)}\in \mathfrak{g}$. See Theorem \ref{ring_stab_trpp_redone} and related Remark \ref{rmk_stab_trapp_ring}.

\textbf{Step 4. (Section \ref{sect_proof_main_th})} Once the black ring eigenvalue problem has been fully understood, we will construct quasimodes on the domain of outer communication of the form
\begin{equation*}
\Psi_m(t,r_*,\theta_*,\phi,\psi)=e^{-i\,\omega_m t}e^{i(m\,\phi+\hat{J}\,\psi)}\chi(r_*,\theta_*)\, u_m(r_*,\theta_*) \, ,
\end{equation*}
where $u_m$ are eigenfunctions of problem \eqref{ring_EVP_intro} with associated frequencies $\omega_m$. The function $\chi$ cuts-off close to the boundary of $\Omega$ and sets the quasimodes equal to zero outside. In this way, the quasimodes $\Psi_m$ are \textit{smooth} functions and solve the wave equation on the whole domain of outer communication with the exception of the cut-off region, where the error will be proved to decay exponentially in the frequency $m$. The exponential decay is intimately connected to the structure of the potential in the cut-off region and it is essential to prove the logarithmic lower bound of Theorem \ref{Main_th_intro}. The quantitative estimate of the error is achieved by applying Agmon distances and proving energy estimates for eigenfunctions $u_m$ in the cut-off region. This step provides the proof of Theorem \ref{quasimodes_ring_intro}.

\textbf{Step 5. (Section \ref{sect_proof_main_th})} The last part of the proof employs the quasimodes constructed in Step 4 to derive the lower bound of Theorem \ref{Main_th_intro}, following an argument already sketched in Section \ref{Quasimodes_to_lower_bound_intro}.

\subsection{Applications of our method and outlook}

Our paper provides a framework to prove a version of Theorem \ref{Main_th_intro} for spacetimes on which the wave equation is not fully-separable. In particular, our approach could generalize the lower bound of Keir \cite{KeirLogLog} to a more general class of static, \textit{axisymmetric} spacetimes exhibiting stable trapping.\footnote{Note that without the spherical symmetry assumption of \cite{KeirLogLog}, one cannot a priori conclude that the wave equation fully-separates.} Our method (as the one in \cite{KeirLogLog}) only relies on the local geometry of the spacetime in a neighbourhood the trapping set, so the more general class of spacetimes would not require any specific asymptotic structure.

Furthermore, the stable trapping phenomenon observed for black rings is likely to characterise some other higher dimensional black hole spacetimes, including
\begin{enumerate}[(i)]
\item Kerr (possibly boosted) black strings ($\text{Kerr}_4\times\mathbb{R}$), Schwarzschild $p$-branes ($\text{Schw}_4\times\mathbb{R}^p$) and Kerr $p$-branes ($\text{Kerr}_4\times\mathbb{R}^p$),
\item Thin, doubly-spinning black rings in five dimensions,
\item Some ultraspinning objects, such as ultraspinning Myers--Perry black holes in six dimensions \cite{Stable_trapping_6MP}. 
\end{enumerate}
In the case of \textit{stationary} black holes, to apply our method and produce an analogue of Theorem \ref{Main_th_intro}, one would have to go through some sort of perturbation argument. Even though, for some of these spacetimes, a more complicated perturbation argument might be needed, some of them are easier to treat in that the wave equation fully-separates.\footnote{This is, for instance, the case of Myers--Perry black holes.} In any case, the presence of stable trapping would already suggest that a result like Theorem \ref{Main_th_intro} might hold and, therefore, nonlinear instabilities of similar nature to the ones discussed in Section \ref{sect_nonlinear_instab} might be conjectured.

As another potential direction, it would be certainly interesting to further investigate uniform boundedness of the energy of linear waves on black rings, inside or outside our class $\mathfrak{g}$. One of the main conceptual difficulties in doing this originates from the fact that the black ring family does not possess a static black hole. Therefore, even the (in principle) easier analysis on slowly rotating black rings cannot be, at first glance, understood in a perturbative fashion. Singly-spinning black rings (especially the thin ones) seem to be good candidates for uniform boundedness to hold, while doubly-spinning (thin) black rings might exhibit a superradiant instability and exponentially growing modes could be rigorously constructed.

\subsection{Acknowledgements}

The author would like to thank his advisors Gustav Holzegel and Claude Warnick for suggesting the problem and for their continuous guidance and support. The author is also grateful to Dejan Gajic and Joe Keir for valuable discussions and to Harvey Reall for comments on the manuscript. Part of the research presented in this work has been carried out while the author was visiting the Department of Pure Mathematics and Mathematical Statistics at the University of Cambridge. The author's research is funded by Imperial College London through an EPSRC/Roth Scholarship for Mathematics.

\section{Notation and conventions}  \label{Notation_section}

We collect here some basic notation and conventions adopted in the paper.
\begin{itemize}
\item \textbf{(Signature).} The signature of a Lorentzian metric $g$ is $(-,+,\ldots,+)$. We denote by $\Box_g$ the usual Laplace-Beltrami operator with respect to $g$. With our signature convention, $\Box_{g_{\text{Mink}}}=-\partial^2_{x^0}+\sum_{i}\partial^2_{x^i}$ for the Minkowski metric $g_{\text{Mink}}$ in rectangular coordinates $(x^0,x^1,\ldots,x^n)$, where $\partial_{x^i}:=\partial/\partial x^i$. We use the notation $\Delta_{(x^1,\ldots,x^n)}:=\sum_{i}\partial^2_{x^i}$ to denote the Cartesian Laplacian in coordinates $(x^1,\ldots,x^n)$.

\item \textbf{(Indices).} Given a coordinate system $(x^0,x^1,\ldots,x^n)$, we will follow the convention for which Greek indices take on all values, e.g. $\mu=0,1,\ldots,n$, and Latin indices only the spatial ones, e.g. $a=1,\ldots,n$. We adopt the usual upper/lower indices conventions and the summation convention. For instance, $g_{\mu\nu}$ are components of the metric tensor $g$, while $g^{\mu\nu}\equiv (g^{-1})^{\mu\nu}$ are components of the inverse metric. We will avoid the abstract index notation. 

\item \textbf{(Volume form).} For a given system of coordinates $(x^0,x^1,\ldots,x^n)$, we denote by $dvol_g$ the volume form associated to $g$, i.e. $dvol_g=\sqrt{-\det g}\,dx^0\cdots dx^n$. On any spacelike hypersurface $\Sigma$, $\overline{dvol}_g$ denotes the volume form associated to the Riemannian metric induced by $g$ on $\Sigma$. The volume form will be often omitted in the integrals.

\item \textbf{(Inequalities).} When we write $f \lesssim h$, we implicitly mean that there exists a constant $C>0$ such that $f \leq C h$. The notation $f \gtrsim h$ is analogous. We have $f\sim h$ when $f \lesssim h$ and $f \gtrsim h$. The notation $f\ll h$ means that there exists a sufficiently small constant $c>0$ such that $|f/h| \leq c$.

\item \textbf{(Constants).} We will not keep track of the constants appearing on the right hand side of the estimates. We adopt the notation $C_k$ to denote a constant $C$ which depends on some parameter $k$.

\item \textbf{(Multi-index).} Given coordinates $(x^0,x^1,\ldots,x^n)$, we will adopt the multi-index notation $\partial^{\alpha}:=\partial^{\alpha_0}_{x^0}\cdots \partial^{\alpha_n}_{x^n}$, with $\alpha=(\alpha_1,\ldots,\alpha_n)$ multi-index of order $|\alpha|=\alpha_1+\cdots+\alpha_n$.

\item \textbf{(Function spaces).} Function spaces $H^k(\Sigma)$ are the usual Sobolev spaces $W^{k,2}(\Sigma)$, with the standard Sobolev norm. With $C^{\infty}_0(\Sigma)$ we denote the space of smooth, compactly supported functions on $\Sigma$. We will write $SC^{\infty}_0(\Sigma)$ when we refer to the space of solutions of the wave equation which are $C^{\infty}_0$ on $\Sigma$.     

\item \textbf{(Black ring metric).} We will use the notation $g_{\text{ring}}$, $g_{(r_0,R)}$ or $g_{(\nu,R)}$ for the black ring metric. The first denotes a black ring metric in general, while the second and third one refer to the black ring metric as a two-parameter family of metrics (these are not components of the metric). The definition of the parameters will depend on the particular coordinate system considered.
\end{itemize}

\section{Metrics}  \label{sect_metrics}

Before introducing the relevant metrics, we define the \textit{domain of outer communication} of a black hole spacetime $(\mathcal{M},g)$ as a subset $\mathcal{D}\subset \mathcal{M}$ such that
\begin{equation*}
\mathcal{D}:= \text{clos}(J^-(\mathcal{I}^+)\cap J^+(\mathcal{I}^-)) \, ,
\end{equation*}
where $\mathcal{I}^+$ and $\mathcal{I}^+$ are \textit{future} and \textit{past null infinity} respectively. The event horizon is defined as $\mathcal{H}:=\partial\mathcal{D}$. We understand $(\mathcal{D},g)$ as a Lorentzian manifold with boundary.

Note that we will often refer to $(\mathcal{D},g)$ as the \textit{black hole exterior}.

\subsection{Static black string} \label{Metric_static_string_sect}

The metric of a five-dimensional \textit{static (Schwarzschild) black string} is
\begin{equation} \label{Schw_string_metric}
ds^2=-f(r)\,dt^2+f(r)^{-1}\,dr^2+r^2\,d\Omega^2_2+dz^2 \, ,
\end{equation}
where $(t,r,\theta,\phi)$ are the Schwarzschild coordinates and $d\Omega^2_2$ the round metric on the unit two-sphere. We have 
\begin{equation*}
f(r)=1-\frac{r_0}{r} 
\end{equation*}
with $r=r_0>0$ corresponding to the event horizon $\mathcal{H}$. For each fixed time $t$, the $z$-coordinate is periodically identified so that 
\begin{equation*}
z=z+2\pi R 
\end{equation*}
with $R>0$ constant independent of $t$.

The spacetime is static and possesses a Killing vectorfield $\partial/\partial z$ in addition to the symmetries of four-dimensional Schwarzschild. In view of the compactification along the $z$-direction, static black strings are asymptotically Kaluza--Klein. The domain of outer communication corresponds to $\mathcal{D}=\mathcal{M}\cap\left\lbrace r\geq r_0 \right\rbrace$.

\subsection{Boosted black string} \label{Metric_boosted_string_sect}

The five-dimensional \textit{boosted (Schwarzschild) black string} is obtained by the change of coordinates (Lorentz-boost in the $z$-direction)
\begin{align*}
t\rightarrow (\cosh \beta) t +(\sinh\beta) z &&
z \rightarrow (\sinh\beta )t +(\cosh\beta) z
\end{align*}
applied to the static black string \eqref{Schw_string_metric}, with $\beta>0$ boost parameter. The metric is
\begin{align} \label{boost_string_metric}
ds^2=& -[1-(1-f)\cosh^2\beta]\,dt^2+2(1-f)\sinh\beta\cosh\beta \,dt \,dz +[1+(1-f)\sinh^2\beta]\,dz^2 \\
&+f^{-1}\,dr^2+r^2\,d\Omega^2_2 \, , \nonumber
\end{align}
where $f=f(r)$. 

The spacetime is stationary (but not static) with respect to the boosted coordinates. The Killing vectorfield $\partial/\partial t$ becomes null at 
$r=r_0\cosh^2\beta$ and spacelike for $r<r_0\cosh^2\beta$. Boosted black strings do \textit{not} present superradiance. In fact, there exists a Killing vectorfield of the form
\begin{align*}
X=(\cosh\beta) \, \frac{\partial}{\partial t}-(\sinh\beta) \, \frac{\partial}{\partial z}  &&  g(X,X)=-f(r) \, ,
\end{align*}
which is causal on the whole domain of outer communication $\mathcal{D}=\mathcal{M}\cap\left\lbrace r\geq r_0 \right\rbrace$. As for the static black string, this spacetime is asymptotically Kaluza--Klein.

\subsection{Singly-spinning black ring}

\subsubsection{Ring coordinates}

We construct ring coordinates following the exposition in \cite{RingsRev}. Consider five-dimensional Minkowski space with coordinates 
\begin{equation*}
(t,r_1,r_2,\phi,\psi) \, ,
\end{equation*}
where $(r_1,\phi)$ and $(r_2,\psi)$ are polar coordinates on two independent rotation planes. The Minkowski metric in these coordinates reads
\begin{equation*}
ds^2= -dt^2 +dr_1^2+r_1^2 \, d\phi^2 +dr_2^2+r_2^2 \, d\psi^2 \, .
\end{equation*}
We define \textit{ring coordinates of radius $R$} as coordinates 
\begin{equation*}
(t,y,x,\phi,\psi)
\end{equation*}
such that
\begin{align*}
y=-\frac{R^2+r_1^2+r_2^2}{\sqrt{(r_1^2+r_2^2+R^2)^2-4R^2r_2^2}}  && x=\frac{R^2-r_1^2-r_2^2}{\sqrt{(r_1^2+r_2^2+R^2)^2-4R^2r_2^2}} 
\end{align*}
with $R>0$ constant. The coordinate ranges are
\begin{align*}
-\infty \leq y \leq -1  && -1 \leq x \leq 1
\end{align*}
and the Minkowski metric in ring coordinates of radius $R$ reads
\begin{equation}  \label{Mink_ring_coords}
ds^2=-dt^2 +\frac{R^2}{(x-y)^2}\left[(y^2-1)d\psi^2+\frac{dy^2}{y^2-1}+\frac{dx^2}{1-x^2} +(1-x^2)d\phi^2   \right] \, .
\end{equation}

\begin{figure}[H]
\centering
\includegraphics[width=0.65\textwidth]{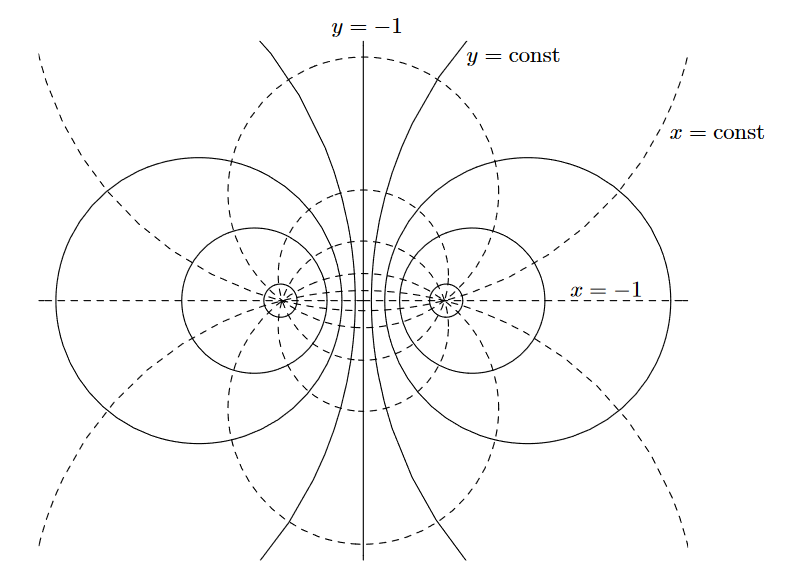}
\caption{Lines of constant $y$ and $x$ on the $(r_1,r_2)$-plane. The set of points with $y=-1$ and $x\neq -1$ are points on the $\psi$-axis of rotation (vertical axis in figure). The equatorial plane is divided into an inner region, for $x=1$, and an outer region, for $x=-1$. The rotational coordinate $\phi$ has to be considered with respect to this plane. The figure is taken from \cite{RingsRev}, courtesy of R. Emparan and H. Reall.}
\end{figure}

In these coordinates, surfaces of constant $y$ (for fixed $t$) have topology $S^1\times S^2$. To see this, we change coordinates to 
\begin{equation}  \label{Mink_ring_coords2}
(t,r,\theta,\phi,\psi)
\end{equation}
such that
\begin{align*}
r=-\frac{R}{y}  &&  \cos\theta=x
\end{align*}
with ranges
\begin{align*}
0\leq r \leq R &&  0\leq \theta \leq \pi \, .
\end{align*}
The Minkowski metric becomes
\begin{equation*}   
ds^2= -dt^2+\frac{1}{\left(  1+\frac{r\cos\theta}{R} \right)^2}  \left[\left( 1-\frac{r^2}{R^2}  \right)R^2 d\psi^2 +\frac{dr^2}{1-\frac{r^2}{R^2}}+r^2\left(d\theta^2+\sin^2\theta d\phi^2 \right)   \right] \, .
\end{equation*}
It is now manifest that surfaces of constant $r$ (for fixed $t$), which correspond to surfaces of constant $y$ (for fixed $t$), have topology $S^1\times S^2$.

\subsubsection{Black ring metrics}

Consider the smooth manifold with boundary
\begin{equation*}
\mathcal{D}=\text{clos}(\mathbb{R}\times \Sigma)
\end{equation*}
with 
\begin{equation*}
\Sigma=\mathbb{R}^4\setminus (S^1\times B^3)
\end{equation*}
and differential structure given by ring coordinates $(t,y,x,\phi,\psi)$, where $B^3$ is the closed $3$-ball. The \textit{black ring exterior} is the Lorentzian manifold with boundary $(\mathcal{D},g_{\text{ring}})$, where $g_{\text{ring}}$ is the five-dimensional, singly-spinning black ring metric\footnote{From now on, we will be frequently referring to singly-spinning black rings as \textit{black rings}. Note that there exist \textit{doubly}-spinning black rings \cite{DoubleBR}.} with line element \cite{EmpReallBR, RingsRev}
\begin{equation} \label{metric}
ds^2=-\frac{F(y)}{F(x)}\left(dt-CR\frac{(1+y)}{F(y)}\,d\psi\right)^2 +\frac{R^2}{(x-y)^2}F(x)\left[-\frac{G(y)}{F(y)}\,d\psi^2-\frac{dy^2}{G(y)}+\frac{dx^2}{G(x)}+\frac{G(x)}{F(x)}\,d\phi^2\right]
\end{equation}
on $\mathcal{D}$. Functions $F(\cdot)$ and $G(\cdot)$ are defined as
\begin{align*}
F(\xi)&=1+\lambda\xi & G(\xi)&=(1-\xi^2)(1+\nu\xi)
\end{align*}
and the positive constant $C$ is
\begin{equation*}
C=\sqrt{\lambda(\lambda-\nu)\frac{1+\lambda}{1-\lambda}} \, .
\end{equation*}
The coordinate ranges are
\begin{align*}
&t\in(-\infty,+\infty) & &y\in(-\infty,-1] & &x\in[-1,1] \\
&\phi\in\left[0,2\pi\sqrt{\frac{1}{1+\nu^2}}\right) & &\psi\in\left[0,2\pi\sqrt{\frac{1}{1+\nu^2}}\right) \, .
\end{align*}
The parameters $\nu$ and $\lambda$ satisfy 
\begin{align*}
\nu,\lambda\in\mathbb{R} && 0<\nu\leq \lambda<1
\end{align*}
and the equilibrium condition
\begin{equation} \label{eq_cond}
\lambda=\frac{2\nu}{1+\nu^2} \, .
\end{equation}
Condition \eqref{eq_cond} is necessary to avoid conical singularities in addition to those introduced by degenerations of our coordinate system. As a result, parameters $\lambda$ and $\nu$ are \textit{not} independent, leaving us with only two independent parameters $\nu$ and $R$. The black ring metric can therefore be understood as a \textit{two-parameter family $(\nu,R)$ of metrics}.

Black rings are \textit{stationary}, \textit{bi-axially symmetric}, \textit{asymptotically flat} spacetimes. The event horizon corresponds to the surface $y=-1/\nu$ and it is a Killing horizon with associated positive surface gravity. For each fixed time $t$, sections of the horizon have topology $S^1\times S^2$. The rotation of the singly-spinning black ring can be interpreted as a rotation along the $S^1$. See \cite{EmpReallBR} for further details. 

The metric admits three independent Killing vector fields, corresponding to stationarity and two rotational symmetries. The stationary Killing vector field $\partial /\partial t$ is timelike for $y>-1/\lambda$, null at $y=-1/\lambda$ and spacelike for $-1/\nu \leq y<-1/\lambda$. The surface $y=-1/\lambda$ is an ergosurface and has sectional topology $S^1\times S^2$.

Spacelike infinity lies at $x=y=-1$, while the limit $y\rightarrow -\infty$ corresponds to a curvature (spacelike) singularity.  The ring metric \eqref{metric} presents coordinate singularities along the axes (including at spacelike infinity) and at the event horizon. The metric is regular at the ergosurface, where $F(y)\rightarrow0$. 

For future convenience, we are interested in re-writing metric \eqref{metric} in coordinates \eqref{Mink_ring_coords2}. After changing from $(\nu,\lambda)$ to new parameters $(r_0,\beta)$ such that 
\begin{align*}
\nu =\frac{r_0}{R} && \lambda=\frac{r_0\cosh^2\beta}{R} \, ,
\end{align*}
the ring metric in coordinates \eqref{Mink_ring_coords2} becomes \cite{RingsRev}
\begin{align} \label{ring_alt_coord}
ds^2=&-\frac{\hat f}{\hat g}\left(dt-
r_0\sinh\beta\cosh\beta\sqrt{\frac{R+r_0\cosh^2\beta}{R- r_0\cosh^2\beta}}\:
\frac{\frac{r}{R}-1}{r\hat f}\:R\:d\psi\right)^2   \\
&
+
\frac{\hat g}{\left(1+\frac{r\cos\theta}{R}\right)^2}
\left[
\frac{f}{\hat f}\left(1-\frac{r^2}{R^2}\right)\, R^2\, d\psi^2
+
\frac{dr^2}{(1-\frac{r^2}{R^2})f}
+ 
\frac{r^2}{g}\,d\theta^2
+
\frac{g}{\hat g}\, r^2 \sin^2\theta\, d\phi^2
\right]			\nonumber
\end{align}
with
\begin{align*}
&f=1-\frac{r_0}{r} && \hat f=1-\frac{r_0\cosh^2\beta}{r} \\
&g=1+\frac{r_0}{R} \cos\theta && \hat g=1+\frac{r_0\cosh^2\beta}{R} \cos\theta \, .
\end{align*}
The event horizon now corresponds to $r=r_0<R$, the ergosurface to $r=r_0\cosh^2\beta$ and spacelike infinity to $(r,\theta)=(R, \pi)$. The equilibrium condition \eqref{eq_cond} becomes
\begin{equation} \label{eq_cond_new_coord}
\cosh\beta=\sqrt{\frac{2R^2}{r_0^2+R^2}} \, .
\end{equation}
As before, we understand black rings as a two-parameter family $(r_0,R)$.

The metric in form \eqref{ring_alt_coord} shows how the parameter $\nu$ can be interpreted. This is the ratio between the radius $r_0$ of the $S^2$ at the event horizon and the radius $R$ of the ring. We refer to \textit{thin black rings} when $\nu$ is \textit{small}. We sometimes use the expression \textit{large radius, thin black rings} to emphasise that we are considering black rings whose parameter $\nu$ is small because $R$ is very large (in contrast to thin black rings whose $R\sim 1$ and $r_0$ very small). 

The event horizon has angular velocity
\begin{equation*}
\Omega_{\mathcal{H}} \sim \frac{1}{R}\sqrt{\frac{\lambda -\nu}{\lambda(1+\lambda)}} \, .
\end{equation*}
In view of the equilibrium condition \eqref{eq_cond}, $\Omega_{\mathcal{H}}$ can be re-written as a quantity depending on $\nu$ and $R$ only. This shows that the rotational velocity of the ring cannot be freely specified, but it is univocally determined by its "thickness". \textit{Fat} black rings ($\nu\sim 1$) need more rotation than thin black rings to prevent the ring from collapsing.

\subsubsection{The boosted black string limit}

Consider the black ring metric \eqref{ring_alt_coord}. In the limit 
\begin{equation} \label{limit}
r,\,r_0,\,r_0\cosh^2\beta\ll R \, ,
\end{equation}
the black ring metric reduces to the one of a boosted black string with boost parameter 
\begin{equation*}
\beta=\cosh^{-1}\sqrt{2} \, .
\end{equation*}
This can be seen by fixing $r$ and $r_0$ in \eqref{ring_alt_coord} and sending $R$ to infinity (see also \cite{Elv_Emp_String}). The boost parameter is fixed in the limit by relation \eqref{eq_cond_new_coord}. The $z$-coordinate of the string is defined as $z=R\,\psi$, with periodic identification $z=z+R\,\Delta\psi$. In limit \eqref{limit}, $\Delta\psi \rightarrow 2\pi$ and one correctly obtains the boosted string metric \eqref{boost_string_metric}.

From a more heuristic point of view, limit \eqref{limit} corresponds to large radius, thin black rings. Condition $r\ll R$ implies that thin black rings approximate the geometry of boosted black strings in a region sufficiently close to the black ring horizon. Note that such limit is not uniform in $r$.\footnote{In fact, the two spacetimes have different asymptotic structure.}

\subsection{Foliation}  \label{sect_foliation}

We define $\Sigma_t$ as the hypersurface of constant $t$ and denote by $\Sigma_0$ the hypersurface $\Sigma_{t=0}$. 

Given any of the black hole spacetimes $(\mathcal{M},g)$ introduced in the previous sections, we fix a Cauchy hypersurface $\Sigma_{\text{Cauchy}}$ of $\mathcal{M}$.\footnote{This is possible because $(\mathcal{M},g)$ is globally hyperbolic.} We define a time function $t^*:J^+(\Sigma_{\text{Cauchy}})\cap \mathcal{D}\rightarrow\mathbb{R}$ such that 
\begin{equation*}
t^*|_{\Sigma_{\text{Cauchy}}\cap \mathcal{D}}=0
\end{equation*}
and $T(t^*)=1$, where $T$ is the stationary Killing vector field. We define the hypersurface
\begin{equation*}
\Sigma_0 := \Sigma_{\text{Cauchy}}\cap \mathcal{D}
\end{equation*} 
and, for any $\tau >0$,
\begin{equation*}
\Sigma_{\tau}:= \Phi_{t^*}(\Sigma_0) \, ,
\end{equation*}  
where $\Phi_{t^*}$ is the one-parameter group of diffeomorphisms generated by the Killing vector field $T$ from $0$ to $\tau$. We have that $\left\lbrace \Sigma_{\tau} \right\rbrace _{\tau\geq 0}$ is a regular spacelike foliation of $J^+(\Sigma_{\text{Cauchy}})\cap \mathcal{D}$. For black rings, hypersurfaces $\Sigma_{\tau}$ are asymptotically flat.

\section{Vectorfield Method and energy currents}  \label{sect_energy_currents}

In the context of the energy methods, the idea of obtaining positive definite quantities (energies) by contracting the energy momentum tensor (defined below) with suitable vector fields is an application of the \textit{vectorfield method}. The construction of vector field multipliers captures the geometry of the problem and its interplay with the analysis. In this section, we illustrate how energies can be rigorously defined via such vector field multipliers. 

Given a scalar field $\Psi:\mathcal{D}\rightarrow\mathbb{C}$ on the black ring exterior $(\mathcal{D},g_{\textup{ring}})$, we define the associated \textit{energy-momentum tensor} by
\begin{equation*}
\mathbb{T}_{\mu\nu}[\Psi]:=\text{Re}\left( \nabla_{\mu}\Psi\cdot \overline{\nabla_{\nu}\Psi}\right)-\frac{1}{2}\, g_{\mu\nu}|\nabla\Psi|^2 \, ,
\end{equation*}
where $g_{\mu\nu}$ are components of the ring metric $g_{\textup{ring}}$. If $\Box_{g_{\textup{ring}}}\Psi=0$, then
\begin{equation*}
\nabla^{\mu}\mathbb{T}_{\mu\nu}[\Psi]=0 \, .
\end{equation*}

Consider a regular, future directed, everywhere \textit{timelike} vector field $N$ on $\mathcal{D}$ and define the associated $N$\textit{-energy current} as
\begin{equation*}
\mathbb{J}_{\mu}^N[\Psi]:=\mathbb{T}_{\mu\nu}[\Psi]N^{\nu} \, .
\end{equation*} 
The (non-degenerate) $N$\textit{-energy} associated to $\Psi$ at time $t^*$ is defined as\footnote{More appropriately, this is the $N$-energy \textit{flux} through $\Sigma_{t^*}$ associated to the scalar field $\Psi$.}
\begin{equation*}
\mathcal{E}^N[\Psi](t^*):=\int _{\Sigma_{t^*}} \mathbb{J}_{\mu}^N[\Psi] n_{\Sigma_{t^*}}^{\mu}\, \overline{dvol}_g \, ,
\end{equation*}
where $n_{\Sigma_{t^*}}^{\mu}$ is the future directed, unit normal to the spacelike hypersurface $\Sigma_{t^*}$ (as defined in Section \ref{sect_foliation}) and the integration is with respect to the volume form associated to the Riemannian metric induced by $g_{\textup{ring}}$ on $\Sigma_{t^*}$ (the volume form will be often dropped). In the way it is defined, the scalar quantity $\mathbb{J}_{\mu}^N[\Psi] n_{\Sigma_{t^*}}^{\mu}$ is \textit{positive definite}.\footnote{This is true by the so-called \textit{positivity property}: If $X$,$Y$ are future directed timelike vectors, then $\mathbb{T}_{\mu\nu}[\Psi]X^{\mu}Y^{\nu}>0$. In particular, $\mathbb{T}_{\mu\nu}[\Psi]X^{\mu}Y^{\nu}\gtrsim \sum_{|\alpha|=1}|\partial^{\alpha}\Psi |^2$.} 

The \textit{local $N$-energy} associated to $\Psi$ at time $t^*$ reads
\begin{equation*}
\mathcal{E}^N_{\Omega}[\Psi](t^*):=\int _{\Sigma_{t^*} \cap \Omega} \mathbb{J}_{\mu}^N[\Psi] n_{\Sigma_{t^*}}^{\mu}\, \overline{dvol}_g \, ,
\end{equation*}
where $\Sigma_{t^*}\cap\Omega$ is some bounded set. The \textit{$k$-th (higher) order $N$-energy} is
\begin{equation*}
\mathcal{E}_k^N[\Psi](t^*):=\sum_{0 \leq |\alpha|\leq k-1} \int _{\Sigma_{t^*}}  \mathbb{J}_{\mu}^N[\partial^{\alpha}\Psi] n_{\Sigma_{t^*}}^{\mu}\, \overline{dvol}_g \, ,
\end{equation*}
with $\partial^{\alpha}$ in multi-index notation. Note that, for us, $\mathcal{E}_1^N[\Psi](t^*)=\mathcal{E}^N[\Psi](t^*)$.

\begin{remark}[\textbf{$\boldsymbol{\mathcal{E}^N_k[\Psi]}$ controls derivatives of $\boldsymbol{\Psi}$ up to order $\boldsymbol{k}$}]
If $N=\partial_{t^*}$ in a neighbourhood of spacelike infinity and hypersurfaces $\Sigma_{t^*}$ are asymptotically flat, then one can easily show that 
\begin{equation*}
\mathcal{E}^N[\Psi](t^*) \sim \sum_{|\alpha|=1}\int _{\Sigma_{t^*}}|\partial^{\alpha}\Psi|^2\, \overline{dvol}_g  \, ,
\end{equation*} 
which means that $\mathcal{E}^N[\Psi](t^*)$ is morally the $\dot{H}^1$ (semi-)norm of $\Psi$ on $\Sigma_{t^*}$. Note also that $\mathcal{E}^N[\, \cdot\,]\sim E[\, \cdot\,]$, where $E[\, \cdot\,]$ are the informal energies adopted in the Introduction. For higher-order energies, we have 
\begin{equation*}
\mathcal{E}^N_k[\Psi](t^*)=\sum_{0\leq |\alpha|\leq k-1} \mathcal{E}^N[\partial^{\alpha}\Psi](t^*) \, ,
\end{equation*}
so $\mathcal{E}^N_k[\Psi](t^*)$ corresponds to the sum of the $\dot{H}^s$ (semi-)norms of $\Psi$ on $\Sigma_{t^*}$ for $1\leq s \leq k$. 
\end{remark}

\section{The main theorem} \label{section_main_th_formal}

We give a more precise statement of Theorem \ref{Main_th_intro} and two additional  remarks. 

\begin{theorem}[\bf{Lower Bound for Uniform Energy Decay Rate, Second Version}] \label{Main_th_formal}
Consider the black ring exterior $(\mathcal{D},g_{(r_0,R)})$, with $g_{(r_0,R)}$ the metric of a singly-spinning black ring \eqref{ring_alt_coord}. Then, for any $r_0>0$, there exists a constant $\mathcal{R}>r_0$ such that the following statement holds for all metrics $g_{(r_0,R)}$ with $R\geq \mathcal{R}$. Fix $t^*_0\geq 0$ and let $N$ be a regular, future directed, everywhere timelike vector field on $J^+(\Sigma_{t_0^*})$ such that $N=\partial_{t^*}$ in a neighbourhood of spacelike infinity. Consider smooth solutions $\Psi:J^+(\Sigma_{t_0^*})\rightarrow\mathbb{C}$ to the linear wave equation
\begin{equation}
\Box_{g_{(r_0,R)}}\Psi=0  \label{WE_main_th_bis}
\end{equation}
and assume that there exists a universal constant $B>0$ (independent of time) such that, for any smooth solution $\Psi$ to \eqref{WE_main_th_bis}, the inequality
\begin{equation*}
\mathcal{E}^N[\Psi](t^*)\leq B \, \mathcal{E}^N[\Psi](t_0^*)
\end{equation*}   
holds for all $t^*\geq t_0^*$. Then, there exists a set $\Omega\subset J^+(\Sigma_{t^*_0})$ and a universal constant $C>0$ (independent of time) such that $\Sigma_{t^*_0}\cap\Omega$ is non-empty and bounded, $\Phi_{t^*}(\Sigma_{t^*_0}\cap\Omega)=\Phi_{t^*}(\Sigma_{t^*_0})\cap\Omega$ remains non-empty and bounded for all $t^*\geq t_0^*$ and
\begin{equation*}
\limsup_{t^*\rightarrow +\infty}\,\,  \sup_{\Psi \in SC^{\infty}_0(\Sigma_{t^*_0}), \Psi\neq 0} \,\, [\log (2+t^*)]^2 \left(\frac{\mathcal{E}^N_{\Omega}[\Psi](t^*) }{ \mathcal{E}^N_2[\Psi](t^*_0) }\right) > C \, ,
\end{equation*}
where the supremum is taken over the space of smooth, non-zero solutions to the wave equation \eqref{WE_main_th_bis} with compactly supported initial data on $\Sigma_{t^*_0}$ and $\Phi_{t^*}$ is the one-parameter group of diffeomorphisms generated by the stationary Killing vector field $\partial_{t^*}$. Moreover, for any $k\in\mathbb{N}$, there exists a universal constant $C_k>0$ (independent of time) such that 
\begin{equation*}
\limsup_{t^*\rightarrow +\infty}\,\,  \sup_{\Psi \in SC^{\infty}_0(\Sigma_{t^*_0}), \Psi\neq 0} \,\, [\log (2+t^*)]^{2k} \left(\frac{\mathcal{E}^N_{\Omega}[\Psi](t^*) }{ \mathcal{E}^N_{k+1}[\Psi](t^*_0) }\right) > C_k \, .
\end{equation*}
\end{theorem}

\begin{remark} [\textbf{Class $\mathfrak{g}$ is not optimal}]
The class of black rings considered in the theorem can be defined as
\begin{align} \label{def_class_g}
\mathfrak{g}:= & \left\lbrace   \textup{$g_{(r_0,R)}$ singly-spinning black ring metric \eqref{ring_alt_coord} such that $\nu=r_0/R<\nu_0$, for} \right. \\
& \left.  \textup{some suitably small constant $0<\nu_0<1$} \right\rbrace \, .  \nonumber
\end{align}
As we shall see later in the paper, the constant $\nu_0$ will be treated as a smallness parameter. For $\nu_0$ sufficiently small but not explicitly identified, we will be able to close the proof of Theorem \ref{Main_th_formal}. In this sense, the class $\mathfrak{g}$ that our proof selects is neither explicitly constructed nor necessarily includes all black rings for which Theorem \ref{Main_th_formal} holds. 
\end{remark}

\begin{remark} [\textbf{Function space and $\boldsymbol{\Sigma_{t^*}}$-foliation}]
Theorem \ref{Main_th_formal} remains true if one takes the supremum over a suitably larger function space. We stated the result for $\Psi \in SC^{\infty}_0(\Sigma_{t^*_0})$ because this is the space that naturally emerges from the quasimode construction. Note also that the statement of the theorem does not depend on the particular choice of time coordinate $t^*$. In fact, the coordinate $t^*$ can be constructed so that it agrees with $t$ in the region $r\geq r_0\cosh^2\beta$. Since our analysis will be mainly carried out on a bounded set contained in such region, we are free to equivalently refer to coordinate $t$ instead of $t^*$ in the proof of Theorem \ref{Main_th_formal}.
\end{remark}

\section{Separation of the wave equation and reduction}  \label{sect_sep_WE}

This section serves as a preliminary discussion to illustrate the logic behind the abstract eigenvalue problems that we are about to consider. In fact, these problems emerge as \textit{reduced equations} when one separates the wave equation 
\begin{equation}  \label{WE_coord_version}
\Box_g\Psi=\frac{1}{\sqrt{-\det g}}\,\partial_{\mu}\left(\sqrt{-\det g}\,g^{\mu\nu}\partial_{\nu}\Psi\right)=0 \, .
\end{equation}
The wave equation fully-separates for all the four-dimensional explicit black hole solutions, including all members of the Kerr family $g_{a,M}$ (see \cite{Carter_Sep_WE}). However, in our paper we will encounter a non-fully-separable wave equation on higher dimensional black holes, namely the wave equation on black rings. The only partial separation will necessarily leave us with a two-variable PDE, which can be reduced to a \textit{Schr\"{o}dinger-type equation}. 

To see this, consider a coordinate system $(t, r, \theta, \phi, z)$, where the particular meaning of the coordinates is not important at this stage, and assume that the metric $g$ depends on $r$ and $\theta$ (but not on $t$, $\phi$ and $z$) and has zero $g_{r\theta}$ components. A formal computation to check the full-separability of the wave equation can be carried out by introducing an ansatz of the form
\begin{equation} \label{ansatz}
\Psi_{\textup{ansatz}}(t,r,\theta,\phi,z)=e^{i(-\omega\, t+m\,\phi+J\,z)}u^{(\omega)}_{m,J}(r,\theta) \, ,
\end{equation}
where $\omega\in\mathbb{R}$ and $m,J\in\mathbb{Z}$ and the choice of the complex exponential factor is determined by the symmetries of the metric $g$. Any solution $\Psi$ to the wave equation \eqref{WE_coord_version} can be formally written as an infinite sum of solutions of the form \eqref{ansatz} as follows
\begin{equation} \label{Full-decomp-sol}
\Psi(t,r,\theta,\phi,z)=\frac{1}{\sqrt{2\pi}}\int^{+\infty}_{-\infty}\sum_{m,J} e^{-i\omega t}e^{i(m\,\phi+J\,z)}\,u^{(\omega)}_{m,J}(r,\theta) \,d\omega \, , 
\end{equation}
where the Fourier transform in time and Fourier series decompositions in $\phi$ and $z$ have been taken.\footnote{To take the Fourier transform in time
\begin{equation*}
\Psi(t,\cdot)=\frac{1}{\sqrt{2\pi}}\int^{+\infty}_{-\infty}e^{-i\omega t}\,\hat{\Psi}(\omega,\cdot) \,d\omega \, ,
\end{equation*} 
one should first know that the solution $\Psi$ is in $L^2_t$. However, this cannot be established a priori, so a cut-off in time is usually needed. This technical issue will not occur in our problem.}\textsuperscript{,}\footnote{The functions $u^{(\omega)}_{m,J}(r,\theta)$ have to be thought as Fourier coefficients.} Therefore, if equation \eqref{WE_coord_version} separates for the ansatz \eqref{ansatz}, then it separates for any solution $\Psi$.

In order for the ansatz \eqref{ansatz} to be a solution to the wave equation, functions $u^{(\omega)}_{m,J}$ must satisfy\footnote{Equation \eqref{WE} is obtained by simply plugging the ansatz in \eqref{WE_coord_version}. For simplicity, we write $u$ instead of $u^{(\omega)}_{m,J}$. If one plugs-in the full solution $\Psi$ in the form \eqref{Full-decomp-sol}, then equation \eqref{WE} has to be understood in an integral sense, i.e. as an equality in $L^2_{\omega}l^2_{m,J}$ for each fixed $(r,\theta)$. Furthermore, some further integrability conditions on $\Psi$ would be needed.} 
\begin{align} \label{WE}
\frac{1}{\sqrt{-\det g}}\,\partial_{r} & \left(\sqrt{-\det g}  \,g^{rr}\partial_{r}u(r,\theta)\right)  \\
&+\frac{1}{\sqrt{-\det g}}\,\partial_{\theta}\left(\sqrt{-\det g}\,g^{\theta\theta}\partial_{\theta}u(r,\theta)\right) 
+V_{(\omega,m,J)}(r,\theta)u(r,\theta)=0  \nonumber
\end{align}
for some \textit{real} function $V_{(\omega,m,J)}(r,\theta)$. If by factorizing
\begin{equation*}
u(r,\theta) = R(r) \, \Theta (\theta)
\end{equation*}
equation \eqref{WE} gives a system of two decoupled ODEs, one for $R(r)$ and one for $\Theta(\theta)$, then the wave equation is fully-separable. If such factorization does not separates \eqref{WE}, then the wave equation is not fully-separable.

Assuming the latter scenario, we aim to reduce the PDE \eqref{WE} to a Schr\"{o}dinger-type equation of the form
\begin{equation*}
-\Delta_{(r_*,\theta_*)}\tilde{u}(r_*,\theta_*)+\tilde{V}_{(\omega,m,J)}(r,\theta)\tilde{u}(r_*,\theta_*)=0 \, ,
\end{equation*} 
with $\Delta_{(r_*,\theta_*)}=\partial_{r_*}^2+\partial_{\theta_*}^2$ and some real function $\tilde{V}_{(\omega,m,J)}(r,\theta)$. To do this, we need to implicitly define new coordinates $(r_*,\theta_*)$ as follows
\begin{align*}
\frac{dr_*}{dr}=a(r) && \frac{d\theta_*}{d\theta}=b(\theta)
\end{align*} 
and define a new function $\tilde{u}(r_*,\theta_*)$ such that
\begin{equation*}
u(r,\theta)=h(r,\theta) \, \tilde{u}(r_*,\theta_*) \, ,
\end{equation*}
for some real functions $a(r),b(\theta)$ and $h(r,\theta)$. If the system of equations 
\begin{equation} \label{syst_change_coord}
\begin{cases}
\partial_r \left(\sqrt{-\det g}\,g^{rr}\,a(r)\right)h(r,\theta)+2\left(\sqrt{-\det g}\,g^{rr}\,a(r)\right)\partial_r h(r,\theta)=0\\
\partial_{\theta} \left(\sqrt{-\det g}\,g^{\theta\theta}\,b(\theta)\right)h(r,\theta)+2\left(\sqrt{-\det g}\,g^{\theta\theta}\,b(\theta)\right)\partial_{\theta} h(r,\theta)=0 \\
a^2(r)g^{rr}=b^2(\theta)g^{\theta\theta}
\end{cases}
\end{equation}
holds, then the wave equation \eqref{WE} correctly reduces to the Schr\"{o}dinger-type equation wanted. Note that the last equation is key, since it gives a necessary condition for the reduction to be possible. Indeed, we have  
\begin{equation*}
a^2(r)=\frac{g^{\theta\theta}}{g^{rr}} \, b^2(\theta) \, ,
\end{equation*}
which holds if and only if there exists $b(\theta)$ such that $\left(g^{\theta\theta}/g^{rr}\right)b^2(\theta)$ is independent of $\theta$, i.e. the function $g^{\theta\theta}/g^{rr}$ needs to be separable. Note that $g^{\theta\theta}/g^{rr}$ is consistently positive because the $(r,\theta)$-block of $g$ is Riemannian and $g_{r\theta}=0$. 

If such condition is satisfied,\footnote{This will be the case for black string and black ring metrics.} then the first two equations of the system give an explicit expression for $h(r,\theta)$, which reads
\begin{equation*}
h(r,\theta)=\left(\sqrt{-\det g}\,\sqrt{g^{rr}g^{\theta\theta}}\right)^{-\frac{1}{2}} \, ,
\end{equation*} 
while the third equation determines $a(r)$ and $b(\theta)$ up to multiplication by a real factor.

\begin{remark}[\textbf{Isothermal coordinates}]
For any two dimensional Riemannian manifold, there always exists a system of locally isothermal coordinates, for which the metric is conformal to the Euclidean metric (see, for instance, \cite{Iso_coords}). In view of this abstract result, the reduction that we have just discussed is, in fact, always possible locally. 
\end{remark}

In what follows, we will consider abstract eigenvalue problems without further referring to the separation of the wave equation.

\section{Eigenvalue problem for the static black string}  \label{sect_static_string}

In this section we discuss the first part of the proof of Theorem \ref{Main_th_formal}. The structure of the exposition and the formulation of the results closely follow Section 4 in \cite{KeirLogLog}.  

With coordinates as in Section \ref{Metric_static_string_sect} and ansatz \eqref{ansatz}, equation \eqref{WE} for static black strings gives the eigenvalue problem 
\begin{equation*}
-\frac{f(r)}{r^2}\partial_r(r^2f(r)\partial_ru)-\frac{f(r)}{r^2\sin\theta}\partial_{\theta}(\sin\theta \partial_{\theta}u)+\left[\frac{f(r)}{r^2\sin^2\theta} m^2+f(r) J^2\right]u=\omega^2 u  \, ,
\end{equation*}
with $u=u(r,\theta)$, $f(r)=1-r_0/r$, $\omega\in \mathbb{R}$ and $m,J\in \mathbb{Z}$. For any $m,J$ such that 
\begin{equation}
m^2>3J^2r_0^2  \, , \label{det_stab_trap} 
\end{equation}
the everywhere positive potential 
\begin{equation*}
\frac{f(r)}{r^2\sin^2\theta} m^2+f(r) J^2
\end{equation*}
has a local maximum and a local minimum at $(r_{\text{max}},\pi/2)$ and $(r_{\text{min}},\pi/2)$ respectively, with $r_0<r_{\text{max}}<r_{\text{min}}$.

Let us now \textit{fix} the scaling between $m$ and $J$ by setting 
\begin{align*}
J=b\, m  \, ,  &&  b^2<\frac{1}{3r_0^2} \, ,
\end{align*}
with $b\in\mathbb{Q}$ positive constant.\footnote{The sign of $b$ does not matter at this stage, since $m$ and $J$ always appear squared.} In this way, condition \eqref{det_stab_trap} is satisfied and, therefore, \textit{the potential has a local minimum}. 

The eigenvalue problem becomes
\begin{equation*}
-\frac{f(r)}{r^2}\partial_r(r^2f(r)\partial_ru)-\frac{f(r)}{r^2\sin\theta}\partial_{\theta}(\sin\theta \partial_{\theta}u)+m^2\left[\frac{f(r)}{r^2\sin^2\theta} +b^2 f(r) \right]u=\omega^2 u \, .
\end{equation*}
We now change coordinates to $(r_*,\theta)$ and define a function $h(r,\theta)$ as follows
\begin{align*}
h(r,\theta):=(r\sin\theta)^{-\frac{1}{2}}f(r)^{-\frac{1}{4}} \, , && \frac{dr_*}{dr}=\frac{1}{r\sqrt{f(r)}} \, .
\end{align*}
Let $u(r,\theta)=h(r,\theta)\, \tilde{u}(r_*,\theta)$. With the abuse of notation $\tilde{u}(r_*,\theta)=u(r_*,\theta)$, the eigenvalue problem reduces to
\begin{equation*}
g(r)(-\partial^2_{r_*}u-\partial^2_{\theta}u)+[V_j(r,\theta)+m^2 V(r,\theta)]u=\omega^2 u  \, ,
\end{equation*}
where 
\begin{gather*}
g(r):= \frac{f(r)}{r^2} \, ,\\
V_j(r,\theta):= -\frac{  (r-r_0 )}{4 r^3\sin^2\theta}-\frac{ r_0 ^2}{16 r^4} \, , \\
V(r,\theta):= \frac{f(r)}{r^2\sin^2\theta} +b^2 f(r) \, .
\end{gather*}
Note that these are smooth, real-valued functions which are bounded away from the axis $r=0$.

We define 
\begin{gather*}
h^{-2}:= m^2 \, ,\\
V_{\textup{eff}}^h(r,\theta):= h^2V_j(r,\theta)+V(r,\theta) \, ,\\
\kappa:= \omega^2h^2 \, ,
\end{gather*}
and write
\begin{equation}
-h^2g(r)\Delta_{(r_*,\theta)}u+V_{\textup{eff}}^h(r,\theta)u=\kappa u \, , \label{lin_eig_problem}
\end{equation}
where the Laplacian is the Cartesian Laplacian and, again, $u=u(r_*,\theta)$.

For the first part of our discussion, we will be considering the eigenvalue problem \eqref{lin_eig_problem} on a bounded domain $\Omega$, for which Dirichlet boundary conditions on $\partial\Omega$ will be imposed. Our interest is to determine the existence of eigenvalues arbitrarily close to some suitably fixed energy level.

\subsection{Continuity of the potential}

By continuity of the potential $V$, we have the following lemma, which defines the domain $\Omega$ on which our eigenvalue problem will be formulated and suitable energy levels $E$. We use the notation $x=(r,\theta)$ for a point in $[r_0,\infty)\times [0,\pi)$.
\begin{lemma}[adapted from \cite{KeirLogLog} Lemma 4.1]  \label{Cont_pot_static_string}
Define $V_{\textup{min}}$ to be the minimum of $V$ and $x_{\textup{min}}\in (r_0,\infty)\times (0,\pi)$ such that $V(x_{\textup{min}})=V_{\textup{min}}$. From a previous observation, $r(x_{\textup{min}})>r_0$. Let $c>0$ be a sufficiently small constant such that there exists $\Omega\subset [r_0,\infty)\times [0,\pi)$ for which
\begin{equation*}
x_{\textup{min}}\in \Omega
\end{equation*} 
and satisfying
\begin{enumerate}[(i)]
\item $V(x)=V_{\textup{min}}+c$ for $x\in \partial\Omega$,
\item there are no local maxima of $V$ in $\Omega$.
\end{enumerate}
Fix some energy level $E>V_{\textup{min}}$ such that $E-V_{\textup{min}}<c$. Then, for any sufficiently small constants $\delta,\delta^{\prime}>0$, there exists some constant $c^{\prime}>0$ such that 
\begin{equation*}
\textup{dist}(x,\partial\Omega)<\delta^{\prime} \implies V(x)-\kappa>c^{\prime}
\end{equation*}
for all $\kappa\in [E-\delta,E+\delta]$, with $\textup{dist}(\cdot,\cdot)$ the Euclidean distance. Note that one can reformulate the lemma for $V_{\textup{eff}}^h$ instead of $V$, where all the constants appearing in the statement are independent of $h$ for $h$ sufficiently small.
\end{lemma}

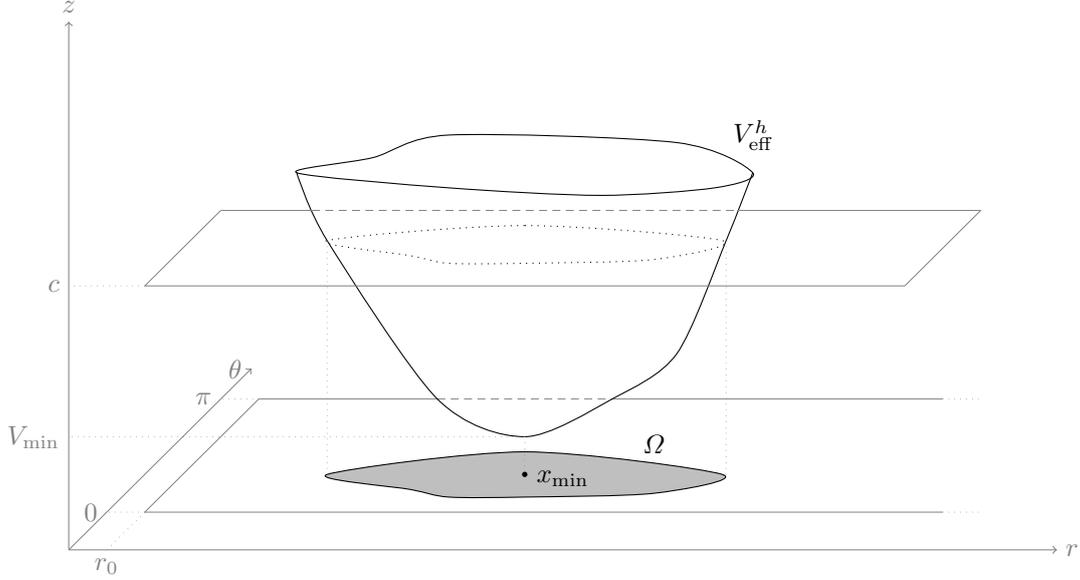
\begin{figure}[H]
\centering
\begin{tikzpicture}
\draw[help lines] (-5,0.5) -- (5.5,0.5);
\draw [dotted, help lines] (5.5,0.5) -- (6,0.5);
\draw [help lines] (-5,0.5) -- (-3.5,2);
\draw [help lines] (-3.5,2) -- (-1.15,2);
\draw [help lines] (1.15,2)--(5.5,2);
\draw [dotted, help lines] (5.5,2) -- (6,2);
\draw [densely dashed, help lines] (-1.15,2) -- (1.15,2);
\draw plot [smooth] coordinates {(3,5) (2.65,4.1) (2,2.6) (1.15,2) (0,1.5) (-1.15,2) (-2.6,4.1) (-3,5) };

\draw  plot [smooth cycle] coordinates {(-3,5) (-2,5.2) (-1,5.5) (2,5.4) (3,5) (2.5,4.8) (1,4.7) (-1,4.8)};

\draw[help lines] (-5,3.5) -- (5,3.5);
\draw[help lines] (-5,3.5) -- (-4,4.5);
\draw[help lines] (-4,4.5) -- (-2.8,4.5);
\draw[help lines] (2.8,4.5) -- (6,4.5);
\draw [densely dashed, help lines] (-2.8,4.5) -- (2.8,4.5);

\draw[fill=lightgray]  plot [smooth cycle] coordinates {(-2.6,1) (0,1.3)  (2.6,1)  (1.7,0.75) (0,0.7) (-1,0.7) (-1.5,0.8) };

\draw[dotted]  plot [smooth cycle] coordinates {(-2.6,4.1) (0,4.3)  (2.6,4.1)  (1.7,3.85) (0,3.8) (-1,3.8) (-1.5,3.9) };

\draw [help lines] (6,4.5)--(5,3.5);

\draw [dotted, help lines] (2.65,4.1)--(2.65,1);

\begin{scope}[yscale=1,xscale=-1] 
 \draw [dotted, help lines] (2.6,4.1)--(2.6,1); 
\end{scope}

\node at (1.7,1.4) {$\Omega$};
\node at (3,5.5) {$V^h_{\textup{eff}}$};
\node at (0.5,0.95) {$x_{\textup{min}}$};

\draw[dotted, help lines] (-5,3.5) -- (-6,3.5) node[left]{$c$};
\draw[dotted, help lines] (0,1.5) -- (-6,1.5) node[left]{$V_{\textup{min}}$};
\draw[dotted, help lines] (-5,0.5) -- (-5.5,0) node[below]{$r_0$};
\draw[dotted, help lines] (-5,0.5) -- (-5.5,0.5) node[left]{$0$};
\draw[dotted, help lines] (-3.5,2) -- (-4,2) node[left]{$\pi$};
\draw[dotted, help lines] (0,1.5) -- (0,1);

\fill (0,1)  circle[radius=1pt];

\draw[->,help lines] (-6,0)--(7,0) node[right]{$r$};
\draw[->,help lines] (-6,0)--(-6,7) node[above]{$z$};
\draw[->,help lines] (-6,0)--(-3.6,2.4) node[left]{$\theta$};
\end{tikzpicture}

\caption{The potential in figure is illustrative, it is not the graph of $V^h_{\textup{eff}}$. The shaded region corresponds to $\Omega$.} 

\end{figure}

\begin{remark}[\textbf{$\Omega$ has smooth boundary}]
The boundary of $\Omega$ is defined as a level set of the smooth function $V$. Since, by definition of $\Omega$, the gradient of $V$ is non zero at each point of the level set, one can conclude that the level set (and therefore $\partial\Omega$) is a smooth curve (this is an application of the Implicit Function Theorem). The set $\Omega$ is therefore a compact set with smooth boundary.  
\end{remark}

The abstract eigenvalue problem that we consider is 
\begin{equation}
\boxed{
\begin{array}{rcl}
-h^2g(r)\Delta_{(r_*,\theta)}u+V_{\textup{eff}}^h(r,\theta)u=&\kappa \, u \quad &\textrm{on} \quad \Omega   \label{Dirichlet_lin_eig_prob}\\
 u=&0 \quad &\textrm{on} \quad \partial\Omega \, .
 \end{array}
 }
\end{equation}

\subsection{Weyl's law for \eqref{Dirichlet_lin_eig_prob}}

In this section we prove the following theorem, which is a version of Weyl's law for the eigenvalue problem \eqref{Dirichlet_lin_eig_prob}. When the operator is a pure Laplacian on a bounded domain, this result is a standard asymptotic property of the eigenvalues. In our case, the presence of a potential requires more work, but without major additional difficulties. Our argument follows the lines of \cite{strauss_textbook, KeirLogLog} and review \cite{Weyl_static}. 

\begin{theorem}[\textbf{Weyl's law}, adapted from \cite{KeirLogLog} Lemma 4.2]  \label{Weyl's_law_static}
Consider the abstract eigenvalue problem \eqref{Dirichlet_lin_eig_prob}. Let $E>V_{\textup{min}}$ be an energy level such that $E-V_{\textup{min}}$ is sufficiently small and fix some positive constant $\delta<E-V_{\textup{min}}$ such that $E+\delta<c$, with $c>0$ the constant introduced in Lemma \ref{Cont_pot_static_string}. Then, the number of eigenvalues of \eqref{Dirichlet_lin_eig_prob} in the interval $[E-\delta,E+\delta]$, denoted by $N[E-\delta,E+\delta]$, tends to infinity as $h\rightarrow 0$. In particular,
 \begin{equation*}
 N[E-\delta,E+\delta] \sim \frac{1}{\pi h^2} \int_{\Omega} \left(E-V\right)\chi_{\left\lbrace V\leq E \right\rbrace} dr_*d\theta \, ,
 \end{equation*}
as $h\rightarrow 0$, with $\chi_{\left\lbrace \cdot \right\rbrace}$ indicator function.
 \end{theorem}

Let $\sigma(\Omega)$ be the set of all possible finite families $\sigma$ of open sets covering $\Omega$ up to a set of measure zero. For any family $\sigma\in\sigma(\Omega)$, we require that
\begin{enumerate}[(i)]
\item each open set of $\sigma$ is an open rectangle $R_i$ of area $A_i=L_{i,1}\times L_{i,2}$,
\item $R_i\cap R_j=\emptyset$ for any $i\neq j$, with $i,j\in\mathcal{I}$,
\item $R_i \cap \Omega \neq \emptyset$ for any $i\in \mathcal{I}$.
\end{enumerate}
Note that, in case condition (ii) is not satisfied for some family of open rectangles satisfying (i) and (iii), one can always consider a refinement of such family for which condition (ii) holds (this being true because the intersection $R_i\cap R_j$ of two rectangles is still a rectangle).

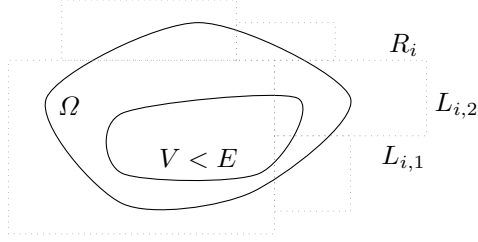
\begin{figure}[H]  
\centering
\begin{tikzpicture}
\draw plot [smooth cycle] coordinates {(-2,2) (0,3) (2,2) (0.7,0.75)  (-1,0.6)  };

\draw plot [smooth cycle] coordinates {(-1,1.8) (1.3,2) (0.8,1) (-1,1)    };

\draw[dotted, help lines] (1,1.5) rectangle (3,2.5);
\draw[dotted, help lines] (0.5,2.5) rectangle (1.8,3);
\draw[dotted, help lines] (1,2.5) rectangle (-2.5,0.2);
\draw[dotted, help lines] (1,1.5) rectangle (2,0.5);
\draw[dotted, help lines] (0.5,2.5) rectangle (-1.8,3.3);

\node at (0,1.2) {$V<E$};
\node at (-1.7,1.9) {$\Omega$};
\node at (2.7,2.7) {$R_i$};
\node at (2.7,1.2) {$L_{i,1}$};
\node at (3.4,1.9) {$L_{i,2}$};
\end{tikzpicture}
\caption{Finite family of recatngles $R_i$ covering $\Omega$.}
\label{fig:open_cover_Omega}
\end{figure}

Let $P$ be an eigenvalue problem. We denote by $N_{\leq E}(P)$ the number of eigenvalues of the problem $P$ which are less or equal to some fixed energy value $E$. 

Let us define the problem $\tilde{P}^i_D$ as
\begin{equation*}
\begin{array}{rcl}
-C^+_{\Omega} h^2 \Delta_{(r_*,\theta)}u+V^i_+ u=& \kappa \, u \quad &\textrm{on} \quad  R_i \cap\Omega   \\
 u=&0 \quad &\textrm{on} \quad \partial (R_i \cap\Omega) \, ,
 \end{array}
\end{equation*}
where 
\begin{align*}
C^+_{\Omega}:=\sup _{\Omega}\left[ 2 r_*^2 + \frac{1}{2}| \partial_{r_*} g(r)|^2  + g(r) \right] \, , &&  V^i_+ := \sup_{R_i\cap\Omega} \left[V_{\textup{eff}}^h(r,\theta) \right] \, .
\end{align*}
Note that the constant $C^+_{\Omega}>0$ is finite.

The two following lemmas, Lemma \ref{lemma_lower_Weyl} and Lemma \ref{lemma_upper_Weyl}, are the key technical ingredients to prove Theorem \ref{Weyl's_law_static}. We will only sketch their proofs, further details can be found in the proof of Lemma 4.2 in \cite{KeirLogLog}.

\begin{lemma}[\textbf{Weyl's lower bound}]  \label{lemma_lower_Weyl}

For any family $\sigma\in\sigma(\Omega)$, we have
\begin{equation}
\sum _{i\in\mathcal{I}} N_{\leq E}(\tilde{P}^i_D) \leq N_{\leq E}(P_D(\Omega))  \, , \label{key_relation}
\end{equation}
where $P_D(\Omega)$ is the eigenvalue problem \eqref{Dirichlet_lin_eig_prob}.

\end{lemma}

\begin{remark}[\textbf{Rectangles with no eigenvalues}]
By fixing the value $E$ as in the statement of Theorem \ref{Weyl's_law_static}, the bounded domain $\Omega$ gets separated into two disjoint regions, namely a region where $V\leq E$ and a region where $V>E$ (see Figure \ref{fig:open_cover_Omega}). The latter region extends up to the boundary $\partial\Omega$. The problem $\tilde{P}^i_D$ formulated on a rectangle $R_i$ intersecting the latter region will involve a constant potential $V^i_+>E$, so $N_{\leq E}(\tilde{P}^i_D)=0$ by classical arguments. In particular, this remains true for each $R_i\in\sigma$ such that $R_i \not\subset \Omega$, i.e. for each $R_i\in\sigma$ exceeding $\Omega$. For this reason, when we consider a particular family $\sigma\in\sigma(\Omega)$, the $R_i$ exceeding $\Omega$ (as well as any $R_i$ intersecting the region where $V>E$) do not contribute to the sum appearing on the left hand side of \eqref{key_relation}.   
\end{remark}

\begin{proof}
From min-max theory, the $n$-th eigenvalue of $P_D(\Omega)$ is given by
\begin{gather*}
\kappa_n=  \inf _{\substack{(f_1,\ldots,f_n), \, f_k \in H^1_0(\Omega) \\ \left\lVert f_k \right\rVert _{L^2} \neq 0, \, \left\langle f_k,f_j \right\rangle=0 \, \forall k \neq j}} \max _{k\leq n} \, Q_{\Omega}[f_k]  \\
Q_{\Omega}[f_k] :=   \frac{\int _{\Omega} \left[h^2f_k (\partial _{r_*}f_k )( \partial_{r_*} g(r))+h^2 g(r)\left( \left\lvert  \partial _{r_*} f_k\right\rvert ^2 +\left\lvert  \partial _{\theta}f_k \right\rvert ^2 \right)+V_{\textup{eff}}^h(r,\theta)|f_k|^2\right]dr_*d\theta}{\left\lVert f_k \right\rVert^2 _{L^2(\Omega)}} \, . 
\end{gather*}
Similarly, the $n$-th eigenvalues for the Dirichlet problem $P^i_D$, defined as
\begin{equation*}
\begin{array}{rcl}
-h^2g(r)\Delta_{(r_*,\theta)}u+V_{\textup{eff}}^h(r,\theta)u=&\kappa u \quad &\textrm{on} \quad R_i\cap\Omega   \\
 u=&0 \quad &\textrm{on} \quad \partial (R_i\cap\Omega) \, ,
 \end{array}
\end{equation*}
is determined by the formula 
\begin{gather*}
\lambda^i_n=\inf _{\substack{(f_1,\ldots,f_n), \, f_k \in H^1_0(R_i\cap\Omega) \\ \left\lVert f_k \right\rVert _{L^2} \neq 0, \, \left\langle f_k,f_j \right\rangle=0 \, \forall k \neq j}} \max _{k\leq n} \, Q^i_{\Omega}[f_k] \\
Q^i_{\Omega}[f_k] := \frac{\int _{R_i\cap\Omega} \left[h^2f_k (\partial _{r_*}f_k )( \partial_{r_*} g(r))+h^2 g(r)\left( \left\lvert  \partial _{r_*} f_k\right\rvert ^2 +\left\lvert  \partial _{\theta}f_k \right\rvert ^2 \right)+V_{\textup{eff}}^h(r,\theta)|f_k|^2\right]dr_*d\theta}{\left\lVert f_k \right\rVert^2 _{L^2(R_i\cap\Omega)}} \, .
\end{gather*}
We order the eigenvalues $\lambda^i_n$ in a non-decreasing sequence $\lambda_1<\lambda_2<\ldots$ \textit{for all $i$}. For each $n$, one has 
\begin{equation*}
\kappa_n\leq \lambda_n \, .
\end{equation*}
The proof of this fact essentially relies on the variational definition of the eigenvalues. The reader interested in the details should refer to Sublemma 4.2.1 in \cite{KeirLogLog} and realise that the argument therein can be easily adapted to our PDE setting.  
 
We observe that, for any $f_k\in H_0^1(R_i\cap\Omega)$, we have 
\begin{align}
&\int _{R_i\cap\Omega}h^2 \left[f_k (\partial _{r_*}f_k )( \partial_{r_*} g(r))+ g(r)\left( \left\lvert  \partial _{r_*} f_k\right\rvert ^2 +\left\lvert  \partial _{\theta}f_k \right\rvert ^2 \right)\right]dr_*d\theta  \label{est_Weyl_static}\\
&\leq \int _{R_i\cap\Omega}h^2 \left[\frac{1}{2} |f_k|^2+ \frac{1}{2} | \partial_{r_*} g(r)|^2|\partial _{r_*}f_k |^2 + g(r)\left( \left\lvert  \partial _{r_*} f_k\right\rvert ^2 +\left\lvert  \partial _{\theta}f_k \right\rvert ^2 \right)\right]dr_*d\theta \nonumber \\
&\leq \int _{R_i\cap\Omega}h^2 \left[\left(2 r_*^2 + \frac{1}{2}| \partial_{r_*} g(r)|^2  + g(r)\right) \left\lvert  \partial _{r_*} f_k\right\rvert ^2 +g(r) \left\lvert  \partial _{\theta}f_k \right\rvert ^2 \right]dr_*d\theta  \nonumber \\
&\leq \int _{R_i\cap\Omega}C^+_{\Omega} h^2 \left(\left\lvert  \partial _{r_*} f_k\right\rvert ^2 + \left\lvert  \partial _{\theta}f_k \right\rvert ^2 \right)dr_*d\theta \, ,  \nonumber
\end{align}
where we used Young's and Hardy's (Poincar\'{e}'s) inequalities and introduced the positive constant
\begin{equation*}
C^+_{\Omega}=\sup _{\Omega}\left[ 2 r_*^2 + \frac{1}{2}| \partial_{r_*} g(r)|^2  + g(r) \right] \, .
\end{equation*}
In particular, note that $C^+_{\Omega}$ does not depend on $i$. Let us now define
\begin{equation}
\tilde{\lambda}^i_n := \inf _{\substack{(f_1,\ldots,f_n), \, f_k \in H^1_0(R_i\cap\Omega) \\ \left\lVert f_k \right\rVert _{L^2} \neq 0, \, \left\langle f_k,f_j \right\rangle=0 \, \forall k \neq j}} \max _{k\leq n} \frac{\int _{R_i\cap\Omega} \left[C^+_{\Omega} h^2 \left(\left\lvert  \partial _{r_*} f_k\right\rvert ^2 + \left\lvert  \partial _{\theta}f_k \right\rvert ^2 \right)+V^i_+|f_k|^2\right]dr_*d\theta}{\left\lVert f_k \right\rVert^2 _{L^2(R_i\cap\Omega)}} \, , \label{weyl_prob}
\end{equation}
where 
\begin{equation*}
V^i_+= \sup_{R_i\cap\Omega} \left[V_{\textup{eff}}^h(r,\theta) \right] \, .
\end{equation*}
From estimate \eqref{est_Weyl_static}, combined with the variational definition of the eigenvalues, we have $\lambda^i_n\leq \tilde{\lambda}^i_n$ for each $i$ because the Rayleigh quotient gets increased. Therefore, after ordering $\tilde{\lambda}^i_n$ in a non-decreasing sequence for all $i$, we have
\begin{equation}
\kappa_n\leq\lambda_n\leq \tilde{\lambda}_n \, .\label{Weyl_ineq_eigenvalue}
\end{equation}

Recall now the definition of the problem $\tilde{P}_D^i$. Formula \eqref{weyl_prob} gives the $n$-th eigenvalue of $\tilde{P}_D^i$. So, from \eqref{Weyl_ineq_eigenvalue}, we can conclude  
\begin{equation*}
\sum _{i\in\mathcal{I}} N_{\leq E}(\tilde{P}^i_D) \leq N_{\leq E}(P_D(\Omega)) \, ,
\end{equation*}
where the result is independent of the family $\sigma\in \sigma(\Omega)$ chosen.

\end{proof} 

\begin{lemma}[\textbf{Weyl's upper bound}]  \label{lemma_upper_Weyl}
For any family $\sigma\in\sigma(\Omega)$, we have
\begin{equation*}
  N_{\leq E}(P_D(\Omega)) \leq  \sum _{i\in\mathcal{I}} N_{\leq E}(\tilde{P}^i_N) \, ,
\end{equation*}
where $\tilde{P}^i_N$ is the problem 
\begin{equation*}
-C^-_{\Omega} h^2 \Delta_{(r_*,\theta)}u+V^i_- u= \kappa u \quad \textrm{on} \quad  R_i \cap\Omega   \, ,
\end{equation*}
with Neumann boundary conditions imposed on $\partial(R_i\cap\Omega)$ and  
\begin{align*}
C_{\Omega}^- :=\frac{1}{2}\inf _{\Omega}g(r) \, , &&  V^i_- := \inf_{R_i\cap\Omega} \left[V_{\textup{eff}}^h(r,\theta)\right]-h^2\frac{C_{\Omega}^+}{4C_{\Omega}^-} \, .
\end{align*}
\end{lemma}

\begin{proof}
Consider the variational definition of the eigenvalues
\begin{equation*}
\mu^i_n=\inf _{\substack{(f_1,\ldots,f_n), \, f_k \in H^1(R_i\cap\Omega) \\ \left\lVert f_k \right\rVert _{L^2} \neq 0, \, \left\langle f_k,f_j \right\rangle=0 \, \forall k \neq j}} \max _{k\leq n} \, Q^i_{\Omega}[f_k] 
\end{equation*}
and order them in a non-decreasing sequence $\mu_1<\mu_2<\ldots$ for all $i$. These are eigenvalues of the problem $P^i_N$, defined as the problem $P^i_D$ but now with Neumann boundary conditions imposed on $\partial(R_i\cap\Omega)$. Note that the space of test functions is now $H^1$ instead of $H^1_0$. Following the argument in \cite{KeirLogLog}, we define the space
\begin{align*}
Y := & \left\lbrace   \textup{$f\in H^1(\Omega)$ such that $\left\lVert f \right\rVert^2 _{L^2(\Omega)}\neq 0$ and $f$ is in the closure in $H^1$ of piecewise $C^2$}\right. \\
&\left. \textup{functions, which are $C^2$ on each $R_i\cap\Omega$}   \right\rbrace
\end{align*}
and eigenvalues 
\begin{equation*}
\hat{\mu}_n=\inf _{\substack{(f_1,\ldots,f_n), \, f_k \in Y \\ \left\lVert f_k \right\rVert _{L^2} \neq 0, \, \left\langle f_k,f_j \right\rangle=0 \, \forall k \neq j}} \max _{k\leq n} \, Q_{\Omega}[f_k] \, .
\end{equation*}
These are eigenvalues of the problem $\hat{P}_N(\Omega)$, defined as $P_D(\Omega)$ but with Neumann boundary conditions on $\partial\Omega$ and test function space $Y$. Since $H^1_0(\Omega)\subset Y$, we can easily conclude that 
\begin{equation*}
\hat{\mu}_n\leq \kappa_n \, .
\end{equation*}
The delicate part of the argument is now proving that $\hat{\mu}_n=\mu_n$ for each $n$ (when counted with multiplicity). See the proof of Sublemma 4.2.1 in \cite{KeirLogLog} and textbook \cite{strauss_textbook} for more details on this. Once the equality $\hat{\mu}_n=\mu_n$ has been established, one has
\begin{equation*}
\mu_n\leq \kappa_n \, .
\end{equation*}
Similar considerations to the ones presented in the proof of Lemma \ref{lemma_lower_Weyl} allow to conclude that 
\begin{equation*}
\tilde{\mu}_n \leq \mu_n\leq \kappa_n  \, ,
\end{equation*}
where $\tilde{\mu}_n$ are eigenvalues of $\tilde{P}^i_N$. The estimate for the Rayleigh quotient now reads
\begin{align*}
&\int _{\Omega} \left[h^2f_k (\partial _{r_*}f_k )( \partial_{r_*} g(r))+h^2 g(r)\left( \left\lvert  \partial _{r_*} f_k\right\rvert ^2 +\left\lvert  \partial _{\theta}f_k \right\rvert ^2 \right)+V_{\textup{eff}}^h(r,\theta)|f_k|^2\right]dr_*d\theta \\
&\geq \int _{R_i \cap \Omega} \left[-h^2\varepsilon |\partial _{r_*}f_k|^2-\frac{h^2}{4\varepsilon}  | \partial_{r_*} g(r)|^2|f_k|^2 |+h^2 g(r)\left( \left\lvert  \partial _{r_*} f_k\right\rvert ^2 +\left\lvert  \partial _{\theta}f_k \right\rvert ^2 \right)+V_{\textup{eff}}^h(r,\theta)|f_k|^2\right]dr_*d\theta \\
& \geq \int _{R_i\cap \Omega} \left[h^2 (g(r)-\varepsilon)\left( \left\lvert  \partial _{r_*} f_k\right\rvert ^2 +\left\lvert  \partial _{\theta}f_k \right\rvert ^2 \right)+\left(V_{\textup{eff}}^h(r,\theta)-\frac{h^2}{4\varepsilon}  |\partial_{r_*} g(r)|^2\right)|f_k|^2\right]dr_*d\theta \, ,
\end{align*}
where one chooses $\varepsilon=C_{\Omega}^-$ and obtains that the last line is greater or equal to
\begin{equation*}
\int _{R_i\cap \Omega} \left[C_{\Omega}^- h^2 \left( \left\lvert  \partial _{r_*} f_k\right\rvert ^2 +\left\lvert  \partial _{\theta}f_k \right\rvert ^2 \right)+\left(V_{\textup{eff}}^h(r,\theta)-h^2\frac{C_{\Omega}^+}{4C_{\Omega}^-}  \right)|f_k|^2\right]dr_*d\theta  \, .
\end{equation*}
This proves the lemma.

\end{proof}

\begin{remark}
In what follows we will show how we can derive an explicit lower bound for $N_{\leq E}(P_D(\Omega))$ from Lemma \ref{lemma_lower_Weyl}. In the same way, Lemma \ref{lemma_upper_Weyl} provides an upper bound that can be explicitly computed. Since the calculations are almost identical, we only present the computation for the former. 
\end{remark}

The key point of inequality \eqref{key_relation} is that we can derive an explicit expression for the sum appearing on the left hand side, which will provide an explicit lower bound for $N_{\leq E}(P_D(\Omega))$.

Recall that the problem $\tilde{P}^i_D$ reads 
\begin{equation*}
-C^+_{\Omega} h^2 \Delta_{(r_*,\theta)}u+V^i_+ u= \kappa u \, ,
\end{equation*}
with Dirichlet boundary conditions imposed on $\partial (R_i\cap\Omega)$. The open set $R_i\cap\Omega$ is an open \textit{rectangle} if $R_i\cap\Omega\equiv R_i$, otherwise it has a more general shape. Remember that, in the latter case where $R_i\cap\Omega\not\equiv R_i$, we have $V^i_+>E$, so 
\begin{equation*}
N_{\leq E}(\tilde{P}^i_D)=0 \, .
\end{equation*}
Therefore, these $R_i$ do not enter the computation of the sum in \eqref{key_relation}, for which it is enough to only consider the $R_i$ such that $R_i\cap\Omega$ are rectangles.

For such $R_i$, the eigenvalue problem $\tilde{P}^i_D$ admits eigenfunctions
\begin{equation*}
u_{n_{i,1},n_{i,2}}(r_*,\theta)=\sin\left(\frac{n_{i,1}\pi }{L_{i,1}}\, r_* \right) \sin\left(\frac{n_{i,2}\pi }{L_{i,2}}\, \theta \right)
\end{equation*} 
with eigenvalues
\begin{equation*}
\tilde{\lambda}^i_{n_{i,1},n_{i,2}}=\left(\frac{n_{i,1}\pi }{L_{i,1}} \right)^2+ \left(\frac{n_{i,2}\pi }{L_{i,2}} \right)^2 \, ,
\end{equation*}
where $n_{i,1}, n_{i,2}\in \mathbb{N}$. Therefore, we have that $N_{\leq E}(\tilde{P}^i_D)$ is the number of eigenvalues $\tilde{\lambda}^i_{n_{i,1},n_{i,2}}$ satisfying
\begin{equation*}
\left(\frac{n_{i,1} }{L_{i,1}} \right)^2+ \left(\frac{n_{i,2} }{L_{i,2}} \right)^2 \leq  \left(\frac{E-V^i_+}{C^+_{\Omega}\pi^2 h^2} \right) \chi_{\left\lbrace V^i_+\leq E \right\rbrace} \, ,
\end{equation*}
which corresponds to the number of integer lattice points $(n_{i,1},n_{i,2})\in\mathbb{N}\times\mathbb{N}$ contained in the portion of the first quadrant defined by
\begin{equation*}
\mathcal{E}_E=\left\lbrace (x,y)\in\mathbb{R}^2: \left(\frac{x}{\sqrt{\frac{E-V^i_+}{C^+_{\Omega}\pi^2 h^2}}\,\chi_{\left\lbrace V^i_+\leq E \right\rbrace}\, L_{i,1}}\right)^2+  \left(\frac{y}{\sqrt{\frac{E-V^i_+}{C^+_{\Omega}\pi^2 h^2}}\,\chi_{\left\lbrace V^i_+\leq E \right\rbrace}\, L_{i,2}}\right)^2 \leq 1, x\geq 0, y\geq 0 \right\rbrace
\end{equation*}
This is a region delimited by an ellipse. By constructing unit-area squares $[n_{i,1}-1,n_{i,1}]\times [n_{i,2}-1,n_{i,2}]$, one can easily see that the number of such squares contained in $\mathcal{E}_E$ equals the number of integer lattices points in $\mathcal{E}_E$ (see \cite{Weyl_static} for more details). Since the squares have unit area, we have
\begin{equation*}
N_{\leq E}(\tilde{P}^i_D) = \sum \text{Area(squares)} \leq \text{Area(}\mathcal{E}_E\text{)} \, ,
\end{equation*} 
where
\begin{align*}
\text{Area(}\mathcal{E}_E\text{)} &= \frac{\pi}{4}\left(\frac{E-V^i_+}{C^+_{\Omega}\pi^2 h^2}\right)\chi_{\left\lbrace V^i_+\leq E \right\rbrace} \, L_{i,1}L_{i,2} \\
&= \frac{1}{4\pi}\left(\frac{E-V^i_+}{C^+_{\Omega} h^2}\right)\chi_{\left\lbrace V^i_+\leq E \right\rbrace} \, \text{Area(}R_i\text{)} \, .
\end{align*}
By an analogous construction (again, see \cite{Weyl_static} for details), it is possible to obtain the lower bound
\begin{equation*}
N_{\leq E}(\tilde{P}^i_D) \geq \frac{1}{2\pi}\left(\frac{E-V^i_+}{C^+_{\Omega} h^2}\right)\chi_{\left\lbrace V^i_+\leq E \right\rbrace} \, \text{Area(}R_i\text{)} - \frac{\text{Perimeter(}R_i\text{)}}{2\pi} \sqrt{\frac{E-V^i_+}{C^+_{\Omega} h^2}} \, \chi_{\left\lbrace V^i_+\leq E \right\rbrace} \, .
\end{equation*}
Therefore, for $h$ sufficiently small, we have
\begin{equation*}
N_{\leq E}(\tilde{P}^i_D) \sim \frac{\text{Area(}R_i\text{)}}{\pi h^2}\left(E-V^i_+\right) \chi_{\left\lbrace V^i_+\leq E \right\rbrace} \, .
\end{equation*}
Note that the scaling in $h$ is different from the one obtained in \cite{KeirLogLog}. This comes from the fact that we are considering an eigenvalue problem on a rectangle, instead of a problem on a segment of the real line.

From \eqref{key_relation}, we can conclude that 
\begin{equation}
\sum _{i\in\mathcal{I}} \frac{\text{Area(}R_i\text{)}}{\pi h^2}\left(E-V^i_+\right)\chi_{\left\lbrace V^i_+\leq E \right\rbrace}  \lesssim N_{\leq E}(P_D(\Omega)) \, . \label{riem_sum}
\end{equation}
We now refine the family $\sigma$ by taking the supremum over all the possible families $\sigma\in\sigma(\Omega)$. In other words, we take a limit which sends the Riemann sum to an integral. From \eqref{riem_sum}, we have 
\begin{equation*}
\sup_{\sigma\in\sigma(\Omega)}   \sum _{i\in\mathcal{I}} \frac{\text{Area(}R_i\text{)}}{\pi h^2}\left(E-V^i_+\right) \chi_{\left\lbrace V^i_+\leq E \right\rbrace} \lesssim N_{\leq E}(P_D(\Omega)) \, ,
\end{equation*}
which gives 
\begin{equation}
N_{\leq E}(P_D(\Omega)) \gtrsim  \frac{\text{Area(}\Omega\text{)}}{\pi h^2} \int_{\Omega} \left(E-V\right)\chi_{\left\lbrace V\leq E \right\rbrace} dr_*d\theta \, .\label{weyl_result}
\end{equation}

\begin{remark}
As already observed, we are sure that $R_i\cap\Omega$ is a rectangle only when $R_i\cap\Omega \equiv R_i$. For problems $\tilde{P}^i_D$, rectangles for which $R_i\cap\Omega \not\equiv R_i$ did not contribute to the sum of the eigenvalues, so we only had to compute the number of eigenvalues for problems formulated on actual rectangles, which is what we are able to do explicitly. However, this is not true for problems $\tilde{P}^i_N$, where one could have a rectangle intersecting both the allowed region and the boundary of $\Omega$ on which $V^i_-<E$. This issue can be avoided by refining the family $\sigma$ after the energy value $E$ has been fixed, in a way that the $R_i$ never intersect both the region where $V<E$ and the boundary $\partial\Omega$.
\end{remark}

In view of Lemma \ref{lemma_upper_Weyl}, an analogous calculation gives
\begin{equation*}
N_{\leq E}(P_D(\Omega)) \lesssim  \frac{\text{Area(}\Omega\text{)}}{\pi h^2} \int_{\Omega} \left(E-V\right)\chi_{\left\lbrace V\leq E \right\rbrace} dr_*d\theta 
\end{equation*}
for $h$ sufficiently small. Combined with \eqref{weyl_result}, this provides the relation
\begin{equation}   \label{final_weyl_relation}
N_{\leq E}(P_D(\Omega)) \sim  \frac{\text{Area(}\Omega\text{)}}{\pi h^2} \int_{\Omega} \left(E-V\right)\chi_{\left\lbrace V\leq E \right\rbrace} dr_*d\theta \, .
\end{equation}

Let us now denote by $N[E-\delta,E+\delta]$ the number of eigenvalues in $[E-\delta,E+\delta]$ for $P_D(\Omega)$. By applying \eqref{final_weyl_relation} to 
\begin{equation*}
N[E-\delta,E+\delta]=N_{\leq E+\delta}(P_D(\Omega))-N_{\leq E-\delta}(P_D(\Omega))
\end{equation*}
we obtain the version of Weyl's law of Theorem \ref{Weyl's_law_static}.

\begin{remark}
Since the result is obtained by considering $h$ sufficiently small, Weyl's law can be equivalently expressed in terms of $V$ instead of $V^h_{\textup{eff}}$. This motivates the presence of $V$ in Theorem \ref{Weyl's_law_static} and in the limit of the Riemann sum considered in our proof.
\end{remark}

\section{Eigenvalue problem for the boosted black string}  \label{sect_boosted_string}

The aim of this section is to prove Theorem \ref{main_theorem_EVP_boost_string}. To do that, we will mainly follow a perturbation argument of \cite{SharpLogHolz}.  

The eigenvalue problem for the boosted black string is \textit{nonlinear}. From \eqref{WE}, we have
\begin{align*}
&-\frac{f(r)}{r^2}\partial_r(r^2f(r)\partial_ru)-\frac{f(r)}{r^2\sin\theta}\partial_{\theta}(\sin\theta \partial_{\theta}u) \\
&+\left\lbrace\left(\frac{f(r)}{r^2\sin^2\theta}\right) m^2-\left[(1-f(r))\sinh^2\beta \right]\omega^2-[2(1-f(r))\sinh\beta\cosh\beta]\, \omega\, J \right. \\
&\left. +[1-(1-f(r))\cosh^2\beta]\, J^2\right\rbrace u 
=\omega^2 u  \, ,
\end{align*}
which reduces to 
\begin{equation*}
-g(r)\Delta_{(r_*,\theta)}u+[V_j(r,\theta)+V^{\beta}_{(\omega,m,J)}(r,\theta)]u=\omega^2 u \, ,
\end{equation*}
with
\begin{align}
V^{\beta}_{(\omega,m,J)}(r,\theta) := &\left(\frac{f(r)}{r^2\sin^2\theta}\right) m^2-\left[(1-f(r))\sinh^2\beta \right]\omega^2-[2(1-f(r))\sinh\beta\cosh\beta]\, \omega\, J  \label{pot_eigen_lin} \\
&+[1-(1-f(r))\cosh^2\beta]\, J^2   \nonumber
\end{align}
and $V_j(r,\theta)$ independent of the frequency parameters $(\omega,m,J)$. The eigenvalue problem is
\begin{equation}
\boxed{
\begin{array}{rcl} \label{nonlinear_eig_prob_1}
-g(r)\Delta_{(r_*,\theta)}u+[V_j(r,\theta)+V^{\beta}_{(\omega,m,J)}(r,\theta)]u=&\omega^2 u \quad &\textrm{on} \quad \Omega  \\
 u=&0 \quad &\textrm{on} \quad \partial\Omega
 \end{array}
 }
\end{equation}
for some \textit{fixed} open set $\Omega$ to be later specified. Remember that $u=u(r_*,\theta)$, with the radial coordinate $r_*$ previously defined. The eigenvalue problem \eqref{nonlinear_eig_prob_1} is \textit{nonlinear} in the sense that the potential $V^{\beta}_{(\omega,m,J)}$ depends on the eigenvalue $\omega^2$. 

\begin{remark}[\textbf{Comparison with Holzegel--Smulevici \cite{SharpLogHolz}}]
Potential \eqref{pot_eigen_lin} has one quadratic term and one linear term in $\omega$. In the nonlinear eigenvalue problem considered in \cite{SharpLogHolz} for Kerr-AdS black holes, $\omega$ only appears quadratically in the potential, as an effect of the assumption that solutions to the wave equation are axisymmetric (in fact, this kills the cross-term linear in $\omega$). This symmetry assumption cannot be reproduced in our setting, because the vanishing of $J$ (which is the frequency parameter that generates the cross term $\omega \, J$) would imply that, at the geodesic level, no stable trapping possibly occurs (see Lemma \ref{lemma_m_J_nonzero}). We are therefore left with a more complicated analysis of the potential and, in turn, a more involved perturbation argument than the ones presented in \cite{SharpLogHolz}. Note that the terms involving $\omega_0$ in the application of the Implicit Function Theorem (see, for instance, inequality \eqref{est1}), which are absent in \cite{SharpLogHolz}, are a manifestation of this fact. 
\end{remark}

As in \cite{SharpLogHolz}, we consider the two-parameter family of eigenvalue problems
\begin{equation}
\boxed{
\begin{array}{rcl} \label{nonlinear_eig_prob_2}
Q(b,\omega)u=&\Lambda (b,\omega)u \quad &\textrm{on} \quad \Omega  \\
 u=&0 \quad &\textrm{on} \quad \partial\Omega \, ,
 \end{array}
 }
\end{equation}
where the operator $Q(b,\omega)$ has the form
\begin{equation*}
Q(b,\omega)u := -g(r)\Delta_{(r_*,\theta)}u+[V_j(r,\theta)+V^{b\beta}_{(\omega,m,J)}(r,\theta)-\omega^2]u \, ,
\end{equation*}
with $b\in \mathbb{R}\cap [0,1]$ dimensionless parameter.

\begin{remark}[\textbf{Notation}]  \label{notation_eigenv_boost}
In what follows, we will provide statements about $\omega$ in the context of problem \eqref{nonlinear_eig_prob_2}. In most of the cases, $\omega$ will necessarily depends on the parameters appearing in \eqref{nonlinear_eig_prob_2}, namely $b$, $\beta$, $m$ and $J$. Since $\beta$ will be considered fixed and the other parameters allowed to vary, we shall introduce the notation $\omega_{b,m,J}$. However, to this notation we prefer $\omega_{b,m}$, because the frequency parameter $J$ will be treated as a rescaled version of $m$. To sum up, we denote $\omega$ in \eqref{nonlinear_eig_prob_2} by $\omega_{b,m}$ and adopt
\begin{align*}
\omega_{\textup{lin}}\equiv \omega_{\textup{lin},m}:=\omega_{0,m}\, , &&  \omega\equiv \omega_m:=\omega_{1,m} \, .
\end{align*}
\end{remark}

We aim to prove that, for $b=1$ and $m$ sufficiently large, there exists $\omega_{1,m}$ such that the abstract eigenvalue problem \eqref{nonlinear_eig_prob_2} admits a zero eigenvalue. This, in turn, would prove that the nonlinear eigenvalue problem \eqref{nonlinear_eig_prob_1} admits eigenvalues $\omega^2$.

The main idea is to make use of what we have already proven for the linear eigenvalue problem \eqref{Dirichlet_lin_eig_prob} in Theorem \ref{Weyl's_law_static} of Section \ref{sect_static_string}, namely that, for $b=0$ and $m$ sufficiently large, there exists $\omega_{0,m}$ such that $Q(0,\omega_{0,m})$ admits a zero eigenvalue, say the $n$-th eigenvalue $\Lambda_n(0,\omega_{0,m})$ is zero (where $n=n(m)$). In view of the Implicit Function Theorem, we will be able to prove that this statement remains true for any $b\in [0,1]$, i.e. for any $b\in [0,1]$ and $m$ sufficiently large, there exists $\omega_{b,m}$ such that the eigenvalue $\Lambda_n(b,\omega_{b,m})$ of $Q(b,\omega_{b,m})$ is equal to zero. In particular, this statement will hold for $b=1$, giving us the existence result for \eqref{nonlinear_eig_prob_1}.

\subsection{Preliminaries}

\subsubsection{Analysis of the potential $V^{\beta}_{(\omega,m,J)}-\omega^2$}

We collect here some properties of the potential $V^{\beta}_{(\omega,m,J)}-\omega^2$, where $V^{\beta}_{(\omega,m,J)}$ was defined in \eqref{pot_eigen_lin}. Let us define 
\begin{align}
\hat{V}^{\beta}_{(\omega,m,J)}(r,\theta) := & \, V^{\beta}_{(\omega,m,J)}(r,\theta)-\omega^2   \nonumber\\
=&\left(\frac{f(r)}{r^2\sin^2\theta}\right) m^2-\left[(1-f(r))\sinh^2\beta \right]\omega^2-[2(1-f(r))\sinh\beta\cosh\beta]\, \omega\, J  \label{prop_pot}\\
&+[1-(1-f(r))\cosh^2\beta]\, J^2   -\omega^2  \nonumber\\
=& \left(\frac{f(r)}{r^2\sin^2\theta}\right) m^2-(1-f(r))(\omega \sinh\beta+J\cosh\beta)^2+J^2-\omega^2  \nonumber \, .
\end{align}
When we do not need to specify any particular dependence on the frequency parameters, we denote $\hat{V}^{\beta}_{(\omega,m,J)}$ by $\hat{V}$.

\begin{remark}[\textbf{Fix $\boldsymbol{\theta=\pi/2}$}]
For most of this section we will be fixing $\theta=\pi/2$ and the potential will be considered as a function of $r$ only.  Note that $\partial_{\theta}\hat{V}\left(r,\pi/2\right)=0$ for any $r\geq r_0$ and $\theta=\pi/2$ is a local minimum (and the only stationary point) in the $\theta$-direction. Furthermore, for any fixed $r=\tilde{r}$ and $(\omega,m,J)$, there exists $\theta_0$ sufficiently close to $0$ (or $\pi$) such that $\hat{V}^{\beta}_{(\omega,m,J)}(\tilde{r},\theta)>0$ for $0\leq \theta\leq \theta_0$ (or $\pi-\theta_0\leq \theta \leq \pi$). Indeed, for $\theta$ small (or close to $\pi$), i.e. $\sin^2\theta$ small, the potential gains more and more positivity from the first term in \eqref{prop_pot}. The constant $\theta_0$ depends on both $\tilde{r}$ and $(\omega,m,J)$.
\end{remark}

\begin{theorem}[\textbf{Analysis of $\boldsymbol{\hat{V}^{\beta}_{(\omega,m,J)}}$}] \label{analysis_potential_boosted}
Consider the potential $\hat{V}$, with $\omega\in\mathbb{R}$ and $m,J\in\mathbb{Z}$. Assume $\cosh^2\beta<3$, where $\beta\in\mathbb{R}_{\geq 0}$ is fixed. Then, the following statements hold.
\begin{enumerate}
\item The potential $\hat{V}$ admits at most two stationary points. It has both one local maximum and one local minimum if and only if
\begin{equation}
m^2>3r_0^2(\omega\sinh\beta+J\cosh\beta)^2 \, . \label{max_min_pot}
\end{equation} 
For any $(\omega,m,J)$ satisfying \eqref{max_min_pot}, we have
\begin{equation*}
r_0<r_{\textup{max}}<3r_0<r_{\textup{min}}  \, ,
\end{equation*}
where $r_{\textup{max}}$ and $r_{\textup{min}}$ are the radial coordinates of the local maximum and local minimum respectively. For $m^2=3r_0^2(\omega\sinh\beta+J\cosh\beta)^2$, the potential $\hat{V}$ has an inflection point at $r=3r_0$, while for $m^2<3r_0^2(\omega\sinh\beta+J\cosh\beta)^2$ there are no stationary points.

\item The potential $\hat{V}$ vanishes as a cubic in $r$. In particular, the potential has (i) one real root, or (ii) three distinct real roots, or (iii) three real roots where one is a multiple root. Without any condition on the frequency parameters $(\omega,m,J)$, the roots lie in general on $(-\infty,\infty)$. 

\item If $\hat{V}^{\beta}_{(\omega_0,m_0,J_0)}$ has three distinct real roots in $(r_0,\infty)$ for some triple $(\omega_0,m_0,J_0)$, then $J_0^2>\omega_0^2$ must hold.

\item If $\omega=0$, then $\hat{V}^{\beta}_{(0,m,J)}$ does not admit three distinct real roots in $(r_0,\infty)$ for any $m,J$.

\item There exists a triple $(\omega_0,m_0,J_0)$ such that the potential $\hat{V}^{\beta}_{(\omega_0,m_0,J_0)}$ admits three distinct real roots in $(r_0,\infty)$. In view of (1),(3) and (4), for any such triple $(\omega_0,m_0,J_0)$, we have (i) $0< |\omega_0| < |J_0|$ and (ii) condition \eqref{max_min_pot} holds. In particular, there exists such a triple $(\omega_0,m_0,J_0)$ which also satisfies the condition $\omega_0 J_0>0$.

\item Consider a triple $(\omega_0,m_0,J_0)$ such that the potential $\hat{V}^{\beta}_{(\omega_0,m_0,J_0)}$ admits three distinct real roots $r_1^{\omega_0},r_2^{\omega_0},r_3^{\omega_0}\in(r_0,\infty)$, with $r_1^{\omega_0}<r_2^{\omega_0}<r_3^{\omega_0}$ and $\omega_0J_0>0$. Then, there exist $\mathcal{E}^-,\mathcal{E}^+$, with $0<\mathcal{E}^-<\mathcal{E}^+<|J_0|$, satisfying the following properties: 
\begin{enumerate}[(i)]
\item $|\omega_0|\in (\mathcal{E}^-,\mathcal{E}^+)$. 
\item for any $|\omega|\in (\mathcal{E}^-,\mathcal{E}^+)$, $\textup{sign}(\omega)=\textup{sign}(\omega_0)$, the potential $\hat{V}^{\beta}_{(\omega,m_0,J_0)}$ admits three distinct real roots in $(r_0,\infty)$.
\item For any $\omega_1,\omega_2$ such that $|\omega_1|,|\omega_2|\in (\mathcal{E}^-,\mathcal{E}^+)$, $|\omega_1|<|\omega_2|$, $\textup{sign}(\omega_1)=\textup{sign}(\omega_2)=\textup{sign}(\omega_0)$, we have $r_2^{\omega_1}>r_2^{\omega_2}$ and $r_3^{\omega_1}<r_3^{\omega_2}$, where $r_i^{\omega_j}$ denotes the $i$-th root of $\hat{V}^{\beta}_{(\omega_j,m_0,J_0)}$.
\end{enumerate}
\item For any fixed $\beta_0$, there exists a triple $(\omega_0,m_0,J_0)$, with $\omega_0J_0>0$, such that $\hat{V}^{\beta}_{(\omega_0,m_0,J_0)}$ admits three distinct real roots in $(r_0,\infty)$ for any $\beta\in [0,\beta_0]$. In particular, $\hat{V}^{0}_{(\omega_0,m_0,J_0)}$ admits three distinct real roots in $(r_0,\infty)$ and we have $r_2^{0}>r_2^{\beta_0}$ and $r_3^{0}<r_3^{\beta_0}$, with the notation $r_i^{\beta}$ for the $i$-th root of $\hat{V}^{\beta}_{(\omega_0,m_0,J_0)}$.
\end{enumerate}
\end{theorem}

\begin{remark}[\textbf{$\boldsymbol{\beta}$ is fixed}]
The boost parameter $\beta$ has to be considered fixed. The statement of the theorem holds for any fixed $\beta$ satisfying $\cosh^2\beta<3$. In particular, it holds true for $\beta=0$, i.e. for the static black string. 
\end{remark}

\begin{proof}
We start by observing two properties of $\hat{V}$:
\begin{gather}
\hat{V}\left(r_0,\theta\right)=-(\omega\cosh\beta+J\sinh\beta)^2 <0 \label{sign_pot_hor} \\
\lim_{r\rightarrow +\infty}\hat{V}\left(r,\theta\right)=J^2-\omega^2 \label{sign_pot_infty}
\end{gather}
for any $(\omega,m,J)$. We present the proof divided in parts, accordingly to the statements of the theorem.
\begin{enumerate}
\item We compute the derivative
\begin{equation*}
\frac{\partial \hat{V}}{\partial r}\left(r,\frac{\pi}{2}\right)=\frac{1}{r^4}\left[ m^2(3r_0-2r)+r_0(J\cosh\beta+\omega\sinh\beta)^2r^2   \right] \, .
\end{equation*}
Note that this is positive at $r=r_0$. By imposing $\frac{\partial \hat{V}}{\partial r}(r,\pi/2)=0$, we find that the potential $\hat{V}$ has at most two stationary points with radial coordinate 
\begin{align*}
r_{\text{max}}&=\frac{m^2-\sqrt{m^2 \left[m^2-3 r_0^2 (\omega \sinh \beta +J \cosh \beta )^2\right]}}{r_0 (\omega \sinh \beta +J \cosh \beta )^2} \\
r_{\text{min}}&=\frac{m^2+\sqrt{m^2 \left[m^2-3 r_0^2 (\omega \sinh \beta +J \cosh \beta )^2\right]}}{r_0 (\omega \sinh \beta +J \cosh \beta )^2} \, ,
\end{align*}
with the condition 
\begin{equation}
m^2>3 r_0^2 (\omega \sinh \beta +J \cosh \beta )^2 \label{con_stat_pt}
\end{equation}
for the stationary points to exist and be distinct. Using \eqref{con_stat_pt}, we have
\begin{equation*}
r_{\text{min}}>3r_0+\text{positive term}>3r_0 \, .
\end{equation*}
Furthermore, we clearly have $r_{\text{max}}<r_{\text{min}}$ and
\begin{align*}
r_{\text{max}}&=\frac{3m^2}{(\omega \sinh \beta +J \cosh \beta )^2 r_{\text{min}}} \\
&\geq \frac{3m^2}{(\omega \sinh \beta +J \cosh \beta )^2}\left[\frac{r_0 (\omega \sinh \beta +J \cosh \beta )^2}{2m^2}\right] = \frac{3}{2}r_0 \\
&\geq r_0 \, .
\end{align*}
Since $\frac{\partial \hat{V}}{\partial r}\left(r_0,\pi/2\right)>0$ and the two stationary points are in $(r_0,\infty)$, we can conclude that $r_{\text{max}}$ is really the radial coordinate of a local \textit{maximum} and $r_{\text{min}}$ of a local \textit{minimum}. For $m^2=3 r_0^2 (\omega \sinh \beta +J \cosh \beta )^2$, we have $r_{\text{max}}=r_{\text{min}}=3r_0$ and $\frac{\partial^2 \hat{V}}{\partial r^2}\left(3r_0,\pi/2\right)=0$, so $r=3r_0$ is the radial coordinate of an inflection point.

\item We set $\sin\theta=1$ and rewrite $\hat{V}$ as 
\begin{equation} 
\frac{1}{r^3}\left[(J^2-\omega^2)r^3-r_0(\omega\sinh\beta +J\cosh\beta)^2r^2+m^2r-m^2r_0\right] \, , \label{pot_cubic_r}
\end{equation} 
which vanishes as a cubic in $r$. 

\item From \eqref{sign_pot_hor}, if $\hat{V}$ has three distinct real roots in $(r_0,\infty)$, then we must have $\lim_{r\rightarrow +\infty}\hat{V}>0$, which implies $J^2>\omega^2$ from \eqref{sign_pot_infty}.  

\item Suppose that $\hat{V}$ admits three distinct real roots in $(r_0,\infty)$, with  $r_1<r_2<r_3$. Then the largest root must satisfy $r_{3}>3r_0$. This is true because, if $\hat{V}$ admits three roots in $(r_0,\infty)$, then it must have both one local maximum and one local minimum, with $r_{\text{min}}>3r_0$. In view of \eqref{sign_pot_hor} and point (3), we need to have $r_{3}>r_{\text{min}}>3r_0$. However, if we set $\omega=0$, the potential becomes
\begin{equation*}
\hat{V}^{\beta}_{(0,m,J)}(r,\theta)= \left(\frac{f(r)}{r^2\sin^2\theta}\right) m^2+[1-(1-f(r))\cosh^2\beta]J^2 \, .
\end{equation*}
\underline{Using assumption $\cosh^2\beta<3$}, potential $\hat{V}^{\beta}_{(0,m,J)}(r,\theta)$ is positive for any $r>3r_0$, contradicting $r_{3}>3r_0$. This implies that we need to have $\omega\neq 0$. In particular, for any fixed $m,J$, the frequency parameter $\omega$ has to be bounded away from zero, where the lower bound depends on $m$ and $J$.

\item From point (2), we know that the potential is a cubic in $r$, whose discriminant  
\begin{align*}
D=&-4  \left(J^2-\omega^2\right)m^4  \\
&+ \left[-27 r_0^2 \left(J^2-\omega^2\right)^2+18 r_0^2 \left(J^2-\omega^2\right) \left(\omega  \sinh \beta +J  \cosh \beta \right)^2  \right. \\
& \left. +r_0^2\left(\omega  \sinh \beta +J  \cosh \beta \right)^4\right]m^2   
   -4  r_0^4 \left(\omega  \sinh \beta +J  \cosh \beta \right)^6  
\end{align*}
is positive if and only if the potential admits three distinct real roots in $(-\infty,+\infty)$. We claim that, for the triple $(\omega,m,J)$ with $J^2=\omega^2+\varepsilon$ and $m^2= 5r_0^2(\omega  \sinh \beta +J  \cosh \beta )^2$, the discriminant is indeed positive for sufficiently small $\varepsilon>0$.\footnote{To ensure that $m,J\in\mathbb{Z}$, one can choose $\omega=\omega(\varepsilon)$ such that $\omega^2+\varepsilon\in\mathbb{Z}$ and take the integer part of $5r_0^2(\omega  \sinh \beta +J  \cosh \beta )^2$.} With such a triple, $D$ becomes
\begin{align*}
D=&  \varepsilon\left[ -4m^4  +18 r_0^2  \left(\omega  \sinh \beta +J  \cosh \beta \right)^2  m^2\right]-27 r_0^2 \varepsilon^2m^2\\
&  +r_0^2\left(\omega  \sinh \beta +J  \cosh \beta \right)^4 m^2  
   -4  r_0^4 \left(\omega  \sinh \beta +J  \cosh \beta \right)^6  \, .
\end{align*} 
The expression for $m^2$, which correctly satisfies \eqref{max_min_pot}, makes the second line positive. By choosing $\varepsilon$ sufficiently small, we conclude $D>0$. This proves the existence of a triple $(\omega,m,J)$ for which $\hat{V}$ admits three distinct real roots. In particular, we do not have to impose a sign for $\omega \, J$, so the existence of the triple is compatible with $\omega\, J$ being positive. We now need to check that the roots all lie in $(r_0,\infty)$. To do that, note that the potential $\hat{V}$ is everywhere negative for $0<r\leq r_0$, $\lim_{r\rightarrow 0^{\pm}}\hat{V}=\mp \infty$ and $\hat{V}$ is everywhere positive for $r<0$ because sum of positive terms (using $J^2-\omega^2=\varepsilon >0$). Thus, the potential can only vanish for $r\in (r_0,\infty)$.

\item Part (i) uses point (5), while part (ii) comes from the continuity of $\hat{V}^{\beta}_{(\omega,m_0,J_0)}$ in $\omega$ (note that $m_0$ and $J_0$ are fixed). For part (iii), we note the following monotonicity property of $\hat{V}^{\beta}_{(\omega,m_0,J_0)}$. If $\omega\, J_0>0$, then, for any fixed $r=\tilde{r}$, \eqref{prop_pot} shows that $\hat{V}^{\beta}_{(\omega,m_0,J_0)}\left(\tilde{r},\pi/2\right)$ increases when $\omega$ decreases, because all the terms involving $\omega$ are negative. 

\item We look at the discriminant $D$ and suppose $\omega_0,J_0>0$ (the case $\omega_0,J_0<0$ is equivalent). We again define $J_0^2=\omega_0^2+\varepsilon$ and $m_0^2= 5r_0^2(\omega_0  \sinh \beta_0 +J_0 \cosh \beta_0 )^2$. As proved in point (5), this gives the existence of three distinct real roots for $\hat{V}^{\beta_0}_{(\omega_0,m_0,J_0)}$ when $\varepsilon$ is small enough. Note that, for any $\beta\in [0,\beta_0)$, we have $m_0^2>5r_0^2(\omega_0  \sinh \beta +J_0 \cosh \beta )^2$, which implies $D>0$ for any $\beta\in [0,\beta_0)$ and $\varepsilon$ sufficiently small ($\varepsilon$ depends on $\beta_0$, but uniform in $\beta$). The monotonicity property of the roots derives form the fact that, for $(\omega_0,m_0,J_0)$ with $\omega_0J_0>0$, all the terms of $\hat{V}^{\beta}_{(\omega_0,m_0,J_0)}$ depending on $\beta$ are negative and decrease in absolute value when $\beta$ decreases.

\end{enumerate}

\end{proof}

\subsubsection{The frequency parameters $m$ and $J$}

\begin{lemma} \label{lemma_m_J_nonzero}
Let $m,J\in\mathbb{Z}$. If $m=0$, then $\hat{V}$ does not admit three distinct real roots in $(r_0,\infty)$. Analogously, if $J=0$, then $\hat{V}$ does not admit three distinct real roots in $(r_0,\infty)$.
\end{lemma}

\begin{proof}
If $m=0$, then condition \eqref{max_min_pot} of Theorem \ref{analysis_potential_boosted} implies that $\hat{V}$ has a local maximum and minimum in $(r_0,\infty)$ if and only if $\omega$ and $J$ satisfy
\begin{equation*}
3r_0^2(\omega\sinh\beta+J\cosh\beta)^2<0 \, .
\end{equation*} 
Therefore, $\hat{V}$ does not admit stationary points in $(r_0,\infty)$, so it cannot have three distinct real roots in such interval. If $J=0$ and $\hat{V}$ has three distinct real roots in $(r_0,\infty)$, then we know from point (3) of Theorem \ref{analysis_potential_boosted} that $J^2>\omega^2$ must hold, which gives a contradiction. 
\end{proof}

Motivated by Lemma \ref{lemma_m_J_nonzero}, we will assume
\begin{align}
m\neq 0, \,\,  m>0 && J\neq 0  \label{sign_m}
\end{align}
throughout the discussion. In fact, the sign of $m$ does not play any role because $m$ only appears squared in both eigenvalue problems \eqref{Dirichlet_lin_eig_prob} and \eqref{nonlinear_eig_prob_1}. As done for the eigenvalue problem \eqref{Dirichlet_lin_eig_prob}, we further assume 
\begin{align}
J=c\, m   && c^2<\frac{1}{3r_0^2}  \label{cond_J}
\end{align}
for some \textit{fixed}, \textit{positive} constant $c>0$. We will not make a substitution $J\rightarrow c\, m$ in our equations, but the reader should keep in mind that, from now on, $J$ is really a rescaled version of the frequency parameter $m$. In view of \eqref{sign_m}, we have $J>0$. We summarize all the assumptions on $m$ and $J$ in the following:
\begin{claim}[\textbf{Assumptions on $\boldsymbol{m}$ and $\boldsymbol{J}$}] \label{claim_m_J}
Let $m,J\in\mathbb{Z}$. We assume
\begin{align*}
m\neq 0 && J=c\, m, \,\, J>0, \, \, c\in\mathbb{Q}
\end{align*}
for some fixed constant $c$ such that $0<c^2<1/3r_0^2$.
\end{claim}

\begin{remark} 
Note that: 
\begin{enumerate}[(i)]
\item In the linear problem \eqref{Dirichlet_lin_eig_prob} the sign of the frequency parameters $(\omega,m,J)$ is not relevant, because they all appear squared in the potential. For the nonlinear problem \eqref{nonlinear_eig_prob_1}, the sign ambiguity becomes relevant when one has to deal with the cross-term $\omega\, J$ in potential $\hat{V}$. See, for instance, the proof of Lemma \ref{IFT_boost_string}. 
\item What we really want to assume is $J>0$, while parameters $m$ and $J$ do not need to have the same sign (one could equivalently assume $m<0$ and $c<0$). Most of the statements in the following sections will require $m^2,J^2$ sufficiently large. Claim \ref{claim_m_J} assumes that, in the limit $J^2\rightarrow\infty$, we have $J\rightarrow +\infty$ (and not $J\rightarrow -\infty$).  
\item The results of Theorem \ref{analysis_potential_boosted} did not require any assumption on the sign of $J$. Therefore, all the statements therein remain true after assuming $J>0$. 
\end{enumerate}
\end{remark}

\begin{remark}[\textbf{The choice of sign for $\boldsymbol{J}$ does not matter}] \label{rmk_sign_J}
One could equivalently assume $J<0$ in Claim \ref{claim_m_J} and go through the same arguments that we are about to present. In other words, one only has to eliminate the sign ambiguity of $J$ from the problem, but the particular choice of sign is not important.  
\end{remark}

\subsubsection{How we choose $\Omega$ for \eqref{nonlinear_eig_prob_1}}

The choice of $\Omega$ and suitable energy levels $E$ for the nonlinear eigenvalue problem \eqref{nonlinear_eig_prob_1} is specified by the following proposition, which has to be seen as the nonlinear analogue of Lemma \ref{Cont_pot_static_string}.
\begin{prop} \label{omega_nonlinear}
Define $\hat{V}^{\beta}_{\textup{min}}$ to be the minimum of $\hat{V}^{\beta}_{(\omega,m,J)}$ and $x_{\textup{min}}\in (r_0,\infty)\times (0,\pi)$ such that $\hat{V}^{\beta}_{(\omega,m,J)}(x_{\textup{min}})=\hat{V}^{\beta}_{\textup{min}}$. For suitably $m,J\in\mathbb{Z}$ (as in Claim \ref{claim_m_J}) and $\beta\in\mathbb{R}_+$ (as assumed in Theorem \ref{analysis_potential_boosted}), consider some constant $\mathcal{E}>0$ such that $\hat{V}^{\beta}_{(\mathcal{E}m,m,J)}$ has a local minimum and such that there exists $\Omega\subset [r_0,\infty)\times [0,\pi)$ satisfying
\begin{enumerate}[(i)]
\item $x_{\textup{min}}\in \Omega$,
\item $\hat{V}^{\beta}_{(\mathcal{E}m,m,J)}(x)=0$ for $x\in \partial\Omega$,
\item there are no local maxima of $\hat{V}^{\beta}_{(\mathcal{E}m,m,J)}$ in $\Omega$,
\item $x\in\Omega \implies r(x)>3\, r_0$. 
\end{enumerate}
Fix now some energy level $E>0$ such that 
\begin{enumerate}[(a)]
\item $\hat{V}^{\beta}_{(Em,m,J)}$ has a local minimum and there exists $\Omega^{\prime}\subset\Omega$ satisfying the same properties (i)-(iv) of $\Omega$, but now with respect to $\hat{V}^{\beta}_{(Em,m,J)}$,
\item $\hat{V}^{0}_{(Em,m,J)}$ has a local minimum and there exists $\Omega^{\prime\prime}\subset\Omega$ satisfying the same properties (i)-(iv) of $\Omega$, but now with respect to $\hat{V}^{0}_{(Em,m,J)}$. 
\end{enumerate}
Then, the final part of Lemma \ref{Cont_pot_static_string} holds for $\frac{1}{m^2}\hat{V}^{0}_{(Em,m,J)}$ with respect to the open set $\Omega$. Furthermore, for any sufficiently small constants $\delta,\delta^{\prime}>0$, there exists some constant $c>0$ such that 
\begin{equation*}
\textup{dist}(x,\partial\Omega)<\delta^{\prime} \implies \frac{1}{m^2} \hat{V}^{\beta}_{(\kappa\, m,m,J)}(x)>c
\end{equation*}
for all $\kappa\in\mathbb{R}$ satisfying $|\kappa^2-E^2| \leq\delta$, with $\textup{dist}(\cdot,\cdot)$ the Euclidean distance.
\end{prop}

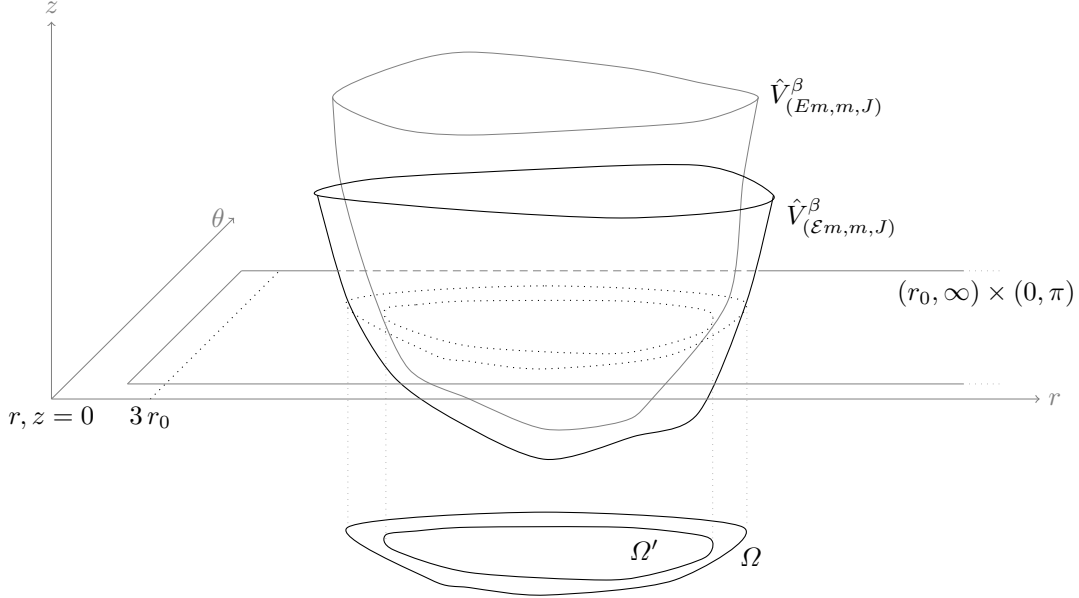
\begin{figure}[H]
\centering
\begin{tikzpicture}

\draw plot [smooth] coordinates {(3,5.5) (2.65,4.1) (2,2.6) (1.15,2.3) (0,2) (-1.15,2.5) (-2,3.1) (-2.6,4.1) (-3,5.5) };

\draw[help lines] plot [smooth] coordinates {(2.8,6.8) (2.6,5.7) (2.4,4.1) (1.4,2.8) (1,2.5) (0,2.4) (-1,2.8) (-1.8,3.2) (-2.3,4.3) (-2.7,5.7) (-2.8,6.8)};

\draw  plot [smooth cycle] coordinates {(-3,5.5) (-2.4,5.7) (-1,5.8) (2,5.9) (3,5.5) (2.5,5.3) (1,5.2) (-1,5.3)};

\draw[help lines]  plot [smooth cycle] coordinates {(-2.8,6.8)(-2,7.2)(-1,7.4)(1,7.2)(2,7) (2.8,6.8) (2.1,6.5) (1,6.4) (-1,6.3) (-2,6.4)};

\draw[dotted]  plot [smooth cycle] coordinates {(-2.1,4)(-1.7,4.05)(-1,4.1)(1,4.1)(2,4) (2.2,3.9)  (2.1,3.7)(1.5,3.5) (0.8,3.4) (-1,3.5) (-2,3.8)};

\draw[help lines] (-5.5,3) -- (5.5,3);
\draw[help lines] (-5.5,3) -- (-4,4.5);
\draw[help lines] (-4,4.5) -- (-2.8,4.5);
\draw[help lines] (2.8,4.5) -- (5.5,4.5);
\draw[dotted, help lines] (5.5,4.5)--(6,4.5);
\draw[dotted, help lines] (5.5,3)--(6,3);
\draw [densely dashed, help lines] (-2.8,4.5) -- (2.8,4.5);

\draw[dotted]  plot [smooth cycle] coordinates {(-2.6,4.1) (0,4.3)  (2.6,4.1)  (1.7,3.4) (0,3.2) (-1,3.3) (-1.5,3.4) };

\draw  plot [smooth cycle] coordinates {(-2.6,1.1) (0,1.3)  (2.6,1.1)  (1.7,0.4) (0,0.2) (-1,0.3) (-1.5,0.4) };

\draw  plot [smooth cycle] coordinates {(-2.1,1)(-1.7,1.05)(-1,1.1)(1,1.1)(2,1) (2.2,0.9)  (2.1,0.7)(1.5,0.5) (0.8,0.4) (-1,0.5) (-2,0.8)};

\draw [dotted, help lines] (2.65,4.1)--(2.65,1.1);
\draw [dotted, help lines] (-2.6,4.1)--(-2.6,1.1); 

\draw [dotted, help lines] (-2.1,0.95)--(-2.1,4);
\draw [dotted, help lines] (2.2,3.9)--(2.2,0.9);

\draw[dotted] (-5.2,2.8)--(-3.5,4.5);

\node at (2.7,0.7) {$\Omega$};
\node at (1.3,0.8) {$\Omega^{\prime}$};
\node at (3.9,5.2) {$\hat{V}^{\beta}_{(\mathcal{E}m,m,J)}$};
\node at (-6.5,2.55) {$r,z=0$};
\node at (-5.2,2.55) {$3\, r_0$};
\node at (3.7,6.8) {$\hat{V}^{\beta}_{(Em,m,J)}$};
\node at (5.8,4.2) {$(r_0,\infty)\times (0,\pi)$};

\draw[->,help lines] (-6.5,2.8)--(6.5,2.8) node[right]{$r$};
\draw[->,help lines] (-6.5,2.8)--(-6.5,7.8) node[above]{$z$};
\draw[->,help lines] (-6.5,2.8)--(-4.1,5.2) node[left]{$\theta$};

\end{tikzpicture}

\caption{Conditions (i)-(iv) on $\Omega$ imply that the potential $\hat{V}^{\beta}_{(\mathcal{E}m,m,J)}$ is negative on $\Omega$, the same being true for $\hat{V}^{\beta}_{(Em,m,J)}$ on $\Omega^{\prime}$. Note also that condition (iv) implies condition (iii) because $r_{\textup{max}}<3\, r_0$, so the latter is in this sense redundant. Similarly, condition (iv) also ensures that $\hat{V}^{\beta}_{(Em,m,J)}$ (and $\hat{V}^{0}_{(Em,m,J)}$) has no local maxima on $\Omega\setminus \Omega^{\prime}$ (and $\Omega\setminus \Omega^{\prime\prime}$). They have therefore positive sign on such sets.}
\label{fig:Omega_boosted_string}
\end{figure}

\begin{remark}
Note that, a priori, it is not obvious that constants $\mathcal{E}$ and $E$ with the properties required by Proposition \ref{omega_nonlinear} do exist. The existence of such constants is the content of parts (5)-(6)-(7) of Theorem \ref{analysis_potential_boosted}. In particular, we have $E<\mathcal{E}$ in view of (6)-(iii) in Theorem \ref{analysis_potential_boosted}. Note also that condition (iv) of Proposition \ref{omega_nonlinear} is always realizable, since by (6)-(iii) in Theorem \ref{analysis_potential_boosted} one can continuously vary $\omega$ and obtain roots of the potential arbitrarily close to the local minimum, for which $r_{\textup{min}}>3\, r_0$ holds.
\end{remark}

\begin{remark} \label{correct_exist_eig_lin_1}
As one can see from Proposition \ref{omega_nonlinear}, the choice of $\Omega$ and the energy level $E$ for the nonlinear problem is slightly more delicate than the one introduced in Lemma \ref{Cont_pot_static_string} for the linear problem. One reason for this is the non-linearity of the problem, namely the fact that the potential varies when we vary the energy level $E$. On the other hand, our formulation of Proposition \ref{omega_nonlinear} (in particular point (b)) aims to 
fix an energy level $E$ which also agrees with the assumptions of Theorem \ref{Weyl's_law_static} for the linear eigenvalue problem with potential $\hat{V}^{0}_{(Em,m,J)}$. This is because we will need to apply Weyl's law of Theorem \ref{Weyl's_law_static} in our perturbation argument. By part (7) of Theorem  \ref{analysis_potential_boosted}, a choice of $E$ with these properties is possible. 

\end{remark}

\begin{remark}[\textbf{A condition on $\boldsymbol{\beta}$}] 
Boosted black strings present an ergosurface at $r=r_0\cosh^2\beta$. From now on, we assume that the real parameter $\beta\geq 0$ satisfies
\begin{equation}
\cosh^2\beta<3 \, .\label{cond_beta}
\end{equation}
In view of condition (iv) of Proposition \ref{omega_nonlinear}, our assumption ensures that the open set $\Omega$ is disjoint from the ergoregion in physical space. This is key for Lemma \ref{lemma_omega_notzero} and, even more importantly, for the proof of Lemma \ref{IFT_boost_string}.
\end{remark}

The open set $\Omega$ is uniquely determined by the choice of the constant $\mathcal{E}$. This concludes the formulation of the eigenvalue problem \eqref{nonlinear_eig_prob_1}. In what follows, the set $\Omega$ will be considered \textit{fixed} throughout.

\subsection{A lower bound for $\omega^2$}

In this section we prove that the claim that the $n$-th eigenvalue $\Lambda_n(b,\omega)$ of $Q(b,\omega)$ is zero determines some compatibility conditions on $\omega_{b,m}$. 

Let $\psi_n(b,\omega)$ be an eigenfunction associated to the eigenvalue $\Lambda_n(b,\omega)=0$. Assuming that $\psi_n(b,\omega)$ is normalized on $\Omega$, i.e. $\int_{\Omega}|\psi_n(b,\omega)|^2\, dr_* d\theta=1$, we have 
\begin{equation}
\Lambda_n(b,\omega)=\int _{\Omega} \psi_n(b,\omega) Q(b,\omega) \psi_n(b,\omega)\, dr_* d\theta=0  \, . \label{low_bound_omega_cond}
\end{equation}
It turns out that $\omega_{b,m}$ needs to satisfy some a priori conditions for \eqref{low_bound_omega_cond} to be possible. The following lemma states that we certainly need to have $\omega_{b,m}\neq 0$.

\begin{lemma} \label{lemma_omega_notzero}
Let $\psi_n(b,\omega)$ be a non identically zero eigenfunction of \eqref{nonlinear_eig_prob_2}, normalized on $\Omega$. If $\beta$ satisfies \eqref{cond_beta} and $m^2,J^2>M$, for some positive constant $M$, then the implication 
\begin{equation*}
\omega_{b,m}=0 \implies \int _{\Omega} \psi_n(b,\omega) Q(b,\omega) \psi_n(b,\omega)\, dr_* d\theta \neq 0
\end{equation*}
holds for any $b\in [0,1]$. 
\end{lemma}

\begin{remark}
In Lemma \ref{lemma_omega_notzero} and Corollary \ref{corollary_lower_bound_omega} we require $m^2$ large. This assumption finds motivation later in our discussion, when we will be looking at certain energy estimates in a high frequency limit. Note also that, in view of \eqref{cond_J}, assuming $m^2$ large implies $J^2$ large as well, so the hypothesis on $J^2$ is in some sense redundant.
\end{remark}

\begin{proof}
Setting $\omega_{b,m}=0$, we have
\begin{align*}
&\int _{\Omega} \psi_n(b,0) Q(b,0) \psi_n(b,0)\, dr_* d\theta \\
&=\int _{\Omega}   -\psi_n(b,0)g(r)\Delta_{(r_*,\theta)}\psi_n(b,0)+ \left[V_j(r,\theta)+V^{b\beta}_{(0,m,J)}(r,\theta)\right]\psi_n^2(b,0) \, dr_* d\theta \\
&=  \int _{\Omega}  g(r) \left(\left\lvert  \frac{\partial \psi_n(b,0)}{\partial r_*} \right\rvert ^2 +\left\lvert  \frac{\partial \psi_n(b,0)}{\partial \theta} \right\rvert ^2 \right)+ \psi_n(b,0)\partial_{r_*}\psi_n(b,0)\partial_{r_*}g(r) \\  
&+\left[V_j(r,\theta)+V^{b\beta}_{(0,m,J)}(r,\theta)\right]\psi_n^2(b,0) \, dr_* d\theta \\
&\geq \int _{\Omega}  \left(g(r)-\varepsilon|\partial_{r_*}g(r)|^2\right) \left\lvert  \frac{\partial \psi_n(b,0)}{\partial r_*} \right\rvert ^2 +g(r)\left\lvert  \frac{\partial \psi_n(b,0)}{\partial \theta} \right\rvert ^2  \\
&+\left[V_j(r,\theta)+V^{b\beta}_{(0,m,J)}(r,\theta)-\frac{1}{4\varepsilon}\right]\psi_n^2(b,0) \, dr_* d\theta \, ,
\end{align*}
where we have integrated by parts, used the boundary condition on $\psi_n(b,\omega)$ and Young's inequality. We first set $b=1$ and we look at  
\begin{equation}
V^{\beta}_{(0,m,J)}(r,\theta)=
\left(\frac{f(r)}{r^2\sin^2\theta}\right) m^2  +[1-(1-f(r))\cosh^2 \beta]\, J^2  \, , \label{pot_omega_zero}
\end{equation}
where the first term is always positive, while the second term becomes negative inside the ergoregion $r<r_0\cosh^2\beta$. We now recall that we are integrating over $\Omega$, which is an open set satisfying the conditions of Proposition \ref{omega_nonlinear}. Furthermore, $\beta$ satisfies condition \eqref{cond_beta}. This implies that the second term of \eqref{pot_omega_zero} is positive on $\Omega$. Note that $\cosh^2 b\beta\leq\cosh^2\beta<3$, so this gives us a positive sign for the second term of \eqref{pot_omega_zero} on $\Omega$ for any $b\in [0,1]$.

By choosing $\varepsilon$ sufficiently small (independently of $m,J$) and $m^2, J^2$ sufficiently large, the integral involves only positive terms ($V_j(r,\theta)$ and $1/4\varepsilon$ can be both absorbed by $V^{b\beta}_{(0,m,J)}$ when $m$ and $J$ are large). The integral is therefore positive for $m^2$ and $J^2$ sufficiently large, unless $\psi_n=0$ identically on $\Omega$. In particular, the integral is nonzero.

\end{proof}

We now derive a second condition for $\omega_{b,m}$, namely a lower bound. The proof follows the same idea of that of Lemma \ref{lemma_omega_notzero}. We state the result as a corollary.

\begin{corollary} \label{corollary_lower_bound_omega}
Let $\psi_n(b,\omega)$ be a non identically zero eigenfunction of \eqref{nonlinear_eig_prob_2}, normalized on $\Omega$. If $\beta$ satisfies \eqref{cond_beta}, then the implication
\begin{gather}
\textup{$\omega^2_{b,m}=o(m^2)$ as $m^2,J^2$ are sufficiently large}   \label{lower_scale_omega}\\
 \implies \int _{\Omega} \psi_n(b,\omega) Q(b,\omega) \psi_n(b,\omega)\, dr_* d\theta \neq 0 \nonumber
\end{gather}
holds for any $b\in [0,1]$.
\end{corollary}

\begin{proof}

The case $\omega_{b,m}^2=0$ is the content of Lemma \ref{lemma_omega_notzero}. We therefore discuss the case $\omega_{b,m}^2>0$. Note that, in view of \eqref{cond_J}, condition \eqref{lower_scale_omega} still holds with $m$ replaced by $J$. 

Condition \eqref{lower_scale_omega} implies that we can always choose a constant $C>0$ sufficiently small and $m^2, J^2$ sufficiently large such that $V^{b\beta}_{(\omega,m,J)}(r,\theta)-\omega_{b,m}^2$ is positive on $\Omega$ for any $b\in[0,1]$. This is true because 
\begin{align*}
V^{b\beta}_{(\omega,m,J)}(r,\theta)-\omega_{b,m}^2\geq & \left(\frac{f(r)}{r^2\sin^2\theta}\right) m^2  +[1-(1-f(r))\cosh^2 b\beta]\, J^2 \\
&-C[1+(1-f(r))\sinh^2b\beta+  2(1-f(r))\sinh b\beta\cosh b\beta ]m^2
\end{align*}
remains positive when $C$ is sufficiently small. The rest of the argument is analogous to the one presented in the proof of Lemma \ref{lemma_omega_notzero}.  

\end{proof}

In view of Corollary \ref{corollary_lower_bound_omega}, we conclude that, if the eigenvalue problem \eqref{nonlinear_eig_prob_1} admits eigenvalues $\omega_m^2$, then, for $m^2$ sufficiently large, there exists a positive constant $C_{r_0,\beta}$, \ul{independent of $m$}, such that
\begin{equation}
\omega_m^2\geq C_{r_0,\beta}\, m^2 \, . \label{lower_scaling_omega}
\end{equation}
In the following section, by an application of the Implicit Function Theorem, we will be able to prove the existence of such eigenvalues and produce an upper bound of the type $\omega_m^2\leq C_{r_0,\beta}\, m^2$ when $m^2$ is sufficiently large.

\subsection{An application of the Implicit Function Theorem}

We now state the key lemma for the nonlinear eigenvalue problem \eqref{nonlinear_eig_prob_1}. In doing this, we will make an assumption on the sign of $\omega_{b_0,m}$, for some $b_0\in [0,1]$. In the application of the lemma to our problem, this sign assumption on $\omega_{b_0,m}$ will become an assumption on the sign of $\omega_{\text{lin},m}$, where $\omega^2_{\text{lin},m}$ are the eigenvalues of the \textit{linear} eigenvalue problem \eqref{Dirichlet_lin_eig_prob}. Since $\omega_{\text{lin},m}$ only appears squared in the linear problem, this choice of sign is free and does not determine any loss of generality.

\begin{lemma}[adapted from \cite{SharpLogHolz} Lemma 4.3] \label{IFT_boost_string}
Let $b_0\in [0,1]$ and $\omega_{b_0,m}> 0$ be such that the $n$-th eigenvalue of $Q(b_0,\omega_{b_0,m})$ is zero. We further assume $m,J\in\mathbb{Z}$ as in Claim \ref{claim_m_J}. Then, for $m^2,J^2$ sufficiently large, there exists a constant $\varepsilon >0$ (independent of $b_0$) such that the following property holds. There exists a differentiable function $\omega_{b,m}(b)$ such that the $n$-th eigenvalue of $Q(b,\omega_{b,m})$ is zero for any $b\in( \max (0,b_0-\varepsilon),b_0+\varepsilon)$.
\end{lemma}

\begin{proof}
Recall that the $n$-th eigenvalue of $Q(b,\omega_{b,m})$ follows the formula
\begin{equation*}
\Lambda_n(b,\omega)=\int _{\Omega} \psi_n(b,\omega) Q(b,\omega) \psi_n(b,\omega)\, dr_* d\theta \, ,
\end{equation*}
where the associated eigenfunction $\psi_n(b,\omega)$ has been renormalized on $\Omega$. Our assumption on the $n$-th eigenvalue of $Q(b_0,\omega_{b_0,m})$ claims that $\Lambda_n(b_0,\omega_{b_0,m})=0$.

The idea of the proof is to apply the Implicit Function Theorem to the equation $\Lambda_n(b,\omega)=0$ in a neighbourhood of the point $(b_0,\omega_{b_0,m})$, where $\Lambda_n$ is a function of $b$ and $\omega_{b,m}$. To apply the Implicit Function Theorem, we first need to compute the derivatives of $\Lambda_n$ with respect to $\omega_{b,m}$ and $b$ at the point $(b_0,\omega_{b_0,m})$ and check that they are non-zero. Note that, at this stage, we are interested in proving the existence of a solving function $\omega_{b,m}(b)$, so only the derivative of $\Lambda_n$ with respect to $\omega_{b,m}$ needs to be checked. We have
\begin{gather}
\frac{\partial\Lambda_n}{\partial\omega}(b_0,\omega_{b_0,m})=\int _{\Omega} \psi_n(b_0,\omega_{b_0,m}) \left(\frac{\partial V^{b_0\beta}_{(\omega,m,J)} }{\partial\omega} (b_0,\omega_{b_0,m})   -2\omega_{b_0,m} \right)\psi_n(b_0,\omega_{b_0,m})\, dr_* d\theta  \label{deriv_lambda_omega} \\
\frac{\partial\Lambda_n}{\partial b}(b_0,\omega_{b_0,m})=\int _{\Omega} \psi_n(b_0,\omega_{b_0,m}) \left(\frac{\partial V^{b_0\beta}_{(\omega,m,J)} }{\partial b} (b_0,\omega_{b_0,m})    \right)\psi_n(b_0,\omega_{b_0,m})\, dr_* d\theta \, . \nonumber
\end{gather}
If integral \eqref{deriv_lambda_omega} is non-zero, then an application of the Implicit Function Theorem allows us to solve for $\omega_{b,m}$ as a function of $b$ in a neighbourhood of the point $(b_0,\omega_{b_0,m})$ and close the argument. To check this, we compute
\begin{align}
\frac{\partial V^{b_0\beta}_{(\omega,m,J)} }{\partial\omega} (b_0,\omega_{b_0,m})   -2\omega_{b_0,m}=& -2\left[(1-f(r))\sinh^2 b_0\beta +1\right]\omega_{b_0,m} \label{est1} \\
&-[2(1-f(r))\sinh b_0\beta\cosh b_0\beta]\, J \nonumber \\
\leq & -2\, \omega_{b_0,m} \, ,  \nonumber
\end{align}
where the estimate holds because $\omega_{b_0,m}>0$ (by assumption of the theorem) and $J>0$ (by Claim \ref{claim_m_J}). 

From Corollary \ref{corollary_lower_bound_omega}, we have $\omega_{b_0,m}\geq C_{b_0,r_0,\beta}\, m$ for some constant $C_{b_0,r_0,\beta}>0$ (independent of $m$) and $m$ sufficiently large. In particular, there exists a constant $B_{r_0,\beta}:=\inf_{b_0\in [0,1]}C_{b_0,r_0,\beta}>0$ independent of $b_0$ such that $\omega_{b_0,m}\geq B_{r_0,\beta}\, m$. We deduce
\begin{equation}
\frac{\partial V^{b_0\beta}_{(\omega,m,J)} }{\partial\omega} (b_0,\omega_{b_0,m})   -2\omega_{b_0,m} \leq -B_{r_0,\beta} \, m \, .\label{est1_bis}
\end{equation}
Therefore, \eqref{est1} is bounded away from zero \textit{uniformly in $b_0$}. This implies that integral \eqref{deriv_lambda_omega} is bounded away from zero uniformly in $b_0$ as well. We can therefore conclude the proof by Implicit Function Theorem. Note that the fact that estimate \eqref{est1_bis} is uniform in $b_0$ implies that the constant $\varepsilon$ is independent of $b_0$.

\end{proof}

\begin{remark}

The Implicit Function Theorem also provides a formula for the derivative of the function $\omega_{b,m}(b)$ at $b=b_0$, namely 
\begin{equation}
\frac{d\omega_{b,m}}{db}(b_0)=-\frac{\frac{\partial\Lambda_n}{\partial b}(b_0,\omega_{b_0,m})}{\frac{\partial\Lambda_n}{\partial\omega}(b_0,\omega_{b_0,m})} \, . \label{deriv_omega}
\end{equation}
By computing
\begin{equation*}
\frac{\partial V^{b\beta}_{(\omega,m,J)}}{\partial b}(b_0,\omega_{b_0,m})= -\beta (\omega^2_{b_0,m}+J^2)(1-f(r))\sinh (2 b_0\beta) -2\beta\,\omega_{b_0,m} \, J(1-f(r))\cosh (2b_0 \beta) \, ,
\end{equation*}
we conclude that, under the assumptions of Lemma \ref{IFT_boost_string}, we have
\begin{equation} \label{deriv_omega_sign}
 -C_{r_0,\beta} \, \omega_{b_0,m}\leq \frac{d\omega}{db}(b_0)   <0 
\end{equation}
for some constant $C_{r_0,\beta}>0$ and any $b_0\in [0,1]$.

\end{remark}

\subsection{The main theorem for the eigenvalue problem \eqref{nonlinear_eig_prob_1}}

This is the main theorem for the nonlinear eigenvalue problem \eqref{nonlinear_eig_prob_1}. Previous sections provided all the preliminary results that we need to prove the following:

\begin{theorem} [\textbf{Eigenvalues boosted black string}]\label{main_theorem_EVP_boost_string}
Consider the fixed energy levels $\mathcal{E},E>0$ and the bounded set $\Omega$ introduced in Proposition \ref{omega_nonlinear}. Let $\beta\in\mathbb{R}_+$ satisfy \eqref{cond_beta} and $m,J\in\mathbb{Z}$ as in Claim \ref{claim_m_J}. Given eigenvalues $\omega^2_{\textup{lin},m}$ for the linear eigenvalue problem \eqref{Dirichlet_lin_eig_prob} on $\Omega$, we further assume $\omega_{\textup{lin},m}>0$. Then, there exists a constant $M>0$ such that the following statement holds for any $m^2>M$. There exists an eigenvalue $\omega^2_m$ and an associated smooth eigenfunction $u_m$ to the nonlinear eigenvalue problem \eqref{nonlinear_eig_prob_1}. Furthermore, we have $\omega_m>0$ and 
\begin{equation}
C_{r_0,\beta}\leq \frac{\omega^2_m}{m^2}\leq E^2+\frac{\mathcal{E}^2-E^2}{10} \label{acc_nonlin_eig}
\end{equation} 
for some constant $C_{r_0,\beta}>0$ independent of $m$.
\end{theorem}

\begin{remark} \label{correct_exist_eig_lin_2}
In view of Proposition \ref{omega_nonlinear}, the energy level $E$ agrees with the assumptions of Theorem \ref{Weyl's_law_static} for the linear eigenvalue problem \eqref{Dirichlet_lin_eig_prob} on $\Omega$  (see also Remark \ref{correct_exist_eig_lin_1}). Therefore, Theorem \ref{Weyl's_law_static} ensures that eigenvalues $\omega^2_{\textup{lin},m}$ do exist and, in particular, $\omega^2_{\textup{lin},m}/m^2$ accumulate in an arbitrarily small strip around $E^2$ for $m^2$ sufficiently large. This last accumulation property of $\omega^2_{\textup{lin},m}$ will be crucial to prove inequality \eqref{acc_nonlin_eig}.
\end{remark}

\begin{proof}
The spirit of this proof is to repeatedly apply Lemma \ref{IFT_boost_string} to prove the existence of eigenvalues for the nonlinear eigenvalue problem \eqref{nonlinear_eig_prob_1}. 

Let us start by choosing $b_0=0$. Then, we know by our study of the linear problem \eqref{Dirichlet_lin_eig_prob} that, for $m^2$ sufficiently large, there exists an $\omega_{0,m}$ such that $Q(0,\omega_{0,m})$ admits a zero eigenvalue. By Lemma \ref{lemma_omega_notzero}, we necessarily have $\omega_{0,m}\neq 0$.

\textit{By assumption in the statement of the theorem, we have $\omega_{0,m}>0$}. Lemma \ref{IFT_boost_string} now ensures that there exists $\varepsilon >0$ such that there exists a continuous function $\omega_{b,m}(b)$ such that, for any $b\in [0,\varepsilon )$, the eigenvalue problem \eqref{nonlinear_eig_prob_2} admits a zero eigenvalue $\Lambda_n(b,\omega_{b,m})=0$. In particular, in view of Lemma \ref{lemma_omega_notzero}, the function $\omega_{b,m}(b)$ cannot vanish for $b\in[0,\varepsilon )$, so $\omega_{b,m}$ has to remain \textit{positive} on $[0,\varepsilon )$. From \eqref{deriv_omega_sign}, we also have $\omega_{b,m}<\omega_{0,m}$ for all $b\in[0,\varepsilon )$.  

For any $b_0\in[0,\varepsilon )$, the assumptions of Lemma \ref{IFT_boost_string} are still satisfied, so we can apply the lemma again and deduce the existence of a function $\omega_{b,m}(b)$ such that $\Lambda_n(b,\omega_{b,m})=0$ in an enlarged interval of $b$. The fact that $\varepsilon$ is independent of $b_0$ is crucial here and ensures that a \textit{finite} number of applications of Lemma \ref{IFT_boost_string} covers the whole interval $b\in[0,1]$. 

Therefore, problem \eqref{nonlinear_eig_prob_2} admits a zero eigenvalue for any $b\in [0,1]$. In particular, for $b=1$, this proves the existence of eigenvalues $\omega^2_m$ for the nonlinear eigenvalue problem \eqref{nonlinear_eig_prob_1}.

From \eqref{deriv_omega_sign}, for $m^2$ sufficiently large, we have 
\begin{equation*}
\omega^2_m=\omega^2_{b,m}(1)\leq \omega^2_{b,m}(0) \leq C \, m^2 \, ,
\end{equation*} 
where the second inequality comes from Weyl's law for the linear problem. Combining with \eqref{lower_scaling_omega}, we have that, for $m^2$ sufficiently large, the eigenvalues $\omega_m^2$ for the nonlinear problem satisfy
\begin{equation*}
c_{r_0,\beta} \leq \frac{\omega^2_m}{m^2}\leq C_{r_0}
\end{equation*}
for some constants $c_{r_0,\beta},C_{r_0}>0$ \ul{independent of $m$}.

We now prove inequality \eqref{acc_nonlin_eig}. Consider the problem
\begin{equation*}
\overline{Q}_b u=E^2_n(b) u \, ,
\end{equation*}
where the operator $\overline{Q}_b$ and the $n$-th eigenvalue $E^2_n(b)$ are defined as  
\begin{gather*}
\overline{Q}_b u := -\frac{g(r)}{m^2}\Delta_{(r_*,\theta)}u+ \frac{1}{m^2}   [V_j(r,\theta)+V^{b\beta}_{(\omega,m,J)}(r,\theta)]u \\
E^2_n(b) := \frac{\omega^2_{b,m}(b)}{m^2}
\end{gather*}
with $b\in[0,1]$. \textit{From part (b) of Proposition \ref{omega_nonlinear}}, which allows the application of Weyl's law (see Remark \ref{correct_exist_eig_lin_2}), we have
\begin{equation*}
E^2_n(0)\in [E^2-\delta,E^2+\delta]
\end{equation*}
for some $\delta>0$ arbitrarily small and $m^2$ sufficiently large. Furthermore, we have the estimate
\begin{align*}
0\leq \int_{\Omega} \left[u(\overline{Q}_0-\overline{Q}_b) u \right] dr_* d\theta =& \int_{\Omega} \left\lbrace c^2\,f(r)+ \left[(1-f(r))\sinh^2 b\beta \right]\frac{\omega_{b,m}^2}{m^2} \right. \\
& +2\,c\, (1-f(r))\sinh b\beta\cosh b\beta\, \frac{\omega_{b,m}}{m}  \\
& \left. -c^2[1-(1-f(r))\cosh^2 b\beta]\right\rbrace |u|^2 dr_* d\theta 
\end{align*}
for any $b\in[0,1]$, where the constant $c$ was defined in Claim \ref{claim_m_J} and the factor $c\,\omega_{b,m}/m$ is positive. Note that $f(r)\geq 1-(1-f(r))\cosh^2b\beta$, so all the terms in the integral are positive. Therefore,
\begin{equation*}
\int_{\Omega} \left[u(\overline{Q}_b u) \right] dr_* d\theta\leq \int_{\Omega} \left[u(\overline{Q}_0 u) \right] dr_* d\theta \, ,
\end{equation*} 
which implies, recalling the min-max definition of the eigenvalue $E_n(b)$, that
\begin{equation*}
E^2_n(b)\leq E^2_n(0)
\end{equation*} 
for any $b\in[0,1]$. In particular,
\begin{equation*}
E^2_n(1)\leq E^2_n(0)\leq E^2+\delta
\end{equation*}
with \ul{$\delta$ arbitrarily small}. The lower bound in inequality \eqref{acc_nonlin_eig} has already been proven in \eqref{lower_scaling_omega}.

\end{proof}

\begin{remark}[\textbf{$\boldsymbol{\hat{V}^{\beta}_{\left(\omega_m,m,J\right)}}$ has a good structure}]  \label{good_structure_pot_boost}
Inequality \eqref{acc_nonlin_eig} ensures that the eigenvalues $\omega^2_m$ for the nonlinear eigenvalue problem determine a potential $\hat{V}^{\beta}_{(\omega_m,m,J)}$ with the same sign properties of $\hat{V}^{\beta}_{(Em,m,J)}$ in Figure \ref{fig:Omega_boosted_string}. When $\omega^2_m/m^2\in[E^2-\delta,E^2+\delta]$, this is rigorously motivated by the final part of Proposition \ref{omega_nonlinear}. For $C\leq \omega^2_m/m^2 \leq E^2-\delta$, one can invoke part (6) of Theorem \ref{analysis_potential_boosted} or, equivalently, remember that $\partial \hat{V}^{\beta}_{(\omega,m,J)}/\partial \omega<0$. Indeed, the potential gains positivity for lower values of $\omega^2_m/m^2$, thus the sign properties deriving from Proposition \ref{omega_nonlinear} still hold. This structure of the potential will be crucial in our quasimode construction. 
\end{remark}

\begin{remark}[\textbf{The choice of sign for $\boldsymbol{J}$ does not matter}]
The reader should now go back to Remark \ref{rmk_sign_J} and realise that all our arguments can be repeated if $J<0$ is assumed. One only needs to be careful with the proof of Lemma \ref{IFT_boost_string}, where the estimate on 
\begin{equation*}
\frac{\partial V^{b\beta}_{(\omega,m,J)} }{\partial\omega} (b_0,\omega_{b_0,m})   -2\omega_{b_0,m}
\end{equation*} 
was possible because $\omega_{b_0,m}$ and $J$ were \ul{both} positive. Note that the estimate still holds if $\omega_{b_0,m}$ and $J$ are both negative, now in the form
\begin{equation*}
\frac{\partial V^{b\beta}_{(\omega,m,J)} }{\partial\omega} (b_0,\omega_{b_0,m})   -2\omega_{b_0,m} \geq 2\, \omega_{b_0,m} \, .
\end{equation*}
In other words, what we really need is $\omega_{b_0,m}J>0$. Therefore, if we assume $J<0$ at the level of Claim \ref{claim_m_J}, we then need to assume $\omega_{b_0,m}<0$ in Lemma \ref{IFT_boost_string}. In terms of our problem, this means that we are assuming $\omega_{\text{lin},m}<0$ for $\omega^2_{\text{lin},m}$ eigenvalues of the linear eigenvalue problem.
\end{remark}

\section{Eigenvalue problem for the black ring}   \label{sect_black_ring}

The aim of this section is to prove Theorem \ref{eigenv_black_ring_final}.

Consider the coordinate system $(t,r,\theta,\phi,\psi)$ and the ring metric $g_{(r_0,R)}$ introduced in \eqref{ring_alt_coord}. Recall that
\begin{align*}
r\in [r_0,R]  &&  \theta\in [0,\pi) \, , 
\end{align*}
where $r=r_0$ corresponds to the event horizon and $(R,\pi)$ to spacelike infinity. 

Equation \eqref{WE} for the black ring becomes
\begin{equation*}
-g_{\text{ring}}(r,\theta)\Delta_{(r_*,\theta_*)}u+\left[V^{\textup{ring}}_j(r,\theta)+V^{\textup{ring}}_{(\omega,m,\hat{J})}(r,\theta)\right]u=0 \, ,
\end{equation*}
where $u=u(r_*,\theta_*)$, the coordinates $(r_*,\theta_*)$ are implicitly defined by
\begin{align*}
\frac{dr_*}{dr}=\sqrt{\frac{R ^2}{r(r-r_0) (R^2-r^2 ) }} && \frac{d\theta_*}{d\theta}=\sqrt{\frac{R   }{r_0 \cos \theta +R  }}  
\end{align*}
and the function $g_{\text{ring}}(r,\theta)$ is
\begin{equation*}
g_{\text{ring}}(r,\theta):=  \frac{(r-r_0)(R +r \cos \theta )^2 }{r^3 R\left(R+r_0   \cosh ^2\beta\cos \theta  \right)} \, .
\end{equation*}
The potential $V^{\textup{ring}}_{(\omega,m,\hat{J})}(r,\theta)$ is given by
\begin{equation}  \label{pot_eigen_ring}
V^{\textup{ring}}_{(\omega,m,\hat{J})}(r,\theta) =  f_{1,R} (r,\theta)\,m^2+ f_{2,R} (r,\theta)\, \omega^2+f_{3,R}(r,\theta) \,\omega\, \hat{J}  + f_{4,R} (r,\theta)\,\hat{J}^2 \, ,
\end{equation}
where $\omega\in\mathbb{R}$, $m,\hat{J}\in\mathbb{Z}$ and the functions $f_{i,R} (r,\theta)$ are defined as follows:
\begin{align*} 
f_{1,R} (r,\theta):= & \,\frac{ (r-r_0)(R    +r \cos \theta )^2}{R \, r^3 (r_0 \cos \theta +R )\sin^2\theta}	\, ,				\\	
f_{2,R} (r,\theta):= & \,	\frac{r_0^2(R-r)(R+r\cos\theta)^2(R\sinh^22\beta+4r_0\cosh^4\beta\sinh^2\beta)}{4rR(r+R)(R-r_0\cosh^2\beta)(r-r_0\cosh^2\beta)(R+r_0\cosh^2\beta\cos\theta)} \\ & -\frac{(r-r_0)(R+r_0\cosh^2\beta\cos\theta)}{R(r-r_0\cosh^2\beta)} \, ,  \\
f_{3,R}(r,\theta) := &	\,\frac{2r_0  \sinh \beta \cosh \beta  (R +r \cos \theta )^2}{r R   (R +r) \left(r_0   \cosh ^2\beta \cos \theta+R \right)}\sqrt{\frac{R +r_0\cosh^2 \beta }{R -r_0 \cosh ^2\beta}}		\, , 				\\					f_{4,R} (r,\theta) := & \,	\frac{(r-r_0\cosh^2 \beta ) (R +r \cos \theta )^2}{r R   \left(R ^2-r^2\right) \left(r_0  \cosh ^2\beta \cos \theta +R \right)} \, .		
\end{align*}
Potential $V^{\textup{ring}}_j(r,\theta)$ is a smooth, real-valued function and does not involve any frequency parameter. We also recall that for a black ring in equilibrium, the following condition must hold:
\begin{equation*}
\cosh^2\beta= \frac{2R^2}{r_0^2+R^2} \, .
\end{equation*}

\begin{remark}[\textbf{$\boldsymbol{\beta}$ is a function of $\boldsymbol{r_0}$ and $\boldsymbol{R}$}]  \label{rmk_beta_0}
The reader should keep in mind that $\beta$ is really a function of the parameters $r_0$ and $R$. Furthermore, note that $\cosh^2\beta$ is an increasing function of $R$ and
\begin{equation*}
1\leq \cosh^2\beta\leq 2
\end{equation*}
for any $r_0$ and $R$ such that $r_0 \leq R$. Crucially, we have 
\begin{equation*}
\cosh^2\beta< 3 
\end{equation*}
for any $r_0\leq R$, which agrees with condition \eqref{cond_beta} introduced for boosted strings if we interpret $\beta$ as a boost parameter (see later why this is important). We also define
\begin{equation*}
\beta_0 := \lim_{R\rightarrow\infty}\beta=\cosh^{-1}\sqrt{2} \, ,
\end{equation*}
which satisfies $\cosh^2\beta_0<3$. In what follows, we sometimes denote $\beta$ by $\beta_R$ to emphasise the dependence of $\beta$ on $R$.
\end{remark}

\begin{remark}[\textbf{$\boldsymbol{f_{i,R}}$ are smooth}]
Functions $f_{i,R}$, $i=1,\ldots,4$, are smooth for $(r,\theta)\in (r_0,R)\times (0,\pi)$. This is not obvious for $f_{2,R}$, for which the denominator goes to zero at the ergosurface $r=r_0\cosh^2\beta$. However, a more careful analysis shows that such limit is finite. 
\end{remark}

Without loss of generality, we fix $r_0=2$ throughout.

\subsection{A local property of the ring potential $V^{\textup{ring}}_{(\omega,m,\hat{J})}$}

As 
\begin{equation*}
r,\,2,\,2\cosh^2\beta_R\ll R \, ,
\end{equation*}
one can rewrite potential \eqref{pot_eigen_ring} in the form
\begin{align*}
V^{\textup{ring}}_{(\omega,m,\hat{J})} & (r,\theta)=  (1+\mathcal{O}_1(R^{-1}))\left(1-\frac{2}{r}\right)\frac{m^2}{r^2\sin^2\theta} + (1+\mathcal{O}_2(R^{-1})) \left(-1-\frac{2}{r}\sinh^2\beta_R\right) \omega^2 \\
&-(1+\mathcal{O}_3(R^{-1}))\left(2\,\frac{2}{r}\sinh\beta_R\cosh\beta_R \right) \,\omega\, \frac{\hat{J}}{R}  +(1+\mathcal{O}_4(R^{-1})) \left(1-\frac{2}{r}\cosh^2 \beta_R\right) \frac{\hat{J}^2}{R^2}  \\
=& (1+\mathcal{O}_1(R^{-1}))\left[\left(1-\frac{2}{r}\right)\frac{m^2}{r^2\sin^2\theta}\right] + (1+\mathcal{O}_2(R^{-1})) \left[-1-\frac{2}{r}\sinh^2\beta_0+\frac{16}{r(4+R^2)}\right] \omega^2 \\
&-(1+\mathcal{O}_3(R^{-1}))\left[2\,\frac{2}{r}\sinh\beta_0\cosh\beta_0-\frac{4\sqrt{2}}{r}\left(1-\frac{\sqrt{R^2(R^2-4)}}{4+R^2}\right) \right] \,\omega\, \frac{\hat{J}}{R}  \\
&+(1+\mathcal{O}_4(R^{-1})) \left[1-\frac{2}{r}\cosh^2 \beta_0+\frac{16}{r(4+R^2)}\right] \frac{\hat{J}^2}{R^2}
\end{align*}
with $\beta_0$ as in Remark \ref{rmk_beta_0} and $\mathcal{O}_i$ in the usual Bachmann-Landau notation (the label $i$ simply denotes different terms). The following lemma has to be regarded as a formal statement about functions of the same real variables $r$ and $\theta$.\footnote{Coordinates $(r,\theta)$ in the black ring metric \eqref{ring_alt_coord} actually differ from the ones that we used for black strings. In the limit $R\rightarrow\infty$, the two different coordinate systems do locally coincide.}

\begin{lemma} \label{from_ring_to_string}
Let $2<r_1<r_2$ and $0<\theta_1<\theta_2<\pi$. Consider the function $\hat{V}^{\beta_0}_{(\omega,m,J)}(r,\theta)$ as defined in \eqref{prop_pot}, with $\cosh^2\beta_0=2$. Then, for any frequency triple $(\omega,m,J)$, with $\omega\in\mathbb{R}$ and $m,J\in\mathbb{Z}$, and constant $\varepsilon>0$, there exist a frequency parameter $\hat{J}$ and a constant $\mathcal{R}$, with $\mathcal{R}>r_2$, such that
\begin{equation*}
 \left\lVert V^{\textup{ring}}_{(\omega,m,\hat{J})}-\hat{V}^{\beta_0}_{(\omega,m,J)} \right\rVert_{L^{\infty}\left( [r_1,r_2]\times [\theta_1,\theta_2]   \right)} \leq \varepsilon
\end{equation*}
for all $R\geq\mathcal{R}$. The frequency parameter $\hat{J}$ satisfies  
\begin{equation}
\hat{J}=R\, J   \label{limit_ring_pot}
\end{equation}
for each $R$. 
\end{lemma}

\begin{proof}
Consider the difference
\begin{align*}
V^{\textup{ring}}_{(\omega,m,\hat{J})}&(r,\theta)  -  \hat{V}^{\beta_0}_{(\omega,m,J)} (r,\theta)= \mathcal{O}_1(R^{-1})\left[\left(1-\frac{2}{r}\right)\frac{m^2}{r^2\sin^2\theta}\right] + (1+\mathcal{O}_2(R^{-1})) \left(\frac{16}{r(4+R^2)}\right) \omega^2 \\
&+\mathcal{O}_2(R^{-1})\left(-1-\frac{2}{r}\sinh^2\beta_0\right) \omega^2
+(1+\mathcal{O}_3(R^{-1}))\left[\frac{4\sqrt{2}}{r}\left(1-\frac{\sqrt{R^2(R^2-4)}}{4+R^2}\right) \right] \,\omega\, J \\
&-\mathcal{O}_3(R^{-1})\left(2\,\frac{2}{r}\sinh\beta_0\cosh\beta_0\right) \omega\, J
+(1+\mathcal{O}_4(R^{-1})) \left(\frac{16}{r(4+R^2)}\right) J^2 \\
&+\mathcal{O}_4(R^{-1})\left(1-\frac{2}{r}\cosh^2 \beta_0\right)J^2 \, ,
\end{align*}
where $r\in [r_1,r_2]$, $\theta\in [\theta_1,\theta_2]$ and $R$ is large. Note that $\hat{V}^{\beta_0}_{(\omega,m,J)}(r,\theta)$ is independent of $R$. For any fixed triple $(\omega,m,J)$ and $\varepsilon>0$, one proves the lemma by choosing $R$ sufficiently large.

\end{proof}

\begin{remark}[\textbf{Lemma \ref{from_ring_to_string} is local}]
It is important to note that Lemma \ref{from_ring_to_string} only holds \ul{locally}, i.e. on a fixed compact region satisfying $r\ll R$. This is a manifestation of the fact that the local geometry close to the horizon of large radius, thin black rings resembles that of suitably boosted black strings. 
\end{remark}

\begin{remark}[\textbf{$\boldsymbol{C^k}$-formulation of Lemma \ref{from_ring_to_string}}]  \label{norm_ring_string_approx}
The same statement of Lemma \ref{from_ring_to_string} can be proven to hold for the $C^k$-norm of $V^{\textup{ring}}_{(\omega,m,\hat{J})}-\hat{V}^{\beta_0}_{(\omega,m,J)}$, for any $k\in \mathbb{N}$. However, one does not need such improvement to prove Theorem \ref{ring_stab_trpp_redone}, which will be the main application of Lemma \ref{from_ring_to_string}. 
\end{remark}

\subsection{Black rings admit stable trapping}

We now want to prove the analogue of part (7) of Theorem \ref{analysis_potential_boosted} for potential $V_{\left(\omega,m,\hat{J}\right)}^{\textup{ring}}$. At the geodesic level, the following statement can be seen as claiming the existence of stably trapped null geodesics for a class of black ring spacetimes.

\begin{theorem}   \label{ring_stab_trpp_redone}
Consider the black ring metric $g_{(r_0,R)}$ defined in \eqref{ring_alt_coord} and fix $r_0=2$. Then, there exists a constant $\mathcal{R}>2$ such that, for any $R\geq \mathcal{R}$, the metric $g_{(2,R)}$ satisfies the following property. There exists a frequency triple $(\omega,m,\hat{J})$, with $\omega\in\mathbb{R}$ and $m,\hat{J}\in\mathbb{Z}$, and \ul{bounded} sets $\Omega^{\prime}\Subset\Omega\subset [2,R]\times [0,\pi)$ such that the potential $V_{\left(\omega,m,\hat{J}\right)}^{\textup{ring}}$ is negative on $\Omega^{\prime}$, vanishes on $\partial \Omega^{\prime}$ and is positive on $\Omega\setminus \overline{\Omega^{\prime}}$.
\end{theorem} 

\begin{proof}
Consider the function $\hat{V}^{\beta_0}_{(\omega,m,J)}$ with $r_0=2$. \ul{Fix} for $\hat{V}^{\beta_0}_{(\omega,m,J)}$ a triple $(\omega,m,J)$ such that there exist $2<r_1<r_2<r_3$ satisfying 
\begin{align*}
\hat{V}^{\beta_0}_{(\omega,m,J)} \left(r_1,\frac{\pi}{2}\right)>0 \, , &&
\hat{V}^{\beta_0}_{(\omega,m,J)} \left(r_2,\frac{\pi}{2}\right)<0 \, , &&
\hat{V}^{\beta_0}_{(\omega,m,J)} \left(r_3,\frac{\pi}{2}\right)>0 \, .
\end{align*}
Such a triple exists by Theorem \ref{analysis_potential_boosted}. For any $r_1\leq r \leq r_3$, we have 
\begin{align*}
\hat{V}^{\beta_0}_{(\omega,m,J)} \left(r,0+\delta\right)>0
&&\hat{V}^{\beta_0}_{(\omega,m,J)} \left(r,\pi-\delta\right)>0
\end{align*}
for $\delta<\pi/4$ sufficiently small. Moreover, in view of the fact that in the $\theta$ direction the only stationary point of $\hat{V}^{\beta_0}_{(\omega,m,J)}$ is a local minimum at $\theta=\pi/2$, we have
\begin{align*}
\hat{V}^{\beta_0}_{(\omega,m,J)} \left(r_1,\frac{\pi}{2}\right)>0 &\implies  \hat{V}^{\beta_0}_{(\omega,m,J)} \left(r_1,\theta\right)>0 \quad \text{for any $\delta<\theta<\pi-\delta$}  \\
\hat{V}^{\beta_0}_{(\omega,m,J)} \left(r_3,\frac{\pi}{2}\right)>0 &\implies  \hat{V}^{\beta_0}_{(\omega,m,J)} \left(r_3,\theta\right)>0 \quad \text{for any $\delta<\theta<\pi-\delta$} \, .
\end{align*} 

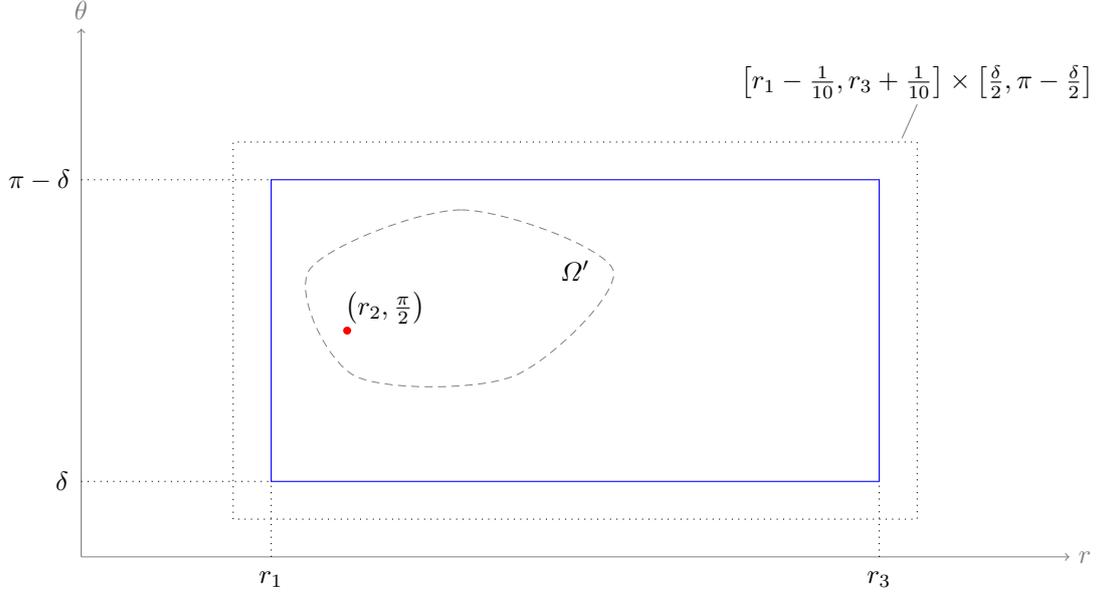
\begin{figure}[t]
\centering

\begin{tikzpicture}

\draw[->,help lines] (-6.5,-3)--(6.5,-3) node[right]{$r$};
\draw[->,help lines] (-6.5,-3)--(-6.5,4) node[above]{$\theta$};

\draw[dotted] (-6.5,-2)--(-4,-2);
\draw[dotted] (-6.5,2)--(-4,2);
\draw[dotted] (-4,-3)--(-4,-2);
\draw[dotted] (4,-3)--(4,-2);

\draw[blue] (-4,-2) rectangle (4,2);
\draw[dotted] (-4.5,-2.5) rectangle (4.5,2.5);

\node at (-7.05,2) {$\pi - \delta$};
\node at (-6.75,-2) {$\delta$};
\node at (4,-3.3) {$r_3$};
\node at (-4,-3.3) {$r_1$};
\node at (-2.5,0.3) {$\left(r_2,\frac{\pi}{2}\right)$};
\node at (4.5,3.3) {$\left[r_1-\frac{1}{10},r_3+\frac{1}{10}\right]\times \left[\frac{\delta}{2},\pi - \frac{\delta}{2}\right]$};
\node at (0,0.8) {$\Omega^{\prime}$};

\fill[red] (-3,0)  circle[radius=1.5pt];

\draw[help lines] (4.3,2.55)--(4.5,3);

\draw[help lines, densely dashed] plot [smooth cycle] coordinates {(-3.5,0.8) (-1.5,1.6) (0.5,0.8) (-0.8,-0.6)  (-2.9,-0.6)  };

\end{tikzpicture}

\caption{Construction for the proof of Theorem \ref{ring_stab_trpp_redone}. The existence of the open set $\Omega^{\prime}$ is deduced by continuity of the potential.}

\end{figure}

By Lemma \ref{from_ring_to_string}, we can deduce that, for $R$ sufficiently large, $V_{\left(\omega,m,\hat{J}\right)}^{\textup{ring}}$ preserves the sign properties of $\hat{V}^{\beta_0}_{(\omega,m,J)}$ for a frequency triple $(\omega,m,\hat{J})$ such that $\hat{J}=R\, J$ ($\omega$ and $m$ remain fixed).
The lemma has to be applied in a compact region containing the points examined above, say on $\left[r_1-1/10,r_3+1/10\right]\times \left[\delta/2,\pi-\delta/2\right]$.

To conclude, note that $V_{\left(\omega,m,\hat{J}\right)}^{\textup{ring}}$ has been proven to be positive along the sides of the rectangle $\left[r_1,r_3\right]\times \left[\delta,\pi-\delta\right]$, while negative at the point $\left(r_2,\pi/2\right)$, which lies inside the rectangle. By continuity, $V_{\left(\omega,m,\hat{J}\right)}^{\textup{ring}}$ has to be negative on a bounded region inside the rectangle.

\end{proof}

\begin{remark} [\textbf{Stable trapping}]  \label{rmk_stab_trapp_ring}
Theorem \ref{ring_stab_trpp_redone} implies the presence of at least one local minimum in the bounded region $\Omega^{\prime}$ where potential $V_{\left(\omega,m,\hat{J}\right)}^{\textup{ring}}$ is negative. In this sense, Theorem \ref{ring_stab_trpp_redone} proves the existence of stably trapped null geodesics. The orbit of the null geodesic corresponding to the local minimum of potential $V_{\left(\omega,m,\hat{J}\right)}^{\textup{ring}}$ lies on the torus $\mathbb{T}^2$ generated by $\partial_{\phi}$ and $\partial_{\psi}$. 
\end{remark}

\begin{remark}   \label{rmk_add_ineq}
After possibly choosing a larger $\mathcal{R}$, the following inequalities 
\begin{gather*}
 \frac{16}{r(4+R^2)}  < \left\lvert -1-\frac{2}{r}\sinh^2\beta_0  \right\rvert  \\   \frac{16}{r(4+R^2)}  < \left\lvert 1-\frac{2}{r}\cosh^2 \beta_0  \right\rvert  \\
\frac{4\sqrt{2}}{r}\left(1-\frac{\sqrt{R^2(R^2-4)}}{4+R^2}\right) <  2\,\frac{2}{r}\sinh\beta_0\cosh\beta_0   \\ \left\lvert  \mathcal{O}_i(R^{-1}) \right\rvert < 1   
\end{gather*}
hold for $R\geq\mathcal{R}$ on the compact region $\Omega$ of Theorem \ref{ring_stab_trpp_redone}, with $i=1,\ldots,4$.  
\end{remark}

From now on, the constant $\mathcal{R}$ will be allowed to be arbitrarily large and such that both Theorem \ref{ring_stab_trpp_redone} and the inequalities of Remark \ref{rmk_add_ineq} hold for $R\geq \mathcal{R}$. In turn, $1/R$ can be thought as a smallness parameter. 

The class of black rings that will prove Theorem \ref{eigenv_black_ring_final} is 
\begin{equation*}
\mathfrak{g}:=\left\lbrace \text{$g_{(r_0,R)}$ such that $r_0=2$ and $R\geq\mathcal{R}$}  \right\rbrace \, .
\end{equation*}
We will not be able to identify the minimum value of $\mathcal{R}$ such that Theorem \ref{eigenv_black_ring_final} holds, so there might exist black rings metrics $g_{(r_0,R)}\notin \mathfrak{g}$ but still satisfying the theorem. In this sense, the class $\mathfrak{g}$ is not optimal.

\subsection{Formulation of the eigenvalue problem}

In analogy with \eqref{nonlinear_eig_prob_2}, we define the two-parameter family of eigenvalue problems for black rings as
\begin{equation}  \label{Eigenvalue_problem_b_ring}
\boxed{
\begin{array}{rcl} 
Q_{\text{ring}}(b,\omega)u=& \Lambda(b,\omega)u \quad &\textrm{on} \quad \Omega  \\
 u=&0 \quad &\textrm{on} \quad \partial\Omega \, ,
 \end{array}
 }
\end{equation}
where the operator $Q_{\text{ring}}(b,\omega)$ has the form
\begin{equation*}
Q_{\text{ring}}(b,\omega) := -g^b_{\text{ring}}(r,\theta)\Delta_{(r_*,\theta_*)}u+\left[V_j^{b,\textup{ring}}(r,\theta)+V_{\left(\omega,m,\hat{J}\right)}^{b,\textup{ring}}(r,\theta)\right]u \, ,
\end{equation*}
with $b\in\mathbb{R}\cap [0,1]$, and $\Omega$ is a bounded set. 

\begin{remark}[\textbf{Notation}]
In the same spirit of Remark \ref{notation_eigenv_boost}, we denote by $\omega_{b,m}$ the frequency appearing in $Q_{\textup{ring}}(b,\omega)$ and
\begin{align*}
\omega_{\textup{boost}}\equiv \omega_{\textup{boost},m}:=\omega_{0,m} \, ,  && \omega\equiv \omega_m := \omega_{1,m} \, .
\end{align*}
Note that $\omega_m$ now refers to the black ring eigenvalue problem, while in Remark \ref{notation_eigenv_boost} it was referring to the boosted black string problem. The $\omega_m$ in Remark \ref{notation_eigenv_boost} is now $\omega_{\textup{boost},m}$. 
\end{remark}

We define the function $g^b_{\text{ring}}(r,\theta)$ as
\begin{equation*}
g^b_{\text{ring}}(r,\theta) := \frac{r-2}{r^3}+b (r-2)\frac{(R+r\cos\theta)^2-R(R+2\cosh^2\beta\cos\theta)}{r^3R(R+2\cosh^2\beta\cos\theta)}        \, ,
\end{equation*}
while the potential $V_{\left(\omega,m,\hat{J}\right)}^{b,\textup{ring}}(r,\theta)$ reads
\begin{align*}
&V_{\left(\omega,m,\hat{J}\right)}^{b,\textup{ring}}(r,\theta) := \\ & (1+b\, \mathcal{O}_1(R^{-1}))\left[\left(1-\frac{2}{r}\right)\frac{m^2}{r^2\sin^2\theta}\right] + (1+b\, \mathcal{O}_2(R^{-1})) \left[-1-\frac{2}{r}\sinh^2\beta_0+b\,\frac{16}{r(4+R^2)}\right] \omega_{b,m}^2 \\
-&(1+b\, \mathcal{O}_3(R^{-1}))\left[2\,\frac{2}{r}\sinh\beta_0\cosh\beta_0-b\, \frac{4\sqrt{2}}{r}\left(1-\frac{\sqrt{R^2(R^2-4)}}{4+R^2}\right) \right] \,\omega_{b,m}\, \frac{\hat{J}}{R}  \\
+&(1+b\, \mathcal{O}_4(R^{-1})) \left[1-\frac{2}{r}\cosh^2 \beta_0+b\, \frac{16}{r(4+R^2)}\right] \frac{\hat{J}^2}{R^2}
\end{align*}
as $r\ll R$. When $b=0$, we have 
\begin{gather*}
g^0_{\text{ring}}(r,\theta)=g(r,\theta) \\
V_j^{0,\textup{ring}}(r,\theta)=V_j(r,\theta) \\
V_{\left(\omega,m,\hat{J}\right)}^{0,\textup{ring}}(r,\theta)=\hat{V}^{\beta_0}_{(\omega,m,J)}(r,\theta)
\end{gather*}
with $J=\hat{J}/R$ and the quantities on the right hand side previously defined. Therefore, $Q_{\text{ring}}(0,\omega)=Q(1,\omega)$, with $\beta=\beta_0$ in $Q(1,\omega)$.

We present here the ring analogue of Lemma \ref{Cont_pot_static_string} and Proposition \ref{omega_nonlinear}. This defines the set $\Omega$ and suitable energy levels $E$, completing the formulation of the eigenvalue problem.

\begin{prop} \label{Omega_black_ring}
Consider a black ring metric $g_{(r_0,R)}$ and fix $(r_0,R)$ such that $r_0=2$ and $R\geq\mathcal{R}$. Let $m,\hat{J}\in\mathbb{Z}$ such that 
\begin{align*}
m\neq 0 && \hat{J}=R(c\, m), \,\, \hat{J}>0
\end{align*}
for some fixed constant $c$, with $0<c^2<\frac{1}{12}$. Define $V^{\textup{ring}}_{\textup{min}}$ to be the minimum of $V_{\left(\omega,m,\hat{J}\right)}^{\textup{ring}}$ and $x_{\textup{min}}\in (2,R)\times (0,\pi)$ such that $V_{\left(\omega,m,\hat{J}\right)}^{\textup{ring}}(x_{\textup{min}})=V^{\textup{ring}}_{\textup{min}}$. Consider some constant $\mathcal{E}>0$ such that $V_{\left(\mathcal{E}m,m,\hat{J}\right)}^{\textup{ring}}
$ has a local minimum and such that there exists $\Omega\subset [2,R]\times [0,\pi)$ satisfying
\begin{enumerate}[(i)]
\item $x_{\textup{min}}\in \Omega\, ,$
\item $V_{\left(\mathcal{E}m,m,\hat{J}\right)}^{\textup{ring}}(x)=0$ for $x\in \partial\Omega \, ,$
\item there are no local maxima of $V_{\left(\mathcal{E}m,m,\hat{J}\right)}^{\textup{ring}}$ in $\Omega \, ,$
\item $x\in\Omega \implies r(x)>6 \, ,$
\item Remark \ref{rmk_add_ineq} holds on $\Omega \, .$ 
\end{enumerate}
Fix now some energy level $E>0$ such that 
\begin{enumerate}[(a)]
\item $V_{\left(Em,m,\hat{J}\right)}^{\textup{ring}}$ has a local minimum and there exists $\Omega^{\prime}\subset\Omega$ satisfying the same properties (i)-(v) of $\Omega$, but now with respect to $V_{\left(Em,m,\hat{J}\right)}^{\textup{ring}} \, ,$
\item $\hat{V}^{\beta_0}_{(Em,m,J)}$ has a local minimum and there exists $\Omega^{\prime\prime}\subset \Omega$ satisfying the same properties (i)-(v) of $\Omega$, but now with respect to $\hat{V}^{\beta_0}_{(Em,m,J)} \, ,$
\item $\hat{V}^{0}_{(Em,m,J)}$ has a local minimum and there exists $\Omega^{\prime\prime\prime}\subset \Omega$ satisfying the same properties (i)-(v) of $\Omega$, but now with respect to $\hat{V}^{0}_{(Em,m,J)} \, .$
\end{enumerate}
Then, the final parts of Lemma \ref{Cont_pot_static_string} and Proposition \ref{omega_nonlinear} hold for $\frac{1}{m^2}\hat{V}^{0}_{(Em,m,J)}$ and $\frac{1}{m^2}\hat{V}^{\beta_0}_{(Em,m,J)}$ with respect to the open set $\Omega$. Furthermore, for any sufficiently small constants $\delta,\delta^{\prime}>0$, there exists some constant $c^{\prime}>0$ such that 
\begin{equation*}
\textup{dist}(x,\partial\Omega)<\delta^{\prime} \implies \frac{1}{m^2} V_{\left(\kappa \, m,m,\hat{J}\right)}^{\textup{ring}}(x)> c^{\prime}
\end{equation*}
for all $\kappa\in\mathbb{R}$ satisfying $|\kappa^2-E^2| \leq\delta$, with $\textup{dist}(\cdot,\cdot)$ the Euclidean distance.
\end{prop}

\begin{proof}
Combining Proposition \ref{omega_nonlinear}, Lemma \ref{from_ring_to_string} and the proof of Theorem \ref{ring_stab_trpp_redone}, one can show that a choice of $\mathcal{E}$ and $E$ with such properties is possible. In particular, to ensure that $\Omega$ satisfies property (iii), one would need a stronger version of Lemma \ref{from_ring_to_string} (see Remark \ref{norm_ring_string_approx}). The result of the theorem follows by continuity arguments.
\end{proof}

Analogous statements of those of Lemma \ref{lemma_omega_notzero} and Corollary \ref{corollary_lower_bound_omega} hold for our class of black rings. 

\begin{lemma} \label{omega_zero_ring}
Let $\psi_n(b,\omega)$ be a non identically zero eigenfunction of \eqref{Eigenvalue_problem_b_ring}, normalized on $\Omega$. If the assumptions of Proposition \ref{Omega_black_ring} are satisfied and $m^2,J^2>M$, for some positive constant $M$, then the implication
\begin{equation*}
\omega_{b,m}=0 \implies \int _{\Omega} \psi_n(b,\omega) Q_{\textup{ring}}(b,\omega) \psi_n(b,\omega)\, dr_* d\theta_* \neq 0
\end{equation*}
holds for any $b\in [0,1]$. 
\end{lemma}

\begin{proof}
One follows the same argument of the proof of Lemma \ref{lemma_omega_notzero}. In particular, properties (iv) (combined with $\cosh^2\beta_0<3$) and (v) of $\Omega$ in Proposition \ref{Omega_black_ring} are crucial to prove that the potential is everywhere positive on $\Omega$ when $\omega_{b,m}=0$.
\end{proof}

\begin{lemma} \label{lower_bound_omega_ring}
Let $\psi_n(b,\omega)$ be a non identically zero eigenfunction of \eqref{Eigenvalue_problem_b_ring}, normalized on $\Omega$. If the assumptions of Proposition \ref{Omega_black_ring} are satisfied, then the implication
\begin{gather*}
\textup{$\omega^2_{b,m}=o(m^2)$ as $m^2,J^2$ are sufficiently large}     \\
 \implies \int _{\Omega} \psi_n(b,\omega) Q_{\textup{ring}}(b,\omega) \psi_n(b,\omega)\, dr_* d\theta_* \neq 0
\end{gather*}
holds for any $b\in [0,1]$.
\end{lemma} 

\begin{proof}
See proof of Corollary \ref{corollary_lower_bound_omega}, combined with considerations in the proof of Lemma \ref{omega_zero_ring}.
\end{proof}

\subsection{A second application of the Implicit Function Theorem}

The perturbation argument to prove Theorem \ref{eigenv_black_ring_final} is almost identical to the one presented for Lemma \ref{IFT_boost_string} and Theorem \ref{main_theorem_EVP_boost_string}, so we only sketch it. As already discussed, to apply the Implicit Function Theorem to $\Lambda(b,\omega)=0$ at $(b_0,\omega_{b_0,m})$, we need 
\begin{equation*}
\frac{\partial\Lambda}{\partial\omega} (b_0,\omega_{b_0,m})= \int_{\Omega}\psi_n(b_0,\omega_{b_0,m})\left(\frac{\partial V_{\left(\omega,m,\hat{J}\right)}^{b,\textup{ring}} }{\partial\omega} (b_0,\omega_{b_0,m})\right)\psi_n(b_0,\omega_{b_0,m}) \, dr_* d\theta_* 
\end{equation*}
bounded away from zero. It is therefore enough to show
\begin{align*}
 \frac{\partial V_{\left(\omega,m,\hat{J}\right)}^{b,\textup{ring}} }{\partial\omega} (b_0,\omega_{b_0,m}) &= 2(1+b_0\, \mathcal{O}_2(R^{-1})) \left[-1-\frac{2}{r}\sinh^2\beta_0+b_0\,\frac{16}{r(4+R^2)}\right] \omega_{b_0,m} \\
&-(1+b_0\, \mathcal{O}_3(R^{-1}))\left[2\,\frac{2}{r}\sinh\beta_0\cosh\beta_0-b_0\, \frac{4\sqrt{2}}{r}\left(1-\frac{\sqrt{R^2(R^2-4)}}{4+R^2}\right) \right] \, \frac{\hat{J}}{R}
\end{align*}
bounded away from zero. Note that, in view of point (v) of Proposition \ref{Omega_black_ring}, we have
\begin{align*}
 b_0\,\frac{16}{r(4+R^2)} &< \left\lvert -1-\frac{2}{r}\sinh^2\beta_0  \right\rvert \\
 b_0\, \frac{4\sqrt{2}}{r}\left(1-\frac{\sqrt{R^2(R^2-4)}}{4+R^2}\right)  &<   2\,\frac{2}{r}\sinh\beta_0\cosh\beta_0   \\
\left\lvert b_0\,  \mathcal{O}_2(R^{-1}) \right\rvert &< 1   \\
\left\lvert b_0\,  \mathcal{O}_3(R^{-1}) \right\rvert &<1
\end{align*}
on $\Omega$, for any $b_0\in [0,1]$. With an assumption on the sign of $\omega_{b_0,m}$ as the one we formulated for Lemma \ref{IFT_boost_string}, i.e. $\omega_{b_0,m}>0$, the derivative becomes the sum of two negative terms and can be bounded away from zero \textit{uniformly in $b_0$} as
\begin{align*}
\frac{\partial V_{\left(\omega,m,\hat{J}\right)}^{b,\textup{ring}} }{\partial\omega} (b_0,\omega_{b_0,m}) &\leq  -C_{R}\, \omega_{b_0,m}  \\ &\leq -C_{R}\, m \, ,
\end{align*}
with constant $C_{R}>0$ independent of $m$, where the first inequality follows from the same argument used in the proof of Lemma \ref{IFT_boost_string}, while the second one makes use of the lower bound on $\omega_{b,m}$ provided by Lemma \ref{lower_bound_omega_ring}.

Furthermore, we have
\begin{align*}
&\frac{\partial \Lambda }{\partial b} (b_0,\omega_{b_0,m}) \\ &=\int_{\Omega}\psi_n(b_0,\omega_{b_0,m})\left(-\frac{\partial g^b_{\text{ring}}}{\partial b}(b_0)\Delta_{(r_*,\theta_*)}+ \frac{\partial V_{j}^{b,\textup{ring}} }{\partial b} (b_0)+ \frac{\partial V_{\left(\omega,m,\hat{J}\right)}^{b,\textup{ring}} }{\partial b} (b_0,\omega_{b_0,m})\right)\psi_n(b_0,\omega_{b_0,m}) \, dr_* d\theta_* \, ,
\end{align*} 
with 
\begin{align*}
\frac{\partial V_{\left(\omega,m,\hat{J}\right)}^{b,\textup{ring}} }{\partial b} & (b_0,\omega_{b_0,m}) = \mathcal{O}_1(R^{-1})\left[\left(1-\frac{2}{r}\right)\frac{m^2}{r^2\sin^2\theta}\right] \\   +&  \left[\mathcal{O}_2(R^{-1})\left(-1-\frac{2}{r}\sinh^2\beta_0\right)+\left(1+2b_0\mathcal{O}_2(R^{-1})\right)\left(\frac{16}{r(4+R^2)}\right)\right] \omega_{b_0,m}^2 \\
-&\left[\mathcal{O}_3(R^{-1})\left(2\,\frac{2}{r}\sinh\beta_0\cosh\beta_0\right)-\frac{4\sqrt{2}}{r}\left(1+2b_0\mathcal{O}_3(R^{-1})\right) \left(1-\frac{\sqrt{R^2(R^2-4)}}{4+R^2}\right) \right] \omega_{b_0,m}\, \frac{\hat{J}}{R}  \\
+& \left[\mathcal{O}_4(R^{-1})\left(1-\frac{2}{r}\cosh^2 \beta_0\right)+\left(1+2b_0\mathcal{O}_4(R^{-1})\right)\left( \frac{16}{r(4+R^2)}\right)\right] \frac{\hat{J}^2}{R^2} \, .
\end{align*}
In view of Lemma \ref{lower_bound_omega_ring}, the bound
\begin{equation*}
\left\lvert \frac{\partial V_{\left(\omega,m,\hat{J}\right)}^{b,\textup{ring}} }{\partial b} (b_0,\omega_{b_0,m}) \right\rvert \leq C_R \, \omega^2_{b_0,m} 
\end{equation*}
holds. Therefore, we have
\begin{equation*}
\left\lvert \frac{d\omega_{b,m}}{db}(b_0) \right\rvert \leq C_R \, \omega_{b_0,m} \, .
\end{equation*}
This gives a bound $\omega_{b,m}\leq C_R \, \omega_{0,m}$ for any $b\in (0,1]$, with constant $C_R>0$ independent of $b$ and $m$. 

\begin{remark}
All the relations derived so far, as well as the proof of Theorem \ref{eigenv_black_ring_final}, are independent of the choice of $r_0$. In fact, all the arguments can be repeated for any $r_0>0$. Theorem \ref{eigenv_black_ring_final} is stated relaxing the assumption $r_0=2$. 
\end{remark}

The main theorem for the black ring eigenvalue problem is the following.

\begin{theorem}[\textbf{Eigenvalues black ring}] \label{eigenv_black_ring_final}
Consider the fixed energy levels $\mathcal{E},E>0$ and the open set $\Omega$, according to Proposition \ref{Omega_black_ring}. Let $m,\hat{J}\in\mathbb{Z}$ as in Proposition \ref{Omega_black_ring}. Given eigenvalues $\omega_{\textup{boost},m}$ for the eigenvalue problem \eqref{nonlinear_eig_prob_1} on $\Omega$, we further assume $\omega_{\textup{boost},m}>0$. Then, there exists a constant $M>0$ such that the following statement holds for any $m^2>M$. There exists an eigenvalue $\omega^2_m$ and an associated smooth eigenfunction $u_m$ to the black ring eigenvalue problem. Furthermore, we have $\omega_m>0$ and 
\begin{equation}  \label{eigenv_ring_ineq}
C_{r_0,R}\leq \frac{\omega^2_m}{m^2}\leq E^2+\frac{\mathcal{E}^2-E^2}{10} 
\end{equation} 
for some constant $C_{r_0,R}>0$ independent of $m$.
\end{theorem}

\begin{proof}
The proof follows the same lines of the proof of Theorem \ref{main_theorem_EVP_boost_string}. We only show how to prove inequality \eqref{eigenv_ring_ineq}, while the first part of the statement is left to the reader.

Recall that we have already concluded that 
\begin{equation} \label{eig_ring_prelim}
c_{r_0,R}\leq \frac{\omega^2_m}{m^2} \leq C_{r_0,R}
\end{equation}
for some constants $c_{r_0,R},C_{r_0,R}>0$ independent of $m$, where $\omega^2_m$ is such that the $n$-th eigenvalue $\Lambda_n(1,\omega)$ of problem \eqref{Eigenvalue_problem_b_ring} is zero. Define
\begin{equation*}
E_n(b):= \frac{\omega_{b,m}(b)}{m}
\end{equation*}
with $b\in[0,1]$. Then,
\begin{align*}
\left\lvert E_n(1)-E_n(0)  \right\rvert &= \left\lvert \frac{1}{m}\int_0^1 \frac{d\omega_{b,m}(b)}{db} db \right\rvert \\
& \leq  \frac{1}{m}\int_0^1 \left\lvert\frac{d\omega_{b,m}(b)}{db}\right\rvert db \\
&\leq \delta^{\prime}_{r_0,R} \, , 
\end{align*}
with $\delta^{\prime}_{r_0,R}>0$ \ul{arbitrarily small constant for $R/r_0$ sufficiently large}. Crucially, $\delta^{\prime}_{r_0,R}$ is \ul{independent of $m$.} The last inequality holds as a consequence of inequality \eqref{eig_ring_prelim} and the fact that, for any fixed $r_0$, on the bounded set $\Omega$ one has 
\begin{equation*}
\frac{1}{m} \left\lvert\frac{d\omega_{b,m}(b)}{db}\right\rvert \rightarrow 0 \quad \text{as} \quad R\rightarrow\infty
\end{equation*} 
uniformly in $b$ and $m$.

From Theorem \ref{main_theorem_EVP_boost_string}, combined with part (b) and (c) of Proposition \ref{Omega_black_ring}, we have
\begin{equation*}
E^2_n(0)\in \left[C_{r_0},E^2+\delta \right]
\end{equation*}
for $m^2$ sufficiently large, where $\delta>0$ is an arbitrarily small constant. We obtain
\begin{equation*}
 C_{r_0}-\delta^{\prime}_{r_0,R} \leq E^2_n(0)-\delta^{\prime}_{r_0,R} \leq E^2_n(1) \leq E^2_n(0) + \delta^{\prime}_{r_0,R} \leq E^2+\delta + \delta^{\prime}_{r_0,R} \, .
\end{equation*}
This concludes the proof of inequality \eqref{eigenv_ring_ineq}.

\end{proof}

\begin{remark}[\textbf{$\boldsymbol{V^{\textup{ring}}_{(\omega_m,m,\hat{J})}}$ has a good structure}]
In view of the same considerations of Remark \ref{good_structure_pot_boost}, potential $V^{\textup{ring}}_{(\omega_m,m,\hat{J})}(r,\theta)$ has the structure that we need to construct quasimodes with exponentially small errors.
\end{remark}

\section{Proof of Theorem \ref{Main_th_formal}}   \label{sect_proof_main_th}

The aim of this section is to construct quasimodes and prove Theorem \ref{Main_th_formal}. We will crucially apply Theorem \ref{eigenv_black_ring_final}. The structure of the presentation and the formulation of the main results closely follow Section 4 of \cite{KeirLogLog}.

\subsection{A lemma for the energy estimate}

We need a preliminary lemma, which will be used to prove an energy estimate for solutions to the black ring eigenvalue problem. We give a statement for a generic open set and smooth functions.

\begin{lemma}[adapted from \cite{KeirLogLog} Lemma 4.3] \label{Lemma_int_by_parts_ring}
Let $\Omega\subset\mathbb{R}^2$ be a bounded set and $h>0$ a real constant. Given smooth functions $u, W, \phi: \Omega\rightarrow\mathbb{R}$ such that $u|_{\partial\Omega}=0$, we have
\begin{align*}
&\int_{\Omega}\left(\left\lvert\frac{\partial}{\partial r_*}\left(e^{\phi/h}u\right)\right\rvert^2 +\left\lvert\frac{\partial}{\partial \theta_*}\left(e^{\phi/h}u\right)\right\rvert^2+h^{-2}\left(W-\left(\frac{\partial\phi}{\partial r_*}\right)^2-\left(\frac{\partial\phi}{\partial \theta_*}\right)^2       \right)e^{2\phi/h}|u|^2 \right)dr_*d\theta_* \\
&=\int_{\Omega}\left(-\frac{\partial^2 u}{\partial r^2_*}-\frac{\partial^2 u}{\partial \theta_*^2}+h^{-2}Wu \right)u\, e^{2\phi/h}dr_*d\theta_* \, .
\end{align*}
\end{lemma}

\begin{proof}
The proof is an integration by parts. We have
\begin{align*}
\int_{\Omega}&\left(-\frac{\partial^2 u}{\partial r^2_*}-\frac{\partial^2 u}{\partial \theta_*^2}+h^{-2}Wu \right)u\, e^{2\phi/h} \, dr_*d\theta_*  \\
=& \int_{\Omega}\frac{\partial u}{\partial r_*}\,\partial_{r_*}\left(u\, e^{2\phi/h}\right)+\frac{\partial u}{\partial \theta_*}\, \partial _{\theta_*}\left(u\, e^{2\phi/h}\right)+h^{-2}W|u|^2     e^{2\phi/h} \, dr_*d\theta_* \\
=& \int_{\Omega}\left(\frac{\partial u}{\partial r_*}\right)^2 e^{2\phi/h}+\frac{\partial u}{\partial r_*}\, u \frac{2}{h}\frac{\partial \phi}{\partial r_*} \,e^{2\phi/h}+\frac{\partial u}{\partial \theta_*}\, \partial _{\theta_*}\left(u\, e^{2\phi/h}\right)+h^{-2}W|u|^2     e^{2\phi/h} \, dr_*d\theta_* \\
=& \int_{\Omega}\left(\frac{\partial u}{\partial r_*} e^{\phi/h}\right)^2+\frac{\partial u}{\partial r_*}\, u \frac{2}{h}\frac{\partial \phi}{\partial r_*}\, e^{2\phi/h}+\left(h^{-1}\frac{\partial \phi}{\partial r_*}e^{\phi/h}u\right)^2   -\left(h^{-1}\frac{\partial \phi}{\partial r_*}e^{\phi/h}u\right)^2   \\
& +\frac{\partial u}{\partial \theta_*}\, \partial _{\theta_*}\left(u\, e^{2\phi/h}\right)+h^{-2}W|u|^2     e^{2\phi/h} \, dr_*d\theta_* \\
=& \int_{\Omega}\left\lvert\frac{\partial}{\partial r_*}\left(e^{\phi/h}u\right)\right\rvert^2 +\frac{\partial u}{\partial \theta_*}\, \partial _{\theta_*}\left(u\, e^{2\phi/h}\right)+h^{-2}\left(W-\left(\frac{\partial\phi}{\partial r_*}\right)^2       \right)e^{2\phi/h}|u|^2  \, dr_*d\theta_*  \\ 
=&\int_{\Omega}\left\lvert\frac{\partial}{\partial r_*}\left(e^{\phi/h}u\right)\right\rvert^2 +\left\lvert\frac{\partial}{\partial \theta_*}\left(e^{\phi/h}u\right)\right\rvert^2+h^{-2}\left(W-\left(\frac{\partial\phi}{\partial r_*}\right)^2-\left(\frac{\partial\phi}{\partial \theta_*}\right)^2       \right)e^{2\phi/h}|u|^2 \,dr_*d\theta_* \, ,
\end{align*}
where we make use of the fact that $u$ vanishes on $\partial\Omega$ in the first equality and we omit the identical calculation for $\partial_{\theta_*}$ in the last equality.
\end{proof}

\subsection{Agmon distance}

Consider
\begin{gather*}
V^{h,E}_{\textup{eff}} := h^2\left[V_j^{\textup{ring}}+ V_{\left(E \, m,m,\hat{J}\right)}^{\textup{ring}}\right] \\
h^2=m^{-2} \, ,
\end{gather*}
where the energy level $E$ is fixed as in Proposition \ref{Omega_black_ring}. The \textit{Agmon distance} between two points $x,y\in \mathbb{R}^2$ associated to the energy level $E$ and potential $V^{h,E}_{\textup{eff}}$ is 
\begin{equation*}
d(x,y):=\inf_{\gamma\in C^{1,pw}([0,1];x,y)}\int_0^1\left(\sqrt{V_{\textup{eff}}^{h,E}(\gamma(t))}\right)|\gamma^{\prime}(t)|\,\chi_{\left\lbrace V_{\textup{eff}}^{h,E}\geq 0\right\rbrace}dt \, ,
\end{equation*} 
where 
\begin{equation*}
C^{1,pw}([0,1];x,y)=\left\lbrace \gamma\in C^{1,pw}([0,1];\mathbb{R}^2),\gamma(0)=x,\gamma(1)=y  \right\rbrace
\end{equation*}
with $C^{1,pw}([0,1];\mathbb{R}^2)$ set of piecewise $C^1$ curves in $\mathbb{R}^2$ (see \cite{Agmon_dist}). The Agmon distance satisfies 
\begin{equation}  \label{deriv_agmon_dist}
|\nabla _x d(x,y)|^2 \leq \max \left\lbrace V_{\textup{eff}}^{h,E}(x),0 \right\rbrace \, .
\end{equation}
For any fixed energy level $E$, we define the distance from the classically allowed region as
\begin{equation*}
d_E(x):=\inf_{y\in \left\lbrace V_{\textup{eff}}^{h,E} \leq 0 \right\rbrace} d(x,y)
\end{equation*}
with $x,y$ points in $\mathbb{R}^2$. We will also make use of the two following sets
\begin{align*}
&\Omega^+_{\varepsilon}(E):=\left\lbrace x:V_{\textup{eff}}^{h,E}(x) > \alpha \, \varepsilon \right\rbrace \cap \Omega \\
&\Omega^-_{\varepsilon}(E):=\left\lbrace x:V_{\textup{eff}}^{h,E}(x) \leq \alpha \, \varepsilon \right\rbrace \cap \Omega
\end{align*}
with constant
\begin{equation*}
\alpha:=\sup _{\Omega} g_{\text{ring}}(r,\theta)  
\end{equation*}
and constant $\varepsilon>0$. Note that $\Omega^+_{\varepsilon}(E)\cap \Omega^-_{\varepsilon}(E)=\emptyset$ and $\Omega^+_{\varepsilon}(E)\cup \Omega^-_{\varepsilon}(E) = \Omega$.

\begin{figure}[H]
\centering
\begin{tikzpicture}
\draw plot [smooth] coordinates {(3,5) (2.65,4.1) (2,2.6) (1.15,2.3) (0,2) (-1.15,2.5) (-2,3.1) (-2.6,4.1) (-3,5) };

\draw  plot [smooth cycle] coordinates {(-3,5) (-2,5.2) (-1,5.5) (2,5.4) (3,5) (2.5,4.8) (1,4.7) (-1,4.8)};

\draw[help lines] (-5,3.5) -- (5,3.5);
\draw[help lines] (-5,3.5) -- (-4,4.5);
\draw[help lines] (-4,4.5) -- (-2.8,4.5);
\draw[help lines] (2.8,4.5) -- (6,4.5);
\draw [densely dashed, help lines] (-2.8,4.5) -- (2.8,4.5);

\draw[help lines] (-5,2.7) -- (5,2.7);
\draw[help lines] (-5,2.7) -- (-4.2,3.5);
\draw [densely dashed, help lines] (-4.2,3.5) -- (-4,3.7);
\draw[help lines] (5.2,3.7) -- (6,3.7);
\draw [densely dashed, help lines] (-4,3.7) -- (5.2,3.7);

\draw[fill=lightgray]  plot [smooth cycle] coordinates {(-2.6,1.2) (0,1.4)  (2.6,1.2)  (1.7,0.55) (0,0.5) (-1,0.5) (-1.5,0.6) };

\draw[dotted]  plot [smooth cycle] coordinates {(-2.6,4.1) (0,4.3)  (2.6,4.1)  (1.7,3.85) (0,3.8) (-1,3.8) (-1.5,3.9) };

\draw[dotted]  plot [smooth cycle] coordinates {(-1.88,3.1) (0,3.3)  (2.15,3.1)  (1.7,2.85) (0,2.8) (-1,2.8) (-1.5,2.9) };

\draw  plot [smooth cycle] coordinates {(-1.88,1.1) (0,1.3)  (2.15,1.1)  (1.7,0.65) (0,0.6) (-1,0.6) (-1.5,0.7) };

\draw  plot [smooth cycle] coordinates {(-3.6,1.6) (0,1.8)  (3.4,1.5)  (2.7,0.25) (0,0.2) (-1,0.2) (-1.8,0.3) (-3.5,1.1) };

\draw [help lines] (6,4.5)--(5,3.5);
\draw [help lines] (6,3.7)--(5,2.7);

\draw [dotted] (2.65,4.1)--(2.65,1.1);
\draw [dotted] (-2.6,4.1)--(-2.6,1.2); 

\draw [dotted] (-1.95,3.1)--(-1.95,1);
\draw [dotted] (2.25,3.1)--(2.25,1);

\node at (3.5,1.7) {$\Omega$};
\node at (-3.3,1.37) {$\Omega^+_{\varepsilon}$};
\node at (-2.09,1.08) {$\Omega^-_{\varepsilon}$};
\node at (3,5.5) {$V^{h,E}_{\textup{eff}}$};
\node at (1.2,0.9) {$V^{h,E}_{\textup{eff}}<0$};

\draw[dotted, help lines] (-5,3.5) -- (-6,3.5) node[left]{$\alpha\, \varepsilon$};
\draw[dotted, help lines] (0,2.7) -- (-6,2.7) node[left]{$0$};

\draw[->,help lines] (-6,0)--(7,0) node[right]{$r$};
\draw[->,help lines] (-6,0)--(-6,7) node[above]{$z$};
\draw[->,help lines] (-6,0)--(-3.6,2.4) node[left]{$\theta$};
\end{tikzpicture}

\caption{The smallest set in the figure is the region where $V^{h,E}_{\textup{eff}}<0$. The shaded region corresponds to $\Omega^-_{\varepsilon}(E)$, while the largest set is $\Omega$.}

\end{figure}
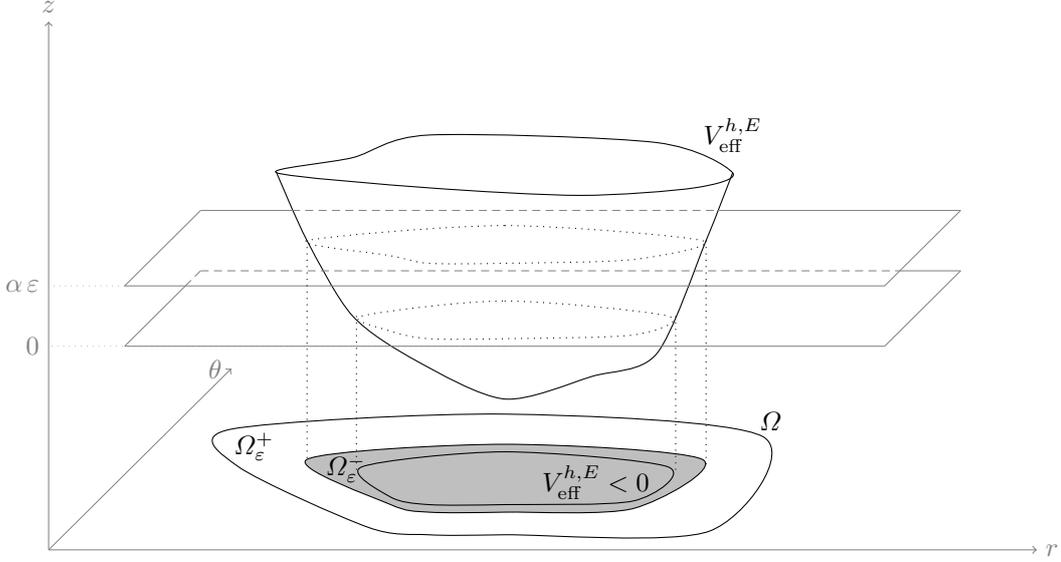

\subsection{The key energy estimate}
 
For any $\varepsilon\in (0,1)$, we define
\begin{equation*}
\phi_{E,\varepsilon}(x):=(1-\varepsilon)d_E(x)
\end{equation*}
and
\begin{equation*}
a_E(\varepsilon):=\sup_{\Omega^-_{\varepsilon}(E)}d_E \, .
\end{equation*}
Consider $u$ smooth eigenfunction to the black ring eigenvalue problem, with associated eigenvalue $\kappa:=h\, \omega_m$ such that $|\kappa^2-E^2|\leq\delta$, with constant $\delta>0$. Then, the following key energy estimate holds.
\begin{claim}[adapted from \cite{KeirLogLog} Lemma 4.4] \label{intermed_en_est_ring}
For any sufficiently small constants $\varepsilon,h>0$, and for sufficiently small constant $\delta_{\varepsilon, h}>0$, we have
\begin{align*}
&\int_{\Omega}h^2\left(\left\lvert\frac{\partial}{\partial r_*}(e^{\phi_{E,\varepsilon}/h}u)\right\rvert^2 +\left\lvert\frac{\partial}{\partial \theta_*}(e^{\phi_{E,\varepsilon}/h}u)\right\rvert^2\right)dr_*d\theta_* +\frac{1}{2}\varepsilon^2\int_{\Omega^+_{\varepsilon}(E)}e^{2\phi_{E,\varepsilon}/h}|u|^2 dr_*d\theta_* \\
&\leq C(\kappa^2+\frac{1}{2}\varepsilon)e^{2a_{E}(\varepsilon)/h}\left\lVert u\right\rVert^2_{L^2(\Omega)}
\end{align*}
for some constant $C>0$ depending only on $\Omega$.
\end{claim}

\begin{figure}[H]
\centering
\begin{tikzpicture}
\draw plot [smooth cycle] coordinates {(-2,2) (0,3) (2,2.9) (3.2,2) (3,0.8) (2,0.4) (0.7,0.3)  (-1,0.6)  };

\draw[densely dashed] plot [smooth cycle] coordinates {(-3,3) (0,4) (3,3.9) (4.5,3) (4,0) (0,-1) (-2,0.3)  (-3,0.6)  };

\draw plot [smooth cycle] coordinates {(-3.3,3.3) (0.3,4.3) (3.3,4.2) (4.8,3) (4.2,-0.3) (0.1,-1.3) (-2.3,0)  (-3.3,0.3)  };

\draw plot [smooth cycle] coordinates {(-1,1.8) (1.3,2) (0.8,1) (-1,1)    };

\fill (3.4,2.6)  circle[radius=1pt];
\fill (1.37,1.95)  circle[radius=1pt];
\fill (2.9,0.7)  circle[radius=1pt];
\fill (0.95,1.1)  circle[radius=1pt];

\draw[dotted] (3.4,2.6) -- (1.37,1.95);
\draw[dotted] (2.9,0.7) -- (0.95,1.1);

\node at (0,1.4) {$V_{\textup{eff}}^{h,E}\leq 0$};
\node at (2.4,1.9) {$d_E(x)$};
\node at (3.6,2.5) {$x$};
\node at (-1,2.2) {$\Omega^-_{\varepsilon}(E)$};
\node at (-2,2.9) {$\Omega^+_{\varepsilon}(E)$};
\node at (4.2,4.2) {$\Omega$};
\node at (3.2,-0.62) {$\delta$};
\node at (4.8,-0.5) {$\Omega_{\delta}$};
\node at (1.4,0.7) {$a_E(\varepsilon)$};

\draw[<->, very thin] (2.9,-0.55)--(3,-0.83);
\draw (3.8,-0.4) -- (4.5,-0.5);
\end{tikzpicture}
\caption{The Agmon distance differs from the Euclidean distance between two points in that it has to be weighted by the square root of $V_{\textup{eff}}^{h,E}(\gamma(t))$. To give some intuition, the figure shows the various quantities introduced so far as if they were defined in the Euclidean sense, i.e. with the weights set equal to one. Although this gives an idea of the construction, the reader should keep in mind that $d_E(x)$ and $a_E(\varepsilon)$ do not necessarily look like the ones in figure.} 
\end{figure}
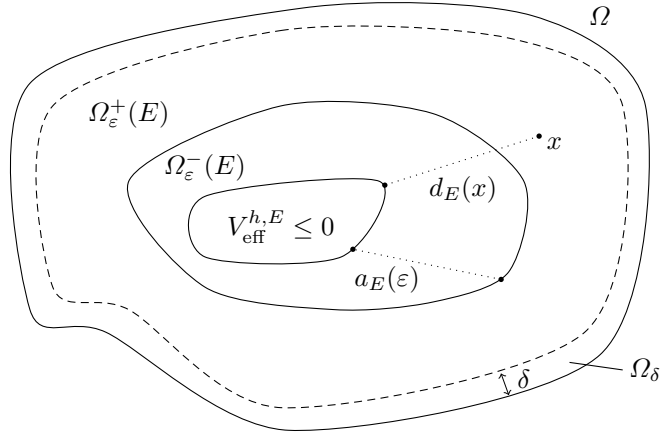

\begin{proof}
By the assumption that $u$ is a smooth solution to the black ring eigenvalue problem, we have
\begin{equation}
-h^2\Delta_{(r_*,\theta_*)} u+\frac{1}{g_{\textup{ring}}(r,\theta)}V_{\textup{eff}}^{h,\kappa}(r,\theta)u=0 \, . \label{eig_mod_ring}
\end{equation}
With the notation of Lemma \ref{Lemma_int_by_parts_ring}, we define 
\begin{align*}
W(r,\theta) :=\frac{1}{g_{\textup{ring}}(r,\theta)}V_{\textup{eff}}^{h,\kappa}(r,\theta) && \phi :=\phi_{E,\varepsilon}
\end{align*}
Together with equation \eqref{eig_mod_ring}, Lemma \ref{Lemma_int_by_parts_ring} gives 
\begin{align*} 
&\int_{\Omega}h^2\left(\left\lvert\frac{\partial}{\partial r_*}(e^{\phi_{E,\varepsilon}/h}u)\right\rvert^2 +\left\lvert\frac{\partial}{\partial \theta_*}(e^{\phi_{E,\varepsilon}/h}u)\right\rvert^2\right)dr_*d\theta_* \\
&+\int_{\Omega^+_{\varepsilon}(E)}\left(\frac{V_{\textup{eff}}^{h,\kappa}(r,\theta)}{g_{\textup{ring}}(r,\theta)}-\left(\frac{\partial\phi_{E,\varepsilon}}{\partial r_*}\right)^2-\left(\frac{\partial\phi_{E,\varepsilon}}{\partial \theta_*}\right)^2       \right)e^{2\phi_{E,\varepsilon}/h}|u|^2 dr_*d\theta_* \\
&=\int_{\Omega^-_{\varepsilon}(E)}\left(\frac{-V_{\textup{eff}}^{h,\kappa}(r,\theta)}{g_{\textup{ring}}(r,\theta)}+\left(\frac{\partial\phi_{E,\varepsilon}}{\partial r_*}\right)^2+\left(\frac{\partial\phi_{E,\varepsilon}}{\partial \theta_*}\right)^2       \right)e^{2\phi_{E,\varepsilon}/h}|u|^2 dr_*d\theta_* \, ,
\end{align*}
where we used $\Omega^+_{\epsilon}(E)\cup \Omega^-_{\epsilon}(E) = \Omega$. For $h$ sufficiently small, we can estimate the right hand side as
\begin{align*}
&\int_{\Omega^-_{\varepsilon}(E)}\left(\frac{-V_{\textup{eff}}^{h,\kappa}(r,\theta)}{g_{\textup{ring}}(r,\theta)}+\left(\frac{\partial\phi_{E,\varepsilon}}{\partial r_*}\right)^2+\left(\frac{\partial\phi_{E,\varepsilon}}{\partial \theta_*}\right)^2       \right)e^{2\phi_{E,\varepsilon}/h}|u|^2 dr_*d\theta_* \\
&\leq C\left(\kappa^2+\varepsilon(1-\varepsilon)\right)\,e^{2a_{E}(\varepsilon)/h}\left\lVert u\right\rVert^2_{L^2(\Omega)}
\end{align*}
with $C$ positive constant independent of $h$. To see this, first recall that $g_{\textup{ring}}(r,\theta)$ is strictly positive and bounded on $\Omega$ and note that, by definition of $a_{E}(\varepsilon)$, one has $\phi_{E,\varepsilon}|_{\Omega^-_{\varepsilon}(E)}\leq a_{E}(\varepsilon)$. Using property \eqref{deriv_agmon_dist} of the Agmon distance, we also have
\begin{align*}
\left(\frac{\partial\phi_{E,\varepsilon}}{\partial r_*}\right)^2+\left(\frac{\partial\phi_{E,\varepsilon}}{\partial \theta}\right)^2&=|\nabla \phi_{E,\varepsilon}|^2 \\
&= (1-\varepsilon)^2|\nabla d_{E}|^2 \\
&\leq (1-\varepsilon)^2 \alpha \, \varepsilon \\
&\leq \alpha (1-\varepsilon) \varepsilon
\end{align*}
where the last inequality holds because $\varepsilon\in (0,1)$. Note the obvious $\left\lVert u\right\rVert^2_{L^2(\Omega^-_{\varepsilon}(E))}\leq \left\lVert u\right\rVert^2_{L^2(\Omega)}$.  

On $\Omega^+_{\varepsilon}(E)$, one has
\begin{align*}
\frac{V_{\textup{eff}}^{h,\kappa}(r,\theta)}{g_{\textup{ring}}(r,\theta)}-\left(\frac{\partial\phi_{E,\varepsilon}}{\partial r_*}\right)^2-\left(\frac{\partial\phi_{E,\varepsilon}}{\partial \theta_*}\right)^2     
& \geq \frac{\alpha \, \varepsilon -\hat{\delta}}{g_{\textup{ring}}(r,\theta)}-(1-\varepsilon)\varepsilon \\
&\geq \varepsilon-\frac{\hat{\delta}}{\min_{\Omega^+_{\varepsilon}(E)}g(r)}-(1-\varepsilon)\varepsilon \\
&= \varepsilon^2-\hat{\delta} \, ,
\end{align*}
where we are using $\alpha/g(r)\geq 1$ and implicitly redefining the constant $\hat{\delta}$ by absorbing a positive factor. \textit{Note that our definitions of $\Omega^{\pm}_{\varepsilon}(E)$ is motivated by this last estimate on $\Omega^+_{\varepsilon}(E)$}. In particular, our definitions allow to effectively absorb the factor $g(r)$. The constant $\hat{\delta}$ continuously depends on $\delta$ and becomes small when $\delta$ is small.

To conclude the proof, we choose $\varepsilon\leq 1/2$ and $\hat{\delta}\leq \varepsilon^2/2$, which can be achieved by choosing $\delta$ sufficiently small.

\end{proof}

We now define the set
\begin{equation*}
\Omega_{\delta}:=\left\lbrace x\in\Omega:\text{dist}(x,\partial\Omega)\leq \delta \right\rbrace \, .
\end{equation*}
The inequality of Claim \ref{intermed_en_est_ring} gives control on each term appearing on the left hand side (since they are both positive), i.e.
\begin{gather}
\int_{\Omega}h^2\left(\left\lvert\frac{\partial}{\partial r_*}(e^{\phi_{E,\varepsilon}/h}u)\right\rvert^2 +\left\lvert\frac{\partial}{\partial \theta_*}(e^{\phi_{E,\varepsilon}/h}u)\right\rvert^2\right)dr_*d\theta_* \leq C (\kappa^2+\frac{1}{2}\varepsilon)e^{2a_{E}(\varepsilon)/h}\left\lVert u\right\rVert^2_{L^2(\Omega)} \label{first_ineq_ring} \\
\frac{1}{2}\varepsilon^2\int_{\Omega^+_{\varepsilon}(E)}e^{2\phi_{E,\varepsilon}/h}|u|^2 dr_*d\theta_* \leq C(\kappa^2+\frac{1}{2}\varepsilon)e^{2a_{E}(\varepsilon)/h}\left\lVert u\right\rVert^2_{L^2(\Omega)} \, . \label{second_ineq_ring}
\end{gather} 
Inequality \eqref{second_ineq_ring} implies that, for any sufficiently small constants $\delta, \delta^{\prime}>0$, we have
\begin{equation*}
\int_{\Omega_{\delta}}|u|^2 dr_*d\theta_* \leq C e^{-C/h}\left\lVert u\right\rVert^2_{L^2(\Omega)}
\end{equation*}
for all $\kappa^2\in [E^2-\delta^{\prime},E^2+\delta^{\prime}]$ and some \textit{positive} constant $C>0$ \textit{independent} of $h$. To see this, remember that $E$ is fixed as in Proposition \ref{Omega_black_ring} and note that there exists a constant $c>0$ such that $\phi_{E,\varepsilon}\geq c$ for any $x\in \Omega_{\delta}$ and $\kappa^2\in [E^2-\delta^{\prime},E^2+\delta^{\prime}]$, with $c$ \textit{uniform} in $\varepsilon$ (this is simply by definition of $\phi_{E,\varepsilon}$). Furthermore, we claim that there exists $\varepsilon>0$ such that $a_E(\varepsilon)\leq c/2$ for any $h$ sufficiently small, which gives a negative exponent on the right hand side. In fact, $a_E(\varepsilon)\rightarrow 0$ as $\varepsilon \rightarrow 0$ uniformly in $h$ when $h$ is sufficiently small. The inequality follows after choosing $\delta$ small enough such that $\Omega_{\delta}\subset \Omega^+_{\varepsilon}(E)$.

By the same argument, combined with Young's inequality, inequality \eqref{first_ineq_ring} gives 
\begin{equation*}
\int_{\Omega_{\delta}}\left(\left\lvert\frac{\partial u}{\partial r_*}\right\rvert^2 +\left\lvert\frac{\partial u}{\partial \theta_*}\right\rvert^2\right)dr_*d\theta_* \leq C h^{-2} e^{-C/h}\left\lVert u\right\rVert^2_{L^2(\Omega)} \, .
\end{equation*}
For $h$ sufficiently small, one can absorb the $h^{-2}$ factor, i.e. there exists a constant $C>0$ independent of $h$ such that 
\begin{equation*}
\int_{\Omega_{\delta}}\left(\left\lvert\frac{\partial u}{\partial r_*}\right\rvert^2 +\left\lvert\frac{\partial u}{\partial \theta_*}\right\rvert^2\right)dr_*d\theta_* \leq C  e^{-C/h}\left\lVert u\right\rVert^2_{L^2(\Omega)} \, .
\end{equation*}
Putting everything together, we get the key estimate
\begin{equation} \label{key_estimate_ring}
\int_{\Omega_{\delta}}\left(\left\lvert\frac{\partial u}{\partial r_*}\right\rvert^2 +\left\lvert\frac{\partial u}{\partial \theta_*}\right\rvert^2+|u|^2\right)dr_*d\theta_* \leq C  e^{-C/h}\left\lVert u\right\rVert^2_{L^2(\Omega)} \, .
\end{equation}

In the following statement, we sum up what we have obtained so far, combining Theorem \ref{eigenv_black_ring_final} with estimate \eqref{key_estimate_ring}.

\begin{theorem}[adapted from \cite{KeirLogLog} Lemma 4.5]  \label{key_theorem_ring}
Consider the assumptions of Theorem \ref{eigenv_black_ring_final} and let $\delta^{\prime}>0$ be a sufficiently small constant such that $\delta^{\prime}<(\mathcal{E}^2-E^2)/10$. Then, for any sufficiently small constant $\delta>0$, there exists a constant $M>0$ and a sequence of solutions $\left\lbrace u_m \right\rbrace _{m\geq M}^{\infty}$ to the black ring eigenvalue problem such that the associated eigenvalues $\kappa_m$ satisfy $\kappa^2_m\in [C^{\prime}_{r_0,R},E^2+\delta^{\prime}]$ and
\begin{equation*}
\int_{\Omega_{\delta}}\left(\left\lvert\frac{\partial u_m}{\partial r_*}\right\rvert^2 +\left\lvert\frac{\partial u_m}{\partial \theta_*}\right\rvert^2+|u_m|^2\right)dr_*d\theta_* \leq C  e^{-C m}\left\lVert u_m\right\rVert^2_{L^2(\Omega)}
\end{equation*}
for some constants $C,C^{\prime}_{r_0,R}>0$ independent of $m$. 
\end{theorem}

In the next sections we aim to apply this result to the quasimode construction, which ultimately leads to the proof of Theorem \ref{Main_th_formal}.

\subsection{Construction of quasimodes}

Define a smooth, real valued cut-off function $\chi : \mathcal{D}\rightarrow \mathbb{R}$ as
\begin{equation*}
\chi(r_*,\theta_*) =
\begin{cases}
1 \quad\quad \text{if $(r_*,\theta_*)\in\Omega\setminus\Omega_{\delta}$}   \\
0 \quad\quad \text{if $(r_*,\theta_*) \notin\Omega$}
 \end{cases} \, .
\end{equation*}
The quasimodes are defined as functions $\Psi_m : \mathcal{D}\rightarrow \mathbb{C}$ such that
\begin{equation}  \label{def_quasimodes_ring_constr_qm}
\Psi_m(t,r_*,\theta_*,\phi,\psi):=e^{-i\,\omega_m\,t}e^{i(m\,\phi+\hat{J}\,\psi)}\chi(r_*,\theta_*)\, \left[\left(-\det g_{\textup{ring}}\right)g_{\textup{ring}}^{rr} \, g_{\textup{ring}}^{\theta\theta}\right]^{-\frac{1}{4}}  \, u_m(r_*,\theta_*)
\end{equation}
with $\omega_m$, $m$, $\hat{J}$ and $u_m(r_*,\theta_*)$ as in Theorem \ref{key_theorem_ring} and $g_{\textup{ring}}\in\mathfrak{g}$ in form \eqref{ring_alt_coord}. Quasimodes do not solve the wave equation on the whole domain of outer communication $\mathcal{D}$, but the error is exponentially small for high frequency $m$ and supported on a bounded region for each fixed time $t$. This is proven by the following lemma.

\begin{lemma}   \label{est_error_quasimodes_ring}
Consider the quasimodes $\Psi_m$ defined in \eqref{def_quasimodes_ring_constr_qm}, which satisfy
\begin{equation*}
\Box_{g_{\textup{ring}}}\Psi_m=\mathfrak{Err}_m(\Psi_m) \, ,
\end{equation*}
where $\mathfrak{Err}_m(\Psi_m)$ is the error. Then, for $m$ sufficiently large, we have the following estimate 
\begin{equation*}
\left\lVert \Box_{g_{\textup{ring}}}\Psi_m \right\rVert_{H^k(\Sigma_{t^*})}\leq C_k e^{-C_k m}\left\lVert \Psi_m\right\rVert_{L^2(\Sigma_0)} \, ,
\end{equation*}
with constant $C_k>0$ independent of $m$. Furthermore, the error is supported on $\Omega_{\delta}$.
\end{lemma}

\begin{remark}
Note the abuse of notation in Lemma \ref{est_error_quasimodes_ring} and in what follows, where $\Omega_{\delta}\equiv \Omega_{\delta}\times [0,\infty) \times [0,\Delta \phi) \times [0,\Delta \psi)$ and $\Omega\equiv \Omega\times [0,\infty) \times [0,\Delta \phi) \times [0,\Delta \psi)$, with $[0,\infty)$ time domain and $\Delta \phi,\Delta \psi$ the periods of the angular coordinates $\phi,\psi$. 
\end{remark}

\begin{proof}
The part of the statement concerning the support of the error immediately follows from the definition of the quasimodes $\Psi_m$. To prove the inequality, first note that given functions $\chi$ and $f$, we have
\begin{equation*}
\Box_{g_{\textup{ring}}}(\chi\, f)=\chi(\Box_{g_{\textup{ring}}} f) +2g_{\textup{ring}}^{\mu\nu}(\partial_{\mu}\chi)(\partial_{\nu}f)+f(\Box_{g_{\textup{ring}}}\chi) \, .
\end{equation*}
Since 
\begin{equation*}
\Box_{g_{\textup{ring}}}\left(e^{-i\,\omega_m\,t}e^{i(m\,\phi+\hat{J}\,\psi)}\, \left[\left(-\det g_{\textup{ring}}\right)g_{\textup{ring}}^{rr} \, g_{\textup{ring}}^{\theta\theta}\right]^{-\frac{1}{4}}\, u_m(r_*,\theta_*)\right)=0
\end{equation*}
on $\Omega$ when $u_m$ is solution to the black ring eigenvalue problem, using the formula above we can deduce 
\begin{align*}
\left\lVert \Box_{g_{\textup{ring}}}\Psi_m \right\rVert_{L^2(\Sigma_{t^*}\cap\Omega_{\delta})} &\lesssim \left\lVert u_m \right\rVert_{H^1(\Sigma_{t^*}\cap\Omega_{\delta})} \\
&\lesssim  \left\lVert u_m \right\rVert_{H^1(\Sigma_0\cap\Omega_{\delta})} \, ,
\end{align*}
where we used the fact that $\chi$ is a smooth function and therefore can be controlled in $L^{\infty}$ on $\Omega$, together with all its derivatives. The second inequality holds because the $H^1$ norm of $u_m$ is constant in time. Note that the $H^1$ norm of $u_m$ involves the $L^2$ norm of $u_m$ and its first derivatives only in the variables $r_*$ and $\theta_*$, so, for $m$ sufficiently large, Theorem \ref{key_theorem_ring} gives
\begin{equation*}
\left\lVert \Box_{g_{\textup{ring}}}\Psi_m \right\rVert_{L^2(\Sigma_{t^*}\cap\Omega_{\delta})} \leq Ce^{-C\, m}\left\lVert \Psi_m\right\rVert_{L^2(\Sigma_0\cap\Omega)} \, ,
\end{equation*}
with constant $C>0$ independent of $m$. In view of the spatial localization of the quasimodes and of the error, the inequality can be rewritten as
\begin{equation*}
\left\lVert \Box_{g_{\textup{ring}}}\Psi_m \right\rVert_{L^2(\Sigma_{t^*})} \leq Ce^{-C\, m}\left\lVert \Psi_m\right\rVert_{L^2(\Sigma_0)} \, .
\end{equation*}
To estimate the $H^k$ norm of $\Box_{g_{\textup{ring}}}\Psi_m$, one commutes the wave equation and controls second derivatives of $u_m$ in $r_*$ and $\theta_*$ using the equation satisfied by $u_m$. Higher order derivatives in $t$, $\phi$ and $\psi$ are controlled by the $L^2$ norm of $u_m$.
      
\end{proof}

\subsection{Lower bound for the uniform energy decay rate}

Define by $\Psi^H_m$ the solution to the homogeneous wave equation
\begin{equation}   \label{hom_lower_bound_ring}
\begin{cases}
\Box_{g_{\textup{ring}}}\Psi^H_m=0  \\
\Psi^H_m(0)=\Psi_m(0)  \\
\partial_t \Psi^H_m(0)=\partial_t\Psi_m(0)
\end{cases} \, ,
\end{equation} 
where the $\Psi_m(0)$ are the quasimodes defined in \eqref{def_quasimodes_ring_constr_qm} at time $t=0$. 

By construction, the quasimodes $\Psi_m$ satisfy
\begin{equation*}
\Box_{g_{\textup{ring}}}\Psi_m=\mathfrak{Err}_m(\Psi_m)
\end{equation*}
with the same initial data of \eqref{hom_lower_bound_ring}, where $\mathfrak{Err}_m$ is the error that we estimated in Lemma \ref{est_error_quasimodes_ring}. By Duhamel's formula for curved spacetimes (see Proposition 6.4 in \cite{MicroKeir}), we can write $\Psi_m$ at time $t$ as
\begin{equation} \label{Duhamel_formula_ring}
\Psi_m (t)=\Psi^H_m (t)+\int _0^t \xi (t,s)\, ds
\end{equation}
with 
\begin{equation}  \label{Duhamel_inhom_term_ring}
\begin{cases}
\Box_{g_{\textup{ring}}}\xi=0  \\
\xi (s)=0  \\
\partial_t \xi (s)= \left(\frac{1}{g_{\textup{ring}}^{tt}}\mathfrak{Err}_m(\Psi_m)\right)(s)
\end{cases} \, ,
\end{equation}
where $\xi (s)$ is $\xi (t=s)$.

We now consider the local energy $\mathcal{E}^N_{\Omega}[\Psi](t)$, as defined in Section \ref{sect_energy_currents}. By \eqref{Duhamel_formula_ring}, we have
\begin{equation*}
\left( \mathcal{E}^N_{\Omega}\left[\Psi_m-\Psi^H_m\right](t) \right)^{\frac{1}{2}} \leq t \sup_{s\in [0,t]} \left( \mathcal{E}^N_{\Omega}[\xi](t) \right)^{\frac{1}{2}} \, .
\end{equation*}
\ul{Using the assumption of uniform boundedness for solutions to the wave equation}, we can bound the energy of the solution to \eqref{Duhamel_inhom_term_ring} as follows
\begin{equation*}
\left( \mathcal{E}^N_{\Omega}\left[\Psi_m-\Psi^H_m\right](t) \right)^{\frac{1}{2}} \leq C t  \left( \mathcal{E}^N_{\Omega}[\xi](0) \right)^{\frac{1}{2}}
\end{equation*}
for some constant $C>0$ independent of time. Since 
\begin{equation*}
\left( \mathcal{E}^N_{\Omega}[\xi](0) \right)^{\frac{1}{2}} \sim \left\lVert  \frac{1}{g_{\textup{ring}}^{tt}}\mathfrak{Err}_m(\Psi_m)  \right\rVert _{L^2(\Omega)}(0) \, , 
\end{equation*}
we have
\begin{align*}
\left( \mathcal{E}^N_{\Omega}\left[\Psi_m-\Psi^H_m\right](t) \right)^{\frac{1}{2}} &\leq C t \left\lVert  \frac{1}{g_{\textup{ring}}^{tt}}\mathfrak{Err}_m(\Psi_m)  \right\rVert _{L^2(\Omega)}(0)  \\
&\leq C t e^{-C\, m} \left\lVert  \Psi_m  \right\rVert _{L^2(\Omega)}(0) \\
&\leq C t e^{-C\, m}  \left( \mathcal{E}^N_{\Omega}\left[\Psi_m\right](0) \right)^{\frac{1}{2}}
\end{align*}
for $m$ sufficiently large, where we have used Lemma \ref{est_error_quasimodes_ring} and Poincar\'{e} inequality for the second and third inequality respectively. 

Note that $\left( \mathcal{E}^N_{\Omega}\left[\Psi_m\right](t) \right)^{\frac{1}{2}}=\left( \mathcal{E}^N_{\Omega}\left[\Psi_m\right](0) \right)^{\frac{1}{2}}$ for the quasimodes. Therefore, for any time 
\begin{equation*}
0< t \leq \frac{e^{C\, m}}{2C} \, ,
\end{equation*} 
we have
\begin{equation*}
\left( \mathcal{E}^N_{\Omega}\left[\Psi^H_m\right](t) \right)^{\frac{1}{2}} \geq \frac{1}{2} \left( \mathcal{E}^N_{\Omega}\left[\Psi_m\right](0) \right)^{\frac{1}{2}}
\end{equation*}
by reverse triangle inequality. Since the $\Psi_m$ are localised in space, the local energy $\mathcal{E}^N_{\Omega}[\Psi_m]$ is equal to the total energy $\mathcal{E}^N[\Psi_m]$, which gives
\begin{align*}
\left( \mathcal{E}^N_{\Omega}\left[\Psi^H_m\right](t) \right)^{\frac{1}{2}} &\geq \frac{1}{2} \left( \mathcal{E}^N\left[\Psi_m\right](0) \right)^{\frac{1}{2}} \\
&\geq  \frac{C}{m}  \left( \mathcal{E}^N_2\left[\Psi_m\right](0) \right)^{\frac{1}{2}} 
\end{align*}
for $m$ sufficiently large, where $\mathcal{E}^N_2[\Psi_m]$ is the second order energy defined in Section \ref{sect_energy_currents}. For the last inequality we made use of the localisation in frequency of $\Psi_m$ and exchanged derivatives in $r_*$ and $\theta_*$ for derivatives in the other variables via the wave equation. Using $\Psi^H_m(0)=\Psi_m(0)$ from system \eqref{hom_lower_bound_ring}, we conclude
\begin{equation}  \label{final_ineq_lower_bound_ring}
\left( \mathcal{E}^N_{\Omega}\left[\Psi^H_m\right](t) \right)^{\frac{1}{2}} \geq  \frac{C}{m}  \left( \mathcal{E}^N_2\left[\Psi^H_m\right](0) \right)^{\frac{1}{2}}
\end{equation}
for any time $0< t_m\leq e^{C\, m}/2C$. By controlling higher order energies $\mathcal{E}^N_{k>2}\left[\Psi^H_m\right](0)$ on the right hand side of \eqref{final_ineq_lower_bound_ring}, one gains powers of $1/m$.

Inequality \eqref{final_ineq_lower_bound_ring} shows that a sequence $\left\lbrace t_m,\Psi^H_m\right\rbrace_{m\geq M}^{\infty}$, where $\Psi^H_m$ is a solution to the homogeneous wave equation with initial data prescribed as in \eqref{hom_lower_bound_ring} and $M>0$ a sufficiently large constant, proves Theorem \ref{Main_th_formal}.

\bibliography{report} 
\bibliographystyle{hsiam}
\end{document}